\documentclass[
aps,
prd,
twocolumn,
eqsecnum,
floatfix,
superscriptaddress,
showpacs
]{revtex4-1} 

\usepackage{graphicx}
\usepackage{dcolumn}
\usepackage{amsmath}
\usepackage{amssymb}
\usepackage{xspace}
\usepackage{color}
\usepackage{subfigure}
\usepackage{braket}
\usepackage{cleveref}
\usepackage[section]{placeins}

\newcommand{\val}[2]{\ensuremath{#1 \; \mathrm{#2}\xspace}}
\newcommand{\sci}[2]{\ensuremath{#1 \times 10^{#2}}\xspace}

\crefname{equation}{Eq.}{Eqs.}
\Crefname{equation}{Equation}{Equations}
\crefname{figure}{Fig.}{Figs.}
\Crefname{figure}{Figure}{Figures}
\crefname{table}{Tab.}{Tabs.}
\Crefname{table}{Table}{Tables}
\crefname{section}{Section}{Sections}
\Crefname{section}{Section}{Sections}
\crefname{asec}{Appendix}{Appendices}
\Crefname{asec}{Appendix}{Appendices}

\newcommand{\superk}    {Super-Kamiokande\xspace}       
\newcommand{\sk}{\mbox{Super-K}\xspace}
\newcommand{\cz}{\ensuremath{\cos \theta_z}\xspace}
\newcommand{\UP}{UP-$\mu$\xspace}

\newcommand{\numu}{\ensuremath{\nu_{\mu}}\xspace}
\newcommand{\nubar}{\ensuremath{\bar{\nu}_{\mu}}\xspace}
\newcommand{\numubar}{\nubar}
\newcommand{\nue}{\ensuremath{\nu_{e}}\xspace}
\newcommand{\nuebar}{\ensuremath{\bar{\nu}_{e}}\xspace}
\newcommand{\nutau}{\ensuremath{\nu_{\tau}}\xspace}

\newcommand{\nus} {\ensuremath{\nu_{s}}\xspace}

\newcommand{\pizero}{\ensuremath{\pi^0}\xspace}

\newcommand{\atom}[2]{\ensuremath{{}^{#1}\textrm{#2}}\xspace}

\newcommand{\dm}{\ensuremath{\Delta m^{2}}\xspace}
\newcommand{\dmsq}[1]{\ensuremath{\dm_{#1}}\xspace}
\newcommand{\dmnew}{\dm}
\renewcommand{\th}[1]{\ensuremath{\theta_{#1}}\xspace}
\newcommand{\sn}[1]{\ensuremath{ \sin^{2}(\theta_{#1}) }\xspace }
\newcommand{\snt}[1]{\ensuremath{ \sin^{2}(2\theta_{#1}) }\xspace }

\newcommand{\dmu}{\ensuremath{d_{\mu}}\xspace}

\newcommand{\Pab}{\ensuremath{P_{\alpha \beta}}\xspace}
\newcommand{\Pee}{\ensuremath{P_{e e}}\xspace}
\newcommand{\Pmm}{\ensuremath{P_{\mu \mu}}\xspace}
\newcommand{\Pem}{\ensuremath{P_{e \mu}}\xspace}
\newcommand{\Pme}{\ensuremath{P_{\mu e}}\xspace}
\newcommand{\Pmt}{\ensuremath{P_{\mu \tau}}\xspace}

\newcommand{\Ptab}{\ensuremath{\tilde{P}_{\alpha \beta}}\xspace}
\newcommand{\Ptee}{\ensuremath{\tilde{P}_{e e}}\xspace}
\newcommand{\Ptmm}{\ensuremath{\tilde{P}_{\mu \mu}}\xspace}
\newcommand{\Ptem}{\ensuremath{\tilde{P}_{e \mu}}\xspace}
\newcommand{\Ptme}{\ensuremath{\tilde{P}_{\mu e}}\xspace}

\newcommand{\ue}{\ensuremath{U_{e4}}\xspace}
\newcommand{\um}{\ensuremath{U_{\mu4}}\xspace}
\newcommand{\ut}{\ensuremath{U_{\tau4}}\xspace}
\newcommand{\us}{\ensuremath{U_{s4}}\xspace}

\newcommand{\uesq}{\ensuremath{|\ue|^2}\xspace}
\newcommand{\umsq}{\ensuremath{|\um|^2}\xspace}
\newcommand{\utsq}{\ensuremath{|\ut|^2}\xspace}
\newcommand{\ussq}{\ensuremath{|\us|^2}\xspace}

\newcommand{\Upmns}{\ensuremath{U_{\rm{PMNS}}}\xspace}
\newcommand{\Ut}{\ensuremath{\tilde{U}\xspace}}

\newcommand{\dmatm}{\val{\sci{2.51}{-3}}{eV^{2}}}
\newcommand{\dmsol}{\val{\sci{7.46}{-5}}{eV^{2}}}
\newcommand{\sonetwo}{0.305}

\newcommand{\stonethree}{0.095}
\newcommand{\stwothree}{0.514}

\newcommand{\degrees}{\ensuremath{^{\circ}}\xspace}
\newcommand{\chisq}{\ensuremath{\chi^{2}}\xspace}
\newcommand{\s}{s}
\newcommand{\diag}[1]{\ensuremath{\textrm{diag}\left(#1\right)}\xspace}

\newcommand{\fitTwo}{no-\nue}

\newcommand{\fitOne}{sterile vacuum\xspace}
\newcommand{\FitOne}{Sterile vacuum\xspace}

\newcommand{\umbest}   {0.016\xspace}
\newcommand{\umlimit}  {0.041\xspace} 
\newcommand{\umlimitnn}{0.054\xspace} 
\newcommand{\utlimit}  {0.18\xspace}
\newcommand{\utlimitnn}{0.23\xspace}

\newcommand{\fwid}{8cm}
\newcommand{\zwid}{5cm}
\renewcommand{\arraystretch}{1.2}
\renewcommand{\tabcolsep}{4pt}

\graphicspath{ {fig/} }

\begin{document}
\title{Limits on Sterile Neutrino Mixing using Atmospheric Neutrinos in Super-Kamiokande}
\newcommand{\AFFicrr}{\affiliation{Kamioka Observatory, Institute for Cosmic Ray Research, University of Tokyo, Kamioka, Gifu 506-1205, Japan}}
\newcommand{\AFFkashiwa}{\affiliation{Research Center for Cosmic Neutrinos, Institute for Cosmic Ray Research, University of Tokyo, Kashiwa, Chiba 277-8582, Japan}}
\newcommand{\AFFipmu}{\affiliation{Kavli Institute for the Physics and
Mathematics of the Universe (WPI), Todai Institutes for Advanced Study,
University of Tokyo, Kashiwa, Chiba 277-8582, Japan }}
\newcommand{\AFFmad}{\affiliation{Department of Theoretical Physics, University Autonoma Madrid, 28049 Madrid, Spain}}
\newcommand{\AFFubc}{\affiliation{Department of Physics and Astronomy, University of British Columbia, Vancouver, BC, V6T1Z4, Canada}}
\newcommand{\AFFbu}{\affiliation{Department of Physics, Boston University, Boston, MA 02215, USA}}
\newcommand{\AFFbnl}{\affiliation{Physics Department, Brookhaven National Laboratory, Upton, NY 11973, USA}}
\newcommand{\AFFuci}{\affiliation{Department of Physics and Astronomy, University of California, Irvine, Irvine, CA 92697-4575, USA }}
\newcommand{\AFFcsu}{\affiliation{Department of Physics, California State University, Dominguez Hills, Carson, CA 90747, USA}}
\newcommand{\AFFcnm}{\affiliation{Department of Physics, Chonnam National University, Kwangju 500-757, Korea}}
\newcommand{\AFFduke}{\affiliation{Department of Physics, Duke University, Durham NC 27708, USA}}
\newcommand{\AFFfukuoka}{\affiliation{Junior College, Fukuoka Institute of Technology, Fukuoka, Fukuoka 811-0295, Japan}}
\newcommand{\AFFgmu}{\affiliation{Department of Physics, George Mason University, Fairfax, VA 22030, USA }}
\newcommand{\AFFgifu}{\affiliation{Department of Physics, Gifu University, Gifu, Gifu 501-1193, Japan}}
\newcommand{\AFFgist}{\affiliation{GIST College, Gwangju Institute of Science and Technology, Gwangju 500-712, Korea}}
\newcommand{\AFFuh}{\affiliation{Department of Physics and Astronomy, University of Hawaii, Honolulu, HI 96822, USA}}
\newcommand{\AFFkanagawa}{\affiliation{Physics Division, Department of Engineering, Kanagawa University, Kanagawa, Yokohama 221-8686, Japan}}
\newcommand{\AFFkek}{\affiliation{High Energy Accelerator Research Organization (KEK), Tsukuba, Ibaraki 305-0801, Japan }}
\newcommand{\AFFkobe}{\affiliation{Department of Physics, Kobe University, Kobe, Hyogo 657-8501, Japan}}
\newcommand{\AFFkyoto}{\affiliation{Department of Physics, Kyoto University, Kyoto, Kyoto 606-8502, Japan}}
\newcommand{\AFFumd}{\affiliation{Department of Physics, University of Maryland, College Park, MD 20742, USA }}
\newcommand{\AFFmit}{\affiliation{Department of Physics, Massachusetts Institute of Technology, Cambridge, MA 02139, USA}}
\newcommand{\AFFmiyagi}{\affiliation{Department of Physics, Miyagi University of Education, Sendai, Miyagi 980-0845, Japan}}
\newcommand{\AFFnagoya}{\affiliation{Solar Terrestrial Environment Laboratory, Nagoya University, Nagoya, Aichi 464-8602, Japan}}
\newcommand{\AFFpol}{\affiliation{National Centre For Nuclear Research, 00-681 Warsaw, Poland}}
\newcommand{\AFFsuny}{\affiliation{Department of Physics and Astronomy, State University of New York at Stony Brook, NY 11794-3800, USA}}
\newcommand{\AFFniigata}{\affiliation{Department of Physics, Niigata University, Niigata, Niigata 950-2181, Japan }}
\newcommand{\AFFokayama}{\affiliation{Department of Physics, Okayama University, Okayama, Okayama 700-8530, Japan }}
\newcommand{\AFFosaka}{\affiliation{Department of Physics, Osaka University, Toyonaka, Osaka 560-0043, Japan}}
\newcommand{\AFFregina}{\affiliation{Department of Physics, University of Regina, 3737 Wascana Parkway, Regina, SK, S4SOA2, Canada}}
\newcommand{\AFFseoul}{\affiliation{Department of Physics, Seoul National University, Seoul 151-742, Korea}}
\newcommand{\AFFshizuokasc}{\affiliation{Department of Informatics in
Social Welfare, Shizuoka University of Welfare, Yaizu, Shizuoka, 425-8611, Japan}}
\newcommand{\AFFskk}{\affiliation{Department of Physics, Sungkyunkwan University, Suwon 440-746, Korea}}
\newcommand{\AFFtohoku}{\affiliation{Research Center for Neutrino Science, Tohoku University, Sendai, Miyagi 980-8578, Japan}}
\newcommand{\AFFtokyo}{\affiliation{The University of Tokyo, Bunkyo, Tokyo 113-0033, Japan }}
\newcommand{\AFFtorront}{\affiliation{Department of Physics, University of Torront, 60 St., Torront, Ontario, M5S1A7, Canada }}
\newcommand{\AFFtriumf}{\affiliation{TRIUMF, 4004 Wesbrook Mall, Vancouver, BC, V6T2A3, Canada }}
\newcommand{\AFFtokai}{\affiliation{Department of Physics, Tokai University, Hiratsuka, Kanagawa 259-1292, Japan}}
\newcommand{\AFFtit}{\affiliation{Department of Physics, Tokyo Institute
for Technology, Meguro, Tokyo 152-8551, Japan }}
\newcommand{\AFFtsinghua}{\affiliation{Department of Engineering Physics, Tsinghua University, Beijing, 100084, China}}
\newcommand{\AFFwarsaw}{\affiliation{Institute of Experimental Physics, Warsaw University, 00-681 Warsaw, Poland }}
\newcommand{\AFFuw}{\affiliation{Department of Physics, University of Washington, Seattle, WA 98195-1560, USA}}

\AFFicrr
\AFFkashiwa
\AFFmad
\AFFbu
\AFFubc
\AFFbnl
\AFFuci
\AFFcsu
\AFFcnm
\AFFduke
\AFFfukuoka
\AFFgifu
\AFFgist
\AFFuh
\AFFkek
\AFFkobe
\AFFkyoto
\AFFmiyagi
\AFFnagoya
\AFFsuny
\AFFokayama
\AFFosaka
\AFFregina
\AFFseoul
\AFFshizuokasc
\AFFskk
\AFFtokai
\AFFtokyo
\AFFipmu
\AFFtorront
\AFFtriumf
\AFFtsinghua
\AFFuw

\author{K.~Abe}
\AFFicrr
\AFFipmu
\author{Y.~Haga}
\AFFicrr
\author{Y.~Hayato}
\AFFicrr
\AFFipmu
\author{M.~Ikeda}
\AFFicrr
\author{K.~Iyogi}
\AFFicrr 
\author{J.~Kameda}
\author{Y.~Kishimoto}
\author{M.~Miura} 
\author{S.~Moriyama} 
\author{M.~Nakahata}
\AFFicrr
\AFFipmu 
% ...

\collaboration{The Super-Kamiokande Collaboration}
\noaffiliation

\date{\today}

\begin{abstract}
We present limits on sterile neutrino mixing using 4,438 live-days of atmospheric neutrino data from the Super-Kamiokande experiment.  We search for fast oscillations driven by an eV$^2$-scale mass splitting and for oscillations into sterile neutrinos instead of tau neutrinos at the atmospheric mass splitting. When performing both of these searches we assume that the sterile mass splitting is large, allowing $\sin^2(\Delta m^2 L/4E)$ to be approximated as $0.5$, and we assume that there is no mixing between electron neutrinos and sterile neutrinos ($|U_{e4}|^2 = 0$).  No evidence of sterile oscillations is seen and we limit $|U_{\mu4}|^2$ to less than 0.041 and $|U_{\tau4}|^2$ to less than 0.18 for $\Delta m^2 > 0.1$ eV$^2$ at the 90\% C.L. in a 3+1 framework.  The approximations that can be made with atmospheric neutrinos allow these limits to be easily applied to 3+N models, and we provide our results in a generic format to allow comparisons with other sterile neutrino models.
\end{abstract}


\maketitle

\section{Introduction}

The flavor oscillations of massive neutrinos have been well established by a wide range of experiments looking at the disappearance of neutrinos produced in the atmosphere~\cite{Fukuda:1998mi,Ashie:2004mr}, in the Sun~\cite{Abe:2010hy,Cleveland:1998nv, Abdurashitov:2009tn, Altmann:2005ix, Hampel:1998xg, Aharmim:2011vm}, in nuclear reactors~\cite{Abe:2008aa, An:2012eh, Abe:2013sxa, Ahn:2012nd}, and at particle accelerators~\cite{Ahn:2006zza, Adamson:2013whj, Abe:2014ugx} where recently the appearance of electron and tau neutrinos were observed in primarily muon neutrino samples~\cite{Abe:2012jj, Abe:2013xua, Agafonova:2013dtp}.   The evidence from these experiments suggest two independent neutrino mass differences, an `atmospheric' $\dm \approx \val{3 \times 10^{-3}}{eV^2}$ and a `solar' $\dm \approx \val{7 \times 10^{-5}}{eV^2}$, requiring that three neutrinos are participating in oscillations. Experiments at the Large Electron-Positron collider (LEP) also probed the number of neutrinos using the width of the $Z^0$ mass peak, which depends on the number of neutrino flavors into which a $Z^0$ can decay. A combined analysis of all the LEP data measured $2.980 \pm 0.0082$ light neutrino families~\cite{ALEPH:2005ab}.

However, not all neutrino experiments are consistent with this three-flavor picture; several hints of another, larger mass splitting have appeared. The LSND experiment observed \nuebar appearance in a \numubar beam consistent with two-flavor oscillations with $\dmnew \approx \val{1}{eV^2}$~\cite{Aguilar:2001ty}. 
The later MiniBooNE experiment saw some possible signs of $\numubar \to \nuebar$ as well as $\numu \to \nue$ oscillations at a similar \dmnew~\cite{Aguilar-Arevalo:2013pmq}. 
Additional anomalies appear in experiments looking at intense \nuebar and \nue sources at distances too short for standard oscillations:
a lower rate of \nuebar's than predicted was seen at several reactor experiments~\cite{Huber:2011wv,Mention:2011rk}, 
and the rate of \nue's from \atom{51}{Cr} and \atom{37}{Ar} calibration sources at Gallium-based solar neutrino experiments was $3\sigma$ lower than the expected rate~\cite{Anselmann:1994ar, Hampel:1997fc, Abdurashitov:1996dp, Abdurashitov:1998ne, Abdurashitov:2005tb}.  
Both of these hints are consistent with oscillations driven by a $\dmnew > \val{1}{eV^2}$~\cite{Kopp:2013vaa}.  In order for the interpretation of these measurements to coexist with the well-established solar and atmospheric mass-splittings, at least one additional neutrino must be introduced. The LEP measurements further require that any additional neutrinos either be heavier than half the $Z^0$ mass, which would make it difficult for them to participate in oscillations, or not couple to the $Z^0$ boson, and hence not participate in weak interactions. These non-interacting neutrinos are called `sterile.' 

Cosmological measurements are also sensitive to the number of neutrinos, albeit in a model-dependent way, by identifying the neutrinos as the relativistic species present in the early universe.  Recent measurements are generally consistent with an effective number of neutrinos a little above three, but not excluding four~\cite{Ade:2013zuv,Das:2013zf,Bennett:2012zja,Story:2012wx}.

If all of the hints and anomalies are interpreted as consistent evidence of a single additional sterile neutrino (called the `3+1' model), they require a \numu disappearance signal with a similar \dmnew, which has not been seen in short-baseline \numu disappearance experiments like CCFR~\cite{Stockdale:1984cg} or MiniBooNE and SciBooNE~\cite{Cheng:2012yy}, or in the long-baseline experiment, MINOS ~\cite{Adamson:2011ku}. Consequently, 3+1 models fit the combined global oscillation data poorly. Theories with additional sterile neutrinos (3+2, 3+3, 1+3+1) have been investigated without a clear consensus interpretation of the experimental data~\cite{Kopp:2013vaa,Conrad:2012qt,Giunti:2013waa}.

The \superk (\sk, SK) atmospheric data sample can provide a useful constraint on sterile neutrinos across a wide variety of proposed sterile neutrino models. The atmospheric neutrino sample covers a wide range in both energy, $E$, and distance traveled, $L$. The signatures of sterile neutrino oscillations in SK data are valid over a range of mass splittings relevant to previous hints and the limits set in the 3+1 framework can be readily extended to models with more than one sterile neutrino.

\section{The Super-Kamiokande Experiment}

\superk is a cylindrical, underground, water-Cherenkov detector, \val{41.4}{m} in height and \val{39.3}{m} in diameter. It is arranged into two optically-separated regions.  The inner detector (ID) is instrumented with 11,129 20-inch PMT's \footnote{There were 11,129 PMT's during SK-III and SK-IV.  During SK-I there were 11,146 PMT's and during SK-II there were 5,182 PMT's.} and an active-veto outer detector (OD) instrumented with 1,885 8-inch PMT's, both filled with ultra-pure water. A fiducial volume is defined \val{2}{m} from the walls of the ID and has a mass of \val{22.5}{kton}. 

Neutrinos are detected by observing the Cherenkov radiation from the highly relativistic charged particles produced in neutrino-nucleus interactions.
The charged particles must have a velocity greater than the speed of light in water, introducing a total energy threshold which depends on particle mass: \val{780}{keV} for electrons, \val{160}{MeV} for muons, and \val{212}{MeV} for charged pions. The particles radiate Cherenkov photons in a cone (42\degrees in water for particles with velocity close to $c$) as long as the particle is above threshold, producing a circular pattern of light which is projected onto the wall of the detector. Particles which stop inside the detector produce a ring while those that exit produce a filled circle. The timing of the Cherenkov light allows the vertex to be reconstructed, and the direction of travel of the particle is estimated from the vertex and the Cherenkov ring pattern. More energetic particles typically produce more total light. Particle types are identified based on the pattern of the hits making up the ring. Electrons and photons
produce electromagnetic showers which create many overlapping rings and appear as a single ring with a fuzzy edge. Non-showering particles (muons, pions, protons) produce concentric light cones as they travel and appear as a single ring with a sharp outer boundary. 

The neutrino oscillation probability depends on the initial neutrino flavor, the distance the neutrino travels, $L$, and the neutrino energy, $E$. We separate our data into samples with enhanced \numu or \nue flavor content and bin it using observables correlated with $L$ and $E$. 
Instead of distance, we bin the data in zenith angle, \cz, defined as the angle between the event direction and the downward vertical direction.  The neutrinos with the shortest path lengths are downward-going (\cz near 1) and the neutrinos with the longest path lengths are upward-going (\cz near -1).  The simulation which predicts the number of neutrino events in each bin includes a distribution of neutrino production heights based on a model of the atmosphere described in more detail in~\cite{Ashie:2005ik}.  This range of production heights introduces a smearing of the oscillation probability for a given zenith angle for downward-going and horizontal events but is negligible for upward-going events which cross most of the Earth.
For events with one visible ring, we bin in momentum and for multi-ring events we bin in visible energy,
defined as the energy of an electron that would produce the total amount of light observed in the detector. 

\section{Data Sample}
\label{sec:datasample}

\sk has had four run periods, summarized in \cref{tab:runperiods}, with a total exposure 4,438 live-days which are each considered separately in the simulation and analysis.  
The previous atmospheric neutrino oscillation paper~\cite{Wendell:2010md} included only the first three run periods.  The current period, SK-IV began with the installation of new front-end electronics (QTC Based Electronics with Ethernet, QBEE) whose key component is a new high-speed Charge-to-Time converter (QTC) ASIC~\cite{Nishino:2009zu}.  The SK-IV data continues to be accumulated, but this analysis includes only data taken until September, 2013.

\begin{figure}
  \includegraphics[width=\fwid,clip]{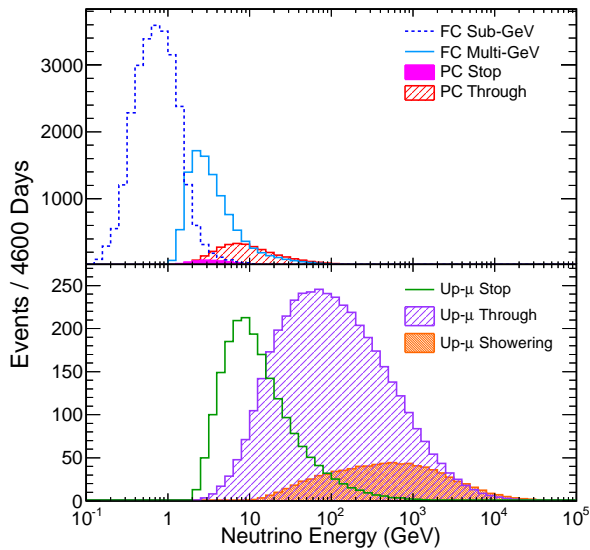}
  \caption{(color online) The true energy distribution from simulation without oscillations of the fully-contained (sub-GeV and multi-GeV), partially-contained (stopping and through-going), and up-going muon (stopping, through-going non-showering, and through-going showering) samples.}
  \label{fig:examples}
\end{figure}

There are three basic event topologies used in the atmospheric neutrino analysis which cover different neutrino energies (plotted in \cref{fig:examples}). The fully-contained (FC) sample includes events with vertices inside the fiducial volume and which stop before leaving the inner detector. It is the lowest-energy sample ranging from a few hundred MeV up to about \val{10}{GeV}. These events have the best energy resolution since all of the energy is contained within the detector. However, they also have the worst direction resolution (from 12\degrees to 100\degrees, depending on energy~\cite{Ishitsuka:2004,Dufour:2009}) since the outgoing lepton direction is less correlated with the incoming neutrino direction. In the oscillation analysis, the FC sample is divided into 13 subsamples, categorized based on visible energy into sub-GeV, below \val{1.33}{GeV}, and multi-GeV, above \val{1.33}{GeV}.  The FC subsamples are then binned in energy and \cz, though a few sub-GeV subsamples with particularly poor direction resolution have only a single \cz bin.  Details of which bins are used in which subsample are shown in \cref{tab:samples}. The sub-GeV events are categorized into $\mu$-like, $e$-like, and neutral-current $\pi^0$-like samples. The $\mu$- and $e$-like subsamples are further divided by number of decay electrons, which can signify the presence of a charged pion produced below Cherenkov threshold and thus help isolate NC backgrounds.  The Multi-GeV subsamples are split into $\mu$-like and $e$-like, with the $e$-like events divided into \nue-like and \nuebar-like. The FC sample selection techniques are described in greater detail in~\cite{Wendell:2010md}.

\begin{table}
  \centering
  \begin{tabular}{llrrc}
      \hline \hline
             &              & \multicolumn{2}{c}{Live-days}  & \multicolumn{1}{c}{Photo-}\\
             &              & \multicolumn{1}{c}{FC/PC} & \multicolumn{1}{c}{\UP}  & \multicolumn{1}{c}{coverage (\%)} \\
      \hline
      SK-I   & 1996--2001   & 1,489 & 1,646          & 40  \\
      SK-II  & 2002--2005   &   799 &   828          & 19  \\
      SK-III & 2006--2008   &   518 &   635          & 40  \\
      SK-IV  & 2008--2013   & 1,632 & 1,632          & 40  \\
      \hline\hline
  \end{tabular}
  \caption{Summary of the four SK data-taking periods. The photo-coverage was reduced during SK-II due to an accident in 2001. SK-IV data taking is continuing, but this analysis includes only data taken until September, 2013. The difference of livetimes between FC/PC and \UP is due to the insensitivity of the \UP reduction to noise such as ``flasher'' PMT's. Unlike the \UP reduction, the FC and PC reductions exclude data close in time to known flashing PMT's to avoid including fake events, reducing the total livetime for those samples.}
  \label{tab:runperiods}
\end{table}

\newcommand{\tsp}{\hspace{1em}}

\begin{table*}
\centering
\begin{tabular}{lll@{\tsp}rr@{\tsp}rr@{\tsp}rr@{\tsp}rr}
\hline\hline
   & & 
   & \multicolumn{2}{c}{SK-I} 
   & \multicolumn{2}{c}{SK-II} 
   & \multicolumn{2}{c}{SK-III} 
   & \multicolumn{2}{c}{SK-IV} \\
   & \multicolumn{1}{c}{Energy Bins} & \multicolumn{1}{c}{\cz Bins}   
   & \multicolumn{1}{c}{Data} & \multicolumn{1}{c}{MC}
   & \multicolumn{1}{c}{Data} & \multicolumn{1}{c}{MC}
   & \multicolumn{1}{c}{Data} & \multicolumn{1}{c}{MC} 
   & \multicolumn{1}{c}{Data} & \multicolumn{1}{c}{MC} \\
\hline
\multicolumn{3}{l}{\textbf{Fully Contained (FC) Sub-GeV}}  \\

\hspace*{4pt} e-like, Single-ring  \\
\hspace*{8pt} 0 decay-e     & 5 $e^\pm$ momentum   & 10 in $[-1, 1]$    & 2987 & 2975.2  & 1573 & 1549.1  & 1091 & 1052.2  & 3074 & 3126.0  \\
\hspace*{8pt} 1 decay-e     & 5 $e^\pm$ momentum   &                    &  301 &  310.5  &  172 &  170.3  &  118 &  108.8  &  402 &  333.8  \\

\hspace*{4pt} $\mu$-like, Single-ring \\
\hspace*{8pt} 0 decay-e     & 5 $\mu^\pm$ momentum & 10 in $[-1, 1]$    & 1025 &  974.1  &  561 &  534.5  &  336 &  338.1  &  583 &  592.8  \\
\hspace*{8pt} 1 decay-e     & 5 $\mu^\pm$ momentum & 10 in $[-1, 1]$    & 2012 & 2042.1  & 1037 & 1068.4  &  742 &  735.0  & 2767 & 2741.2  \\
\hspace*{8pt} 2 decay-e     & 5 $\mu^\pm$ momentum &                    &  147 &  145.4  &   86 &   76.7  &   61 &   60.7  &  245 &  255.0  \\

\hspace*{4pt} \pizero-like  & \\
\hspace*{8pt} Single-ring   & 5 $e^\pm$  momentum  &                    &  181 &  183.6  &  111 &  109.1  &   59 &   60.7  &  194 &  167.7  \\
\hspace*{8pt} Two-ring      & 5 \pizero  momentum  &                    &  493 &  492.4  &  251 &  265.8  &  171 &  175.3  &  548 &  546.3  \\
\\
\multicolumn{3}{l}{\textbf{Fully Contained (FC) Multi-GeV}} \\

\hspace*{4pt} Single-ring  \\
\hspace*{8pt} \nue-like     & 4 $e^\pm$ momentum   & 10 in $[-1, 1]$ &  191 &  170.3  &   79 &   82.4  &   68 &   59.8  &  238 &  221.3  \\
\hspace*{8pt} \nuebar-like  & 4 $e^\pm$ momentum   & 10 in $[-1, 1]$ &  665 &  664.4  &  317 &  338.2  &  206 &  230.3  &  626 &  641.3  \\
\hspace*{8pt} $\mu$-like    & 2 $\mu^\pm$ momentum & 10 in $[-1, 1]$ &  712 &  730.5  &  400 &  384.1  &  238 &  250.5  &  788 &  794.4  \\

\hspace*{4pt} Multi-ring \\
\hspace*{8pt} \nue-like     & 3 visible energy     & 10 in $[-1, 1]$ &  216 &  222.2  &  143 &  138.3  &   65 &   77.3  &  269 &  267.5  \\
\hspace*{8pt} \nuebar-like  & 3 visible energy     & 10 in $[-1, 1]$ &  227 &  224.3  &  134 &  132.4  &   80 &   76.9  &  275 &  264.8  \\
\hspace*{8pt} $\mu$-like    & 4 visible energy     & 10 in $[-1, 1]$ &  603 &  596.4  &  337 &  328.7  &  228 &  219.6  &  694 &  705.3  \\
\\

\multicolumn{3}{l}{\textbf{Partially Contained (PC)}} \\
\hspace*{4pt} Stopping      & 2 visible energy     & 10 in $[-1, 1]$ &  143 &  144.4  &   77 &   73.2  &   54 &   55.4  &  188 &  187.9  \\
\hspace*{4pt} Through-going & 4 visible energy     & 10 in $[-1, 1]$ &  759 &  777.3  &  350 &  370.1  &  290 &  306.0  &  919 &  948.4  \\
\\

\multicolumn{3}{l}{\textbf{Upward-going Muons (\UP)}} \\
\hspace*{4pt} Stopping      & 3 visible energy     & 10 in $[-1, 0]$ &  432 &  444.7  &  206 &  216.2  &  194 &  172.1  &  416 &  417.1  \\
\hspace*{4pt} Through-going \\
\hspace*{8pt} Non-showering &                      & 10 in $[-1, 0]$ & 1564 & 1532.4  &  726 &  741.4  &  613 &  569.5  & 1467 & 1435.8  \\
\hspace*{8pt} Showering     &                      & 10 in $[-1, 0]$ &  272 &  325.0  &  110 &  117.1  &  110 &  142.7  &  446 &  393.1  \\
\hline\hline
\end{tabular}
\caption{Summary of the atmospheric neutrino data and simulated event samples. The oscillated MC has been calculated assuming three-flavor mixing with $\dmsq{32} = \dmatm$, $\dmsq{21} = \dmsol$, $\sn{12} = \sonetwo$, $\snt{13} = \stonethree$, $\sn{23} = \stwothree$~\cite{Abe:2014ugx,PDG,Abe:2010hy}. Visible energy is defined as the energy of an electron required to produce all the Cherenkov light seen in the event. The distribution of 0-, 1-, and 2-decay electron $\mu$-like sub-samples changes significantly in SK-IV compared to earlier periods due to the improved decay-e tagging efficiency of the upgraded electronics. The fraction of \UP events classified as showering in the SK-IV data is large relative to SK-I due to the slow increase in the gain of the PMT's over time.
}
\label{tab:samples}
\end{table*}

The Partially contained (PC) sample contains events that have vertices in the fiducial volume, but produce leptons that leave the inner detector. They have long tracks and so are almost exclusively from \numu interactions and range in energy from a few GeV up to tens of GeV.  These events have better direction resolution (9\degrees-16\degrees~\cite{Ishitsuka:2004}) than FC events due to their higher energy, but worse energy resolution since the exiting muon carries some energy out of the detector.  They are divided into two subsamples based on their energy deposition in the OD: stopping, which stop in the outer detector, and through-going, which pass through the outer detector out into the rock~\cite{Ashie:2004mr}. They are binned in both visible energy (based on light observed in the ID) and \cz.  

Up-going muon (\UP) events contain muons that start in the surrounding rock and then enter and pass through the outer detector into the inner detector. This sub-sample also starts at a few GeV but extends up to hundreds of TeV. These events are only included if they are up-going, where the bulk of the Earth has shielded the detector from the otherwise overwhelming cosmic-ray muon background. They are split into the lower-energy stopping (stops in the inner detector) and the higher-energy through-going (exits out the far side of the detector) subsamples. The through-going events are further sub-divided into non-showering (minimum-ionizing) and showering subsamples based on the method described in~\cite{Desai:2007ra}. The critical energy at which the muon's energy loss by radiative processes (primarily pair production and bremsstrahlung) equals energy loss by ionization is \val{900}{GeV}~\cite{PDG} so evidence of showering allows us to select a sample with higher average energy despite an unknown fraction of the muon's energy being deposited in the rock before reaching and after leaving the detector. The \UP through-going subsamples are binned only in \cz since the measured energy is only a rough lower bound on the initial neutrino energy.


A summary of all the event samples used in this analysis, including the binning used, the number of observed events, and the number of events predicted by simulation, is shown in \cref{tab:samples}.

Several improvements to the simulation have been included since the last atmospheric neutrino oscillation publication~\cite{Wendell:2010md}. 
The neutrino interaction generator, NEUT~\cite{Hayato:2009zz}, includes an updated tau-neutrino cross section and a more accurate calculation of the NC elastic scattering cross section~\cite{Bradford:2006yz}.  This version also includes an improved model of photon emission from excited nuclei based on recent experimental data~\cite{Kobayashi:2006gb,Yosoi:2003} and improved spectroscopic factor simulation~\cite{Mori:2011un}.  The pion interaction model was also improved: interaction probabilities were tuned to existing pion scattering data~\cite{Lee:2002eq}, particularly at low momentum, $< \val{500}{MeV/c}$, while at higher momenta the model, including both interaction probabilities and kinematics, was updated to the SAID partial wave analysis of world data~\cite{Oh:1997eq,Workman:SAID,Arndt:2003if}.  In all energy regimes, nucleon ejection after pion absorption in the nucleus was implemented with multiplicity determined by data in~\cite{Rowntree:1999dp} and the kinematics of the two-body ejection modeled with the data-based parameterization in~\cite{Ritchie:1991mu}.  

The detector simulation includes a model of the new electronics and software triggers as well as an updated tuning of the PMT response in the ID. Improved models of the PMT geometry and reflective Tyvek surfaces, as well as tube-by-tube dark noise rates and saturation curves based on in-situ measurements have been implemented into the OD simulation. Additionally, low momentum pion interactions in the water are now simulated using the pion interaction model from NEUT. 
The atmospheric neutrino flux model is taken from~\cite{Honda:2011nf}. The momentum reconstruction algorithms have also been updated with some minor improvements. More details on the event generator, Monte Carlo simulation (MC) and reconstruction can be found in~\cite{Ashie:2005ik} and more details on the recent improvements can be found in~\cite{Abe:2013gga}.

\section{Sterile Neutrino Phenomenology}\label{sec:theory}

The neutrino oscillation probabilities in this analysis are based on the framework developed in~\cite{Maltoni:2007zf}. With $N$ additional sterile neutrinos, the PMNS mixing matrix~\cite{Pontecorvo:1967fh,Maki:1962mu} must be expanded to a $(3+N)\times(3+N)$ matrix:
\begin{equation}
U = \left(\begin{array}{ccccc}
U_{e1}    & U_{e2}    & U_{e3}    & U_{e4}    & \cdots \\
U_{\mu1}  & U_{\mu2}  & U_{\mu3}  & U_{\mu4}  & \cdots \\
U_{\tau1} & U_{\tau2} & U_{\tau3} & U_{\tau4} & \cdots \\
U_{s1}    & U_{s2}    & U_{s3}    & U_{s4}    & \cdots \\
\vdots    & \vdots    & \vdots    & \vdots    & \ddots \\
\end{array}
\right)
\end{equation}
This larger mixing matrix then appears in the completely generic $3+N$ Hamiltonian,
\begin{equation}
H        = U M^{(3+N)} U^\dagger + V_e + V_{\s}.
\end{equation}
The matrix $M^{(3+N)}$ is the neutrino mass matrix, 
\begin{equation}
M^{(3+N)} = \frac{1}{2E} \diag{0, \dmsq{21}, \ldots, \dmsq{(3+N)1}}, 
\end{equation}
which also depends on the neutrino energy $E$. $V_e$ and $V_{\s}$ are the potentials experienced by the electron and sterile neutrinos respectively, 
\begin{align}
V_e     &= \pm  (G_F/ \sqrt{2}) \diag{2N_e, 0, \ldots}   \\
V_{\s}  &= \pm  (G_F/ \sqrt{2}) \diag{0, 0, 0, N_n, N_n, \ldots} 
\end{align}
which depend on Fermi's constant, $G_F$, and the average electron and neutron densities along the neutrino path, $N_e$ and $N_n$, respectively.  Depending on their type, neutrinos experience one of three different potentials: \nue's have charged-current (CC) interactions with electrons and neutral-current (NC) interactions with electrons and nucleons, \numu's and \nutau's have only NC interactions, and any \nus's have no interactions.  The NC interactions depend only on the neutron density because the $Z^0$ couplings to electrons and protons are equal and opposite and their densities are identical in neutral matter.  The factor of two between $N_e$ and $N_n$ comes from the difference between the two currents in the standard model.  We have taken advantage of the freedom to arbitrarily set the zero of the potential energy to define lack of NC interactions as a potential experienced by the sterile neutrinos.

In order to simplify the calculation in the analysis of atmospheric neutrino data, we introduce a few assumptions.  We assume the sterile mass splittings are sufficiently large that oscillations in all samples are `fast' and the $L/E$ term, $\sin^2(\dmnew L/4E)$, can be approximated as $\Braket{\sin^2} = 0.5$. For the SK data, this assumption is good for $\dmnew > \val{10^{-1}}{eV^2}$.  The complex phases introduced by the additional neutrinos are also neglected in this treatment since they were shown in~\cite{Maltoni:2007zf} to have a negligible impact on the atmospheric neutrino sample.  We also assume that there are no $\nue-\nus$ oscillations. While hints of these \uesq-driven oscillations have been seen in short-baseline \nue/\nuebar disappearance~\cite{Huber:2011wv}, SK is not very sensitive to this parameter.  We estimate that allowing non-zero \uesq values of the size allowed by these other experiments reduces SK's sensitivity to \umsq by between 3\% and 40\%, depending on which experiment is used and what \dmnew value is assumed.  In order to avoid introducing a complex multi-experiment fit, we assume $\uesq = 0$.  All of these assumptions are discussed further in \cref{sec:assumptions}.  

The assumptions are chosen to eliminate features in the oscillation probability to which the atmospheric neutrinos are not sensitive and focus only on the parameters that can be measured.  The assumptions do not generally limit the applicability of the results (e.g. results can be compared against experiments where different assumptions are made in the theory) except in certain specific cases like the valid range of \dm's discussed in \cref{sec:assumptions}.

We then define,
\begin{equation}
\dmu = \sum_{i \ge 4} |U_{\mu i}|^2
\end{equation}
and divide the mixing matrix $U$ into a standard neutrino model part with only the $3\times3$ \Upmns surrounded by zeros and an $(3+N) \times (3+N)$ sterile part, \Ut:
\begin{equation}
U = \left( \begin{array}{cc}
\Upmns & \bf{0} \\
\bf{0} & \bf{0} 
\end{array}\right) \Ut.
\end{equation}
With these assumptions and definitions, we can calculate the \numu/\nue oscillation probabilities following the method of \cite{Maltoni:2007zf},
\begin{align}
\Pee &= \Ptee \label{eq:pee}\\ 
\Pem &= \left(1-d_\mu\right) \Ptem \label{eq:pem}\\
\Pme &= \left(1-d_\mu\right) \Ptme \label{eq:pme}\\ 
\Pmm &= \left(1-d_\mu\right)^2  \Ptmm + \sum_{i \ge 4} |U_{\mu i}|^4 \label{eq:pmm}
\end{align}
where \Ptab is the probability derived from a three-neutrino Hamiltonian,
\begin{align}\label{eq:hamiltonian}
\tilde{H} =&\; 
\Upmns M^{(3)} \Upmns^\dagger + V_{e} \nonumber \\
&\pm \frac{ G_F N_n}{\sqrt{2}} \sum_{\alpha = \rm{sterile}}
\left(\begin{array}{ccc}
0 & 0 & 0 \\
0 & |\Ut_{\alpha 2}|^2 & \Ut_{\alpha2}^*\Ut_{\alpha3} \\
0 & \Ut_{\alpha2} \Ut_{\alpha3}^* & |\Ut_{\alpha 2}|^2
\end{array}\right),
\end{align}
where the first term is the standard neutrino Hamiltonian in vacuum, the second term is the matter potential in the Earth from \nue CC interactions, and the third term gives the component of the sterile matter potential which is rotated into the three active flavors by the sterile mixing matrix \Ut. The scale of the sterile potential is set by Fermi's constant $G_F$ and the average neutron density along the neutrino's path $N_n$, calculated using the four-layer PREM model of the density profile of the Earth~\cite{Dziewonski:1981xy}.  
\Crefrange{eq:pee}{eq:pmm} and \cref{eq:hamiltonian} show that there are two dominant signatures introduced by sterile neutrino mixing. The first is the reduction of the \numu survival probability at all lengths and energies from the $(1-\dmu)^2$ term in \cref{eq:pmm}.  The second signature is the distortion of the oscillation probabilities when passing through significant amounts of matter due to the matter effects proportional to $N_n$ in \cref{eq:hamiltonian}.

It is not feasible to calculate the oscillation probabilities generated by generic $3+N$ models since there are too many free parameters introduced into $\tilde{H}$ by the sum over several $\alpha$'s: as many as $2N$ magnitudes and $N$ phases. So, following the technique of~\cite{Maltoni:2007zf}, we reduce the parameter space by introducing further approximations. These approximations will allow us to perform the fit in the simpler 3+1 case, described below, and then extend those result into more generic 3+N models in \cref{sec:extension}.

We examine two approximations, appropriate in different circumstances: the \emph{\fitTwo approximation} which assumes electron neutrinos are fully decoupled from $\mu-\tau-s$ oscillations, and the \emph{\fitOne approximation} which includes \nue appearance via standard three-neutrino oscillations but assumes no sterile matter effects by setting the neutron density in the Earth to be zero.  The former approximation includes both sterile oscillation signatures but produces a biased estimate of \dmu while the latter is only sensitive to the \dmu signature, but produces an unbiased estimate of it.  

Note that in both of the 3+1 approximations, one explicit parameter has already been eliminated because $d_\mu = \umsq$ and $\sum |U_{\mu i}|^4 = \dmu^2 = |U_{\mu 4}|^4$.  In the following sections we will use \umsq, but will return to using $d_\mu$ in \cref{sec:extension}.

\subsection{No-\nue oscillation probabilities} \label{sec:nonue}

\begin{figure*}
  \subfigure[No sterile neutrinos]{
        \label{fig:oscillogram:no}
        \includegraphics[width=\fwid,clip]{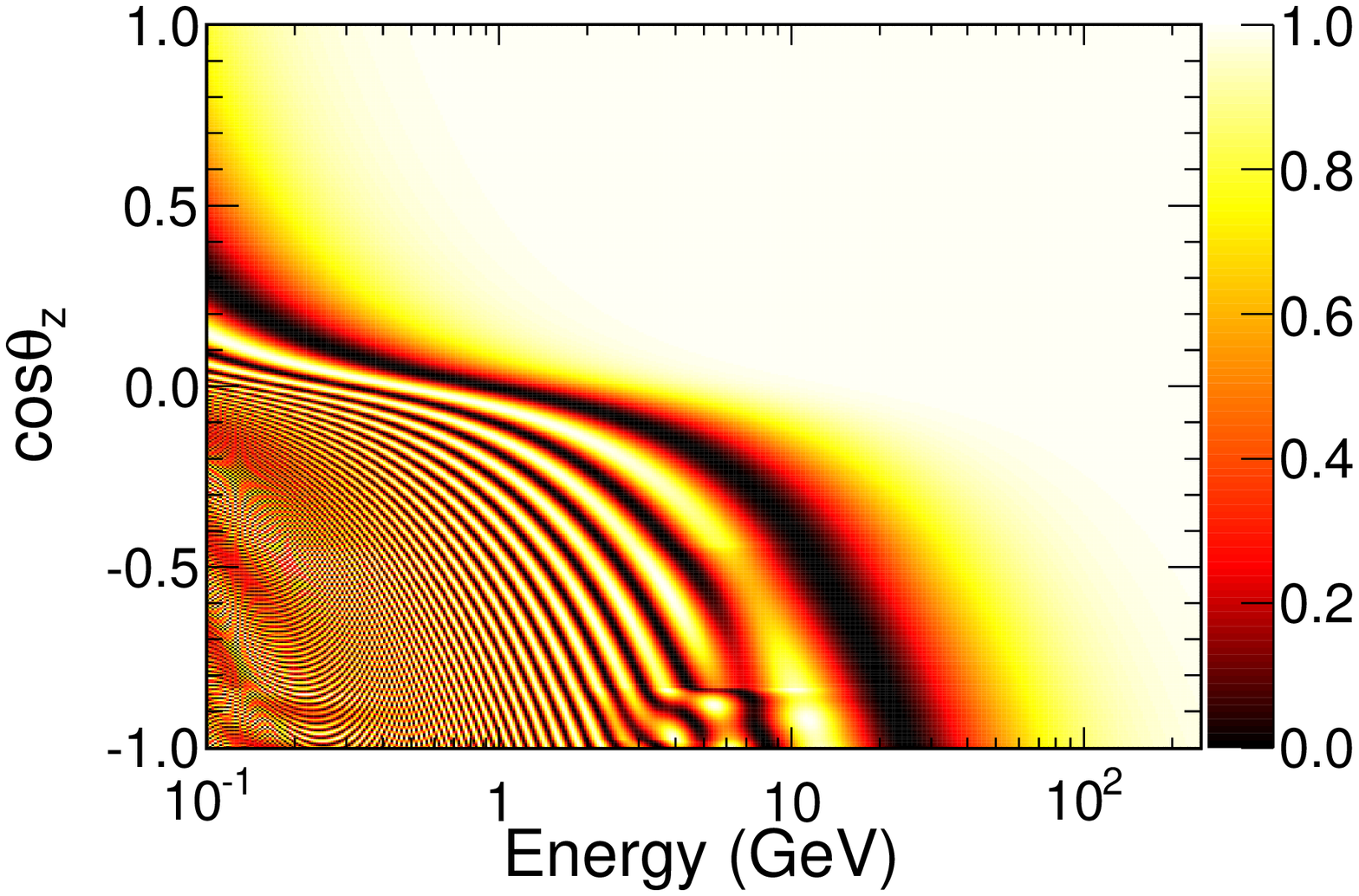}
  }
  \subfigure[$\utsq = 0.33$ and $\umsq = 0.0018$]{
        \label{fig:oscillogram:nc}
        \includegraphics[width=\fwid,clip]{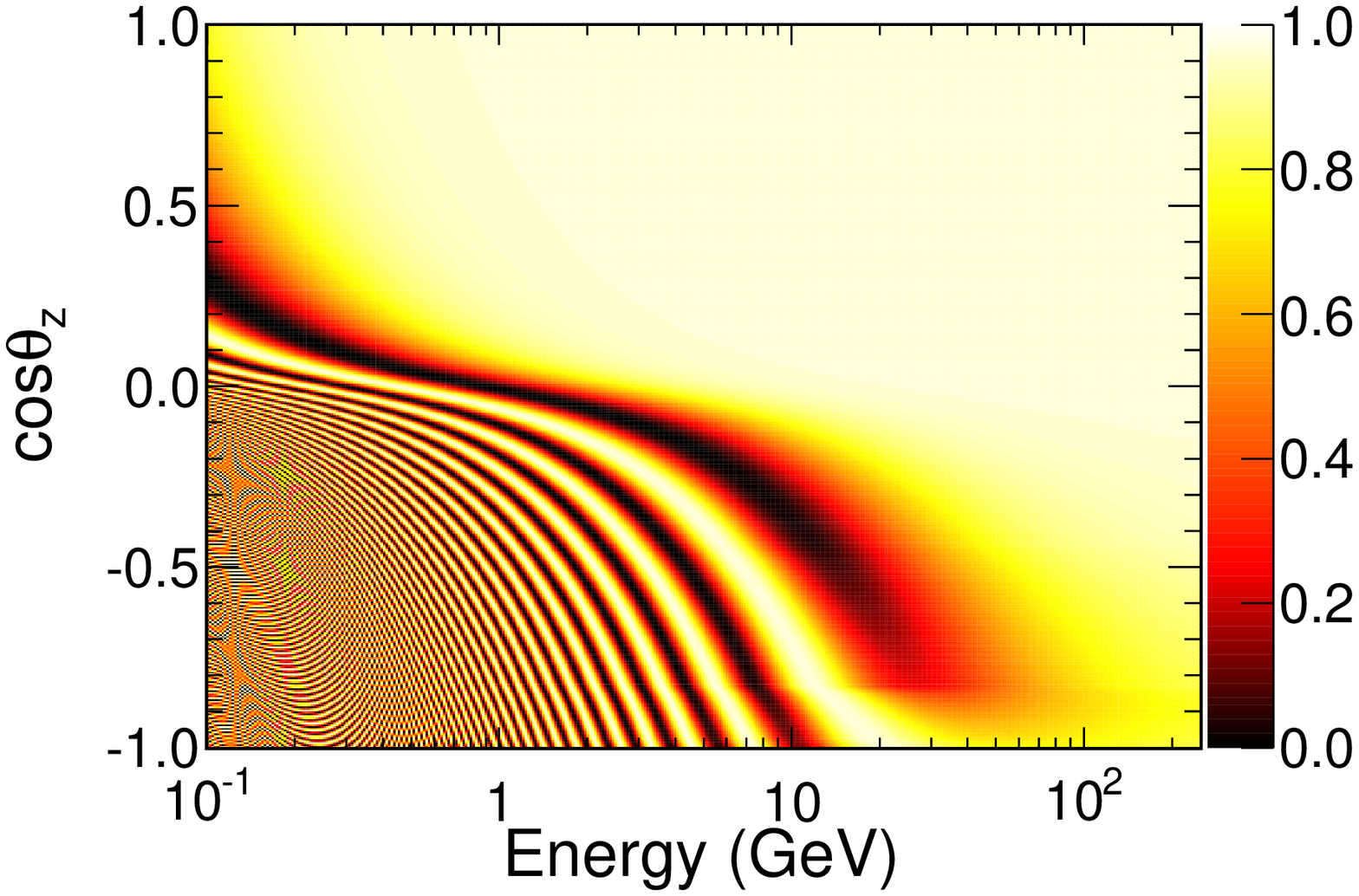}
  }
  \caption{(color online) 
  \subref{fig:oscillogram:no} The \numu survival probability without sterile neutrinos, plotted versus zenith angle and neutrino energy. This includes the standard \nue CC matter effect, which creates the distortion around a few GeV in the most upward-going zenith angles (\cz near -1), which correspond to neutrinos that pass through the core of the Earth.
  \subref{fig:oscillogram:nc}  The \numu survival probability calculated using the \fitTwo approximation with $\umsq = 0.0018$ and $\utsq = 0.33$.  The distortion due to the \nue CC matter effects is gone, but there is now a more pronounced distortion introduced by the sterile matter effects which reduces the amount of \numu disappearance for the most upward-going bins in the 10's of GeV region. There is also a small amount of extra disappearance away from the standard oscillations introduced by the non-zero \umsq which is most visible in the slight darkening of the upper-right part of the plot corresponding to the higher energy downward-going events (\cz near 1).}
  \label{fig:oscillogram}
\end{figure*}

The \nue's are fully decoupled from oscillations by setting $\th{13} = \th{12} = 0$, which allows \cref{eq:hamiltonian} to be reduced to a two-level system:
\begin{align}
\tilde{H} =  & 
\frac{\dmsq{32}}{4E} \left(\begin{array}{cc}
-\cos 2 \th{23} & \sin 2 \th{23} \\
 \sin 2 \th{23} & \cos 2 \th{23} 
\end{array}\right) \nonumber\\
& \pm
\frac{G_FN_n}{\sqrt{2}} \left(\begin{array}{cc}
|\tilde{U}_{s2}|^2 & \tilde{U}_{s2}^*\tilde{U}_{s3} \\ 
\tilde{U}_{s2}\tilde{U}_{s3}^*  & |\tilde{U}_{s3}|^2 
\end{array}\right) \label{eq:htilde}
\end{align}
Noting that the second matrix is Hermitian, it can be diagonalized and then parameterized by one real eigenvalue, $A_s$, and one angle, $\theta_s$, 
\begin{equation}
\frac{G_FN_n}{\sqrt{2}} A_s 
\left(\begin{array}{cc}
- \cos 2\th{s} & \sin 2\th{s} \\
  \sin 2\th{s} & \cos 2\th{s}
\end{array}\right) \label{eq:hermitian}
\end{equation}
which in the $3+1$ model can be expressed in terms of the only two independent sterile matrix elements, \umsq and \utsq:
\begin{align}
A_s    &= \frac{(\umsq + \utsq)}{2} \label{eq:as}\\
\sin 2\th{s} &= \frac{2\sqrt{\umsq\utsq(1 - \umsq - \utsq)}}{(1-\umsq)(\umsq + \utsq)} \label{eq:sins}\\
\cos 2\th{s} &= \frac{\utsq-\umsq(1 - \umsq - \utsq)}{(1-\umsq)(\umsq + \utsq)}. \label{eq:coss}
\end{align}
The complete system, which is itself Hermitian as the sum of two Hermitian matrices, can also be diagonalized to produce new effective two-neutrino oscillation probabilities which are a function of the atmospheric mixing parameters and the sterile parameters above:
\begin{align}
E_m^2        &= A_{32}^2+A_s^2 + 2 A_{32} A_s \cos(2\th{23} - 2\th{s}) \label{eq:em}\\
\sin 2\th{m} &= \frac{A_{32}\sin(2\th{23}) + A_s\sin(2\th{s})}{E_m} \label{eq:sm}\\
\cos 2\th{m} &= \frac{A_{32}\cos(2\th{23}) + A_s\cos(2\th{s})}{E_m} \label{eq:cm}
\end{align}
where $\pm E_m$ are the eigenvalues of the new system,  \th{m} is the new mixing angle, and $A_{32} = \dmsq{32}/4E$ is the magnitude of the eigenvalue of the Hamiltonian for two-flavor oscillations in the atmospheric sector without any sterile neutrinos.

Pulling together these pieces, the oscillation probabilities in the \fitTwo approximation are:
\begin{widetext}
\begin{align}
\Pee =& 1 \label{eq:peematter}\\ 
\Pem =& \Pme = 0 \\
\Pmm =& \left(1-\umsq\right)^2  \left(1 - \snt{m}\sin^{2}(E_m L)  \right)
     + |\um|^4 \label{eq:pmmmatter}\\
\Pmt =& \left(2 A_s + 2 A_s \umsq - |U_{\mu4}|^4 - 1\right)
     \left(1 - \snt{m}\sin^{2}(E_m L)\right) \nonumber\\
     &-(1-\umsq)A_s\sin(2\theta_s) \sin(4\th{m}) \sin^2(E_m L)
     + (1-\umsq)(1+\umsq-2A_s) \\
P_{\textrm{NC}\alpha} =& P_{\alpha e} + P_{\alpha \mu} + P_{\alpha \tau} \label{eq:pncmatter}
\end{align}
\end{widetext}
where $P_{\textrm{NC}\alpha}$ is the probability for a $\nu_\alpha$ to remain any active species and is applied to the NC events in our simulation.  Note that these probabilities depend on \umsq, $A_s$, and $\theta_s$, which in turn depend only on \umsq and \utsq (plus the atmospheric oscillation parameters).  The main signature of \umsq is the reduction in the survival probability of \numu's due to fast oscillations introduced by the $\left(1-\umsq\right)^2$ coefficient in \cref{eq:pmmmatter}.  The primary effect of \utsq comes through $A_s$ which scales the size of the sterile matter effects in the matter Hamiltonian, as can be seen in \cref{eq:hermitian}.  A non-zero \utsq also makes \th{s} non-zero, which enhances the matter effects further.  While \umsq also contributes to $A_s$, for densities available on Earth the fast oscillation effect is always much stronger than the matter effect, so any measurement of \umsq will come primarily from fast oscillations.  If, however, sterile matter effects are seen without accompanying fast oscillations, then that must be caused by \utsq. The effects of both \umsq and \utsq on the \numu survival probability vs. zenith angle and energy in SK are shown in \cref{fig:oscillogram}.

Since the signature of non-zero \umsq is a lower \numu survival probability independent of distance and energy, it manifests itself in the atmospheric neutrino data as a reduction in the normalization of all the $\mu$-like samples. Since there are significant systematic uncertainties on the absolute neutrino flux but much smaller uncertainties on the relative flux of \numu's to \nue's, the constraint on the $\mu$-like normalization depends on the normalization of the $e$-like samples. While the \nue appearance signal is not very large (approximately 7\% of the multi-GeV \nue samples), completely ignoring it does introduce a bias towards lower measured \umsq. 

The sterile matter effect signature, on the other hand, changes the shape of the zenith distribution in the PC and \UP samples.  Consequently, it is not dependent on the \nue samples to control systematic uncertainties and so is not biased by the \fitTwo assumption. The sterile matter effects alter the zenith distribution since the sterile term in \cref{eq:htilde} is enhanced by the high average $N_n$ experienced by the most upward-going neutrinos that pass through the core of the Earth.  The distortion is most pronounced in the higher-energy samples because the large neutrino energy $E$ suppresses the standard model part of $\tilde H$.

\subsection{\FitOne oscillation probabilities}\label{sec:hydrogen}

\begin{figure}
  \includegraphics[width=\fwid,clip]{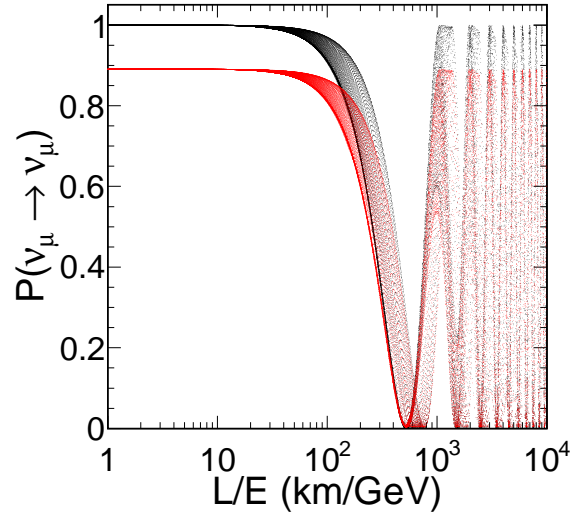}
  \caption{(color online) The \numu survival probability without sterile neutrinos (black) and with $\umsq = 0.058$ (red), calculated using the \fitOne oscillation probability, plotted versus $L/E$. The oscillation probability is not unique for a given $L/E$ since the \nue CC matter effect dependence on $L$ and $E$ is more complicated. So, many points corresponding to simulated neutrino events are plotted versus $L/E$, but with oscillation probabilities calculated using the individual simulated $L$ and $E$ values, to show the band of possible oscillation probabilities.  While the standard atmospheric oscillation pattern and the smaller variation due to \nue CC matter effects are persistent, introducing a sterile neutrino reduces the maximum survival probability at all values of $L/E$. The effect is most visible in regions without standard atmospheric oscillations }
  \label{fig:lovere}
\end{figure}

Under the alternative \fitOne assumption, $N_n$ in \cref{eq:hamiltonian} goes to 0, so $\tilde H \to H_{\rm{SM}}$ and the sterile neutrinos experience only vacuum oscillations. (This assumption is also called the `hydrogen-Earth' approximation in~\cite{Kopp:2013vaa}.)  Then, the \Ptab terms in \crefrange{eq:pee}{eq:pmm} become the standard, three-flavor oscillation probabilities, $\Pab^{(3)}$, which are calculated following~\cite{Barger:1980tf}, consistent with previous SK analyses. Then, the oscillation probabilities can be recalculated as,
\begin{align}
\Pee &= \Pee^{(3)} \label{eq:peethree}\\ 
\Pem &= \left(1-\umsq\right) \Pem^{(3)} \label{eq:pemthree}\\
\Pme &= \left(1-\umsq\right) \Pme^{(3)} \label{eq:methree}\\ 
\Pmm &= \left(1-\umsq\right)^2  \Pmm^{(3)} + |\um|^4 \label{eq:pmmthree}\\
\Pmt &= \left(1-\umsq\right)\left(1-\Pmm^{(3)}\right)\label{eq:pmtthree}\\
P_{\textrm{NC}\alpha} &= P_{\alpha e} + P_{\alpha \mu} + P_{\alpha \tau} \label{eq:pncthree}
\end{align}
where $\Pab^{(3)}$ is the standard three-flavor oscillation probability and $P_{\textrm{NC}\alpha}$ gives the survival probability for NC events.  \Cref{fig:lovere} shows the effect of a non-zero \umsq on the \numu survival probability as a function of $L/E$ in the atmospheric sample.

Since the \nue appearance is included in this approximation, there is no bias introduced in the estimation of \umsq.  However, without the sterile matter effects, there is no sensitivity to \utsq.

\section{Oscillation Analyses with One Sterile Neutrino}\label{sec:threepone}

The data samples described in \cref{sec:datasample} are fit simultaneously to search for evidence of sterile neutrinos using the same technique as in~\cite{Wendell:2010md} with some updates, including adding the SK-IV data and updating some systematic uncertainties.  Each run period, SK-I, SK-II, SK-III, and SK-IV, has its own 500 years-equivalent sample of MC to reflect the different physical and operational conditions during the four run periods. 

The oscillation fit minimizes a ``pulled'' \chisq~\cite{Fogli:2002pt} which compares the MC expectation for a particular set of oscillation parameters with the data based on a Poisson probability distribution:
\newcommand{\skn}{\textrm{SK} n}
\newcommand{\ob}{ \ensuremath{\mathcal{O}^{\skn}_{i} }\xspace } 
\newcommand{\ex}{ \ensuremath{E^{\skn}_{i}(\vec \theta) }\xspace } 
\newcommand{\sy}{ \ensuremath{\tilde{E}^{\skn}_{i}(\vec \theta, \vec \epsilon) }\xspace } 
\begin{widetext}
\begin{equation}
\chisq = 2 \sum_i \left( \sum_n \sy - \sum_n \ob 
       + \sum_n \ob \ln \frac{\sum_n \ob}{ \sum_n \sy} \right)
       + \chisq_{\textrm{penalty}}(\vec \epsilon)
\label{eq:chisq}
\end{equation}
\end{widetext}
where $n$ indexes the four SK run periods, $i$ indexes the analysis bins, \ob is the number of observed events in bin $i$ during SK$n$, and \sy is the MC expectation in bin $i$ in SK$n$ with the oscillation parameters being tested, $\vec \theta$, and systematic parameters, $\vec \epsilon$. The data and expectation are divided into 480 bins of \cz and/or energy, depending on sample, as detailed in \cref{tab:samples}. The binning has been chosen to ensure enough events are in each bin to have a stable \chisq calculation. While the expectation in each bin is calculated separately for each run period, the four run periods are summed together for the comparison between data and MC.

The effects of the systematic errors on the expectation are approximated as linear:
\begin{equation}
\sy    =  \ex \left(1 + \sum_j f^{\skn}_{i,j} \frac{\epsilon_j}{\sigma_j}\right) 
\label{eq:linsys}
\end{equation}
where $j$ indexes the systematic errors, \ex is the MC expectation in bin $i$ in SK$n$ without systematic shifts, and $f^{\skn}_{i,j}$ is the fractional change in bin $i$ in SK$n$ due to $\sigma_j$, the 1-sigma change in systematic $j$.  The constraints on these parameters are included as a penalty term in \cref{eq:chisq}:
\begin{equation}
 \chisq_{\textrm{penalty}}(\vec \epsilon) =  \sum_{j} \left(\frac{\epsilon_j}{\sigma_j}\right)^2.
\end{equation}

The two analyses consider 155 systematic error parameters; some of them are common to all four SK run periods and some are calculated separately and treated as independent for each period. The common errors include uncertainties in the atmospheric neutrino flux, neutrino interaction cross-sections, particle production within nuclei, and the standard PMNS oscillation parameters.  They come from the Honda flux calculation~\cite{Honda:2011nf}, external neutrino interaction measurements as well as model comparisons, and other oscillation measurements, respectively.  For these uncertainties, $f^{\skn}_{i,j}$ is the same in SK-I, SK-II, SK-III, and SK-IV. The period-specific errors are generally related to detector performance: uncertainties on reconstruction, particle identification, energy scale, and fiducial volume differ between run periods since they depend on the specific geometry and hardware of the detector, which are determined using control samples and simulation studies. For these uncertainties, $f^{\skn}_{i,j}$ will be non-zero in one run period and zero in all the others. All the systematic uncertainties and their sizes are listed in \cref{sec:systematics}.

\Cref{eq:chisq} is minimized with respect to $\vec \epsilon$ for each choice of $\vec \theta$ in a fit's parameter space. A set of linear equations in the $\epsilon_j$'s are derived from \cref{eq:chisq} using the fact that the derivative $\partial \chisq/\partial \epsilon_i$ is zero at the minimum~\cite{Fogli:2002pt}. These equations can then be solved iteratively to find the minimum profile likelihood for that set of oscillation parameters, building up a map of \chisq vs. $\vec \theta$. The best fit point is defined as the global minimum of this map.  Tests performed with high-statistics simulation, both without and with simulated sterile neutrino signals, showed no significant biases in the extracted best fit points.

In order to focus the analysis on the sterile neutrino parameters, the standard oscillation parameter values were constrained to external measurements and their uncertainties taken as systematic uncertainties. The T2K measurement of \numu disappearance, 
$|\dmsq{32}| = \val{\sci{(2.51 \pm 0.10)}{-3}}{eV^2}$ and 
$\sn{23} = \stwothree \pm 0.055$~\cite{Abe:2014ugx}, 
is used because its narrow-band beam makes it less sensitive to the sterile effects being measured in this analysis.  The mixing angle 
$\snt{13} = \stonethree \pm 0.01$ 
is taken from the PDG world-average~\cite{PDG}, the solar terms are taken from the global fit performed by the SK solar+KamLAND analysis, 
$\dmsq{21} = \val{\sci{(7.46 \pm 0.19)}{-5}}{eV^2}$, 
$\sn{12} = \sonetwo \pm 0.021$~\cite{Abe:2010hy},
and we assume $\dmsq{32} > 0$ and $\delta_{cp} = 0$, though the precise value of these choices have negligibly small effects on this analysis.

\subsection{No-\nue analysis}\label{sec:threeponetwo}

\begin{figure}
 \begin{center}
 \includegraphics[width=\fwid,clip]{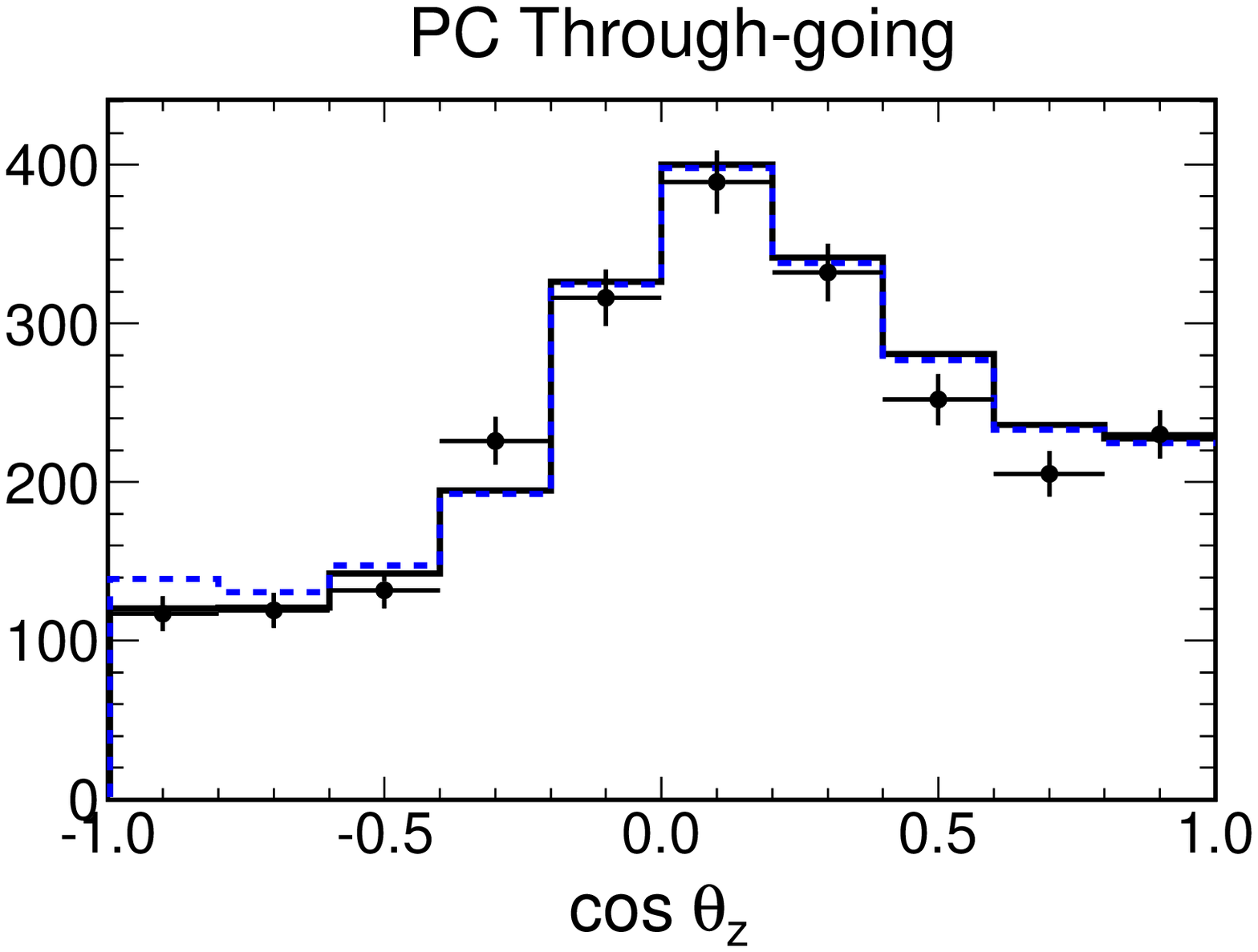}
 \includegraphics[width=\fwid,clip]{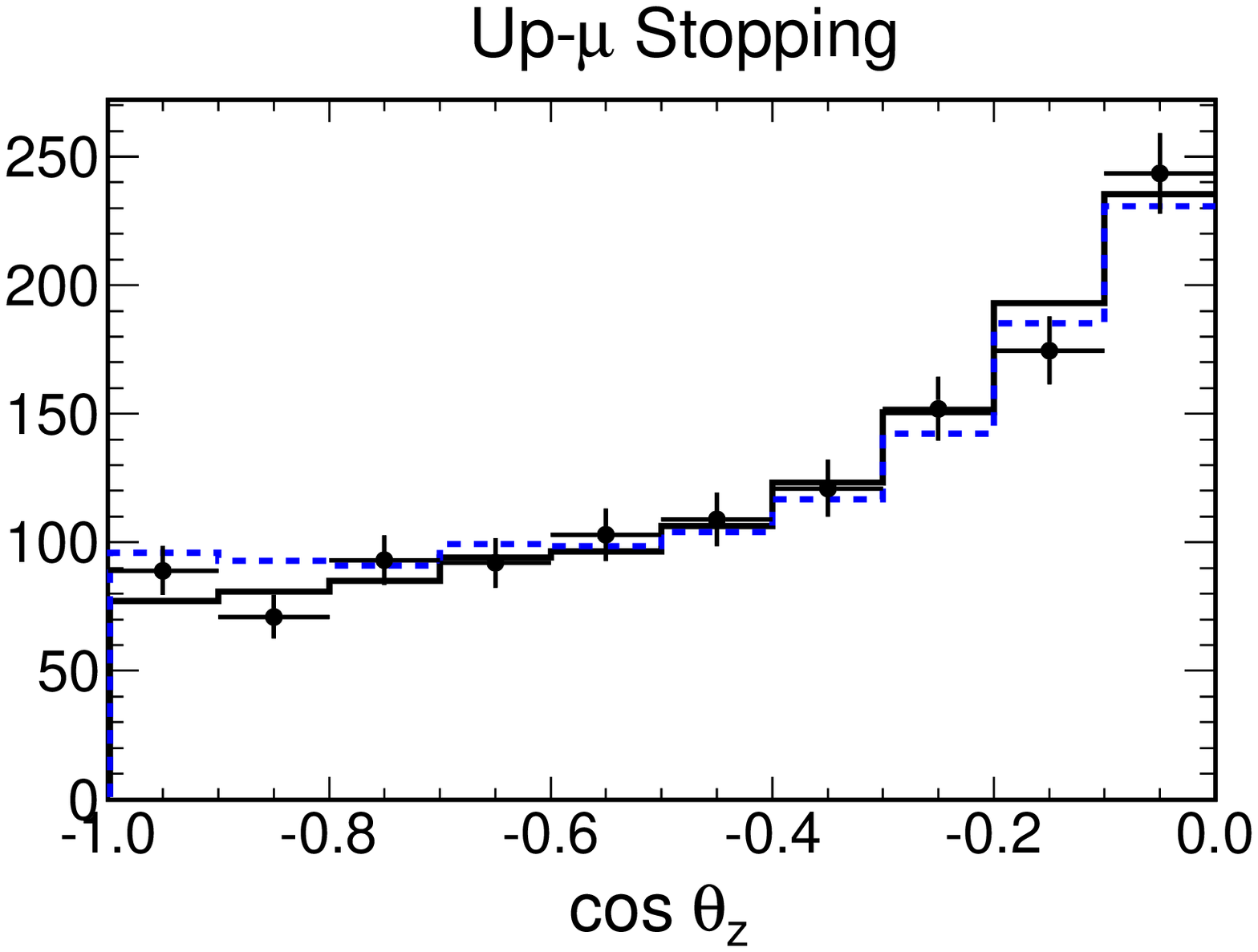}
 \caption{ (color online) Zenith angle distributions summed across SK-I through SK-IV of the PC through-going and \UP stopping sub-samples shown for the data (black points with statistical error bars), the MC prediction without sterile neutrinos (black solid line), and the MC prediction with a large (approximately $5\sigma$ sensitivity) sterile signal of \utsq = 0.31. Both MC predictions are shown after fitting the systematic uncertainties to the data.  These two subsamples are shown because they contain the 10's of GeV neutrinos most sensitive to the sterile matter effect.  The prediction with a sterile component shows an up-turn for the most up-going events which corresponds to the distortion of the oscillogram shown in \cref{fig:oscillogram:nc}.
 }
 \label{fig:zenith_2p1}
 \end{center}
\end{figure}

\begin{figure}
 \begin{center}
 \includegraphics[width=\fwid,clip]{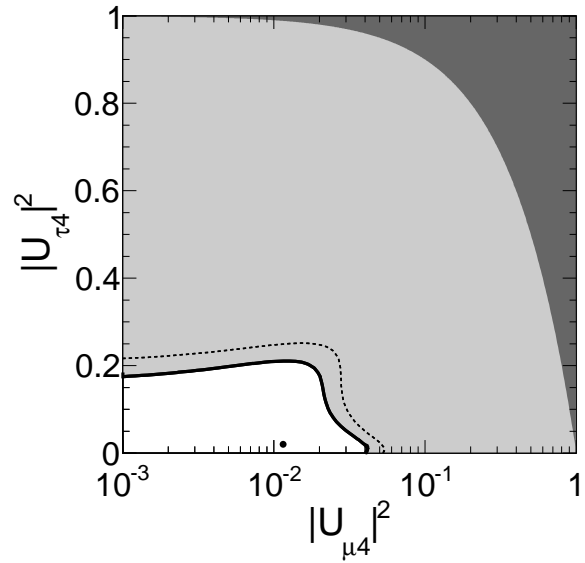}
 \caption{The 90\% and 99\% upper limits on \utsq vs. \umsq from the \fitTwo fit are shown by the solid and dashed lines, respectively.  The best fit point is marked by a black dot. The light gray region is excluded at 90\% and the dark gray region is disallowed by unitarity.} 
 \label{fig:2p1global}
 \end{center}
\end{figure}

As described in \cref{sec:nonue}, the analysis with the \fitTwo approximation fits both \umsq and \utsq. Since it does not include normal \nue matter effects it is systematically biased towards smaller \umsq values than the CC matter effect fit.
The fit is performed on a two-dimensional grid of 200 points, 50 points in \umsq distributed logarithmically between $10^{-3}$ and $10^{-1}$ and 40 points in \utsq distributed linearly between 0 and 0.4. 

The best fit is at $\umsq = 0.012$ and $\utsq = 0.021$ with $\chisq_{\rm{min}} = 531.1$ over 480 bins (goodness-of-fit $p = 0.05$). 
\Cref{fig:zenith_2p1} show the zenith angle distributions for the sub-samples most sensitive to the \utsq parameter and an example of what a large sterile contribution would look like.  The $\Delta \chisq$ to the no-sterile prediction is 1.1, consistent with no sterile neutrinos at the $1\sigma$ level with two degrees of freedom.
We limit \utsq to less than \utlimit at 90\% and less than \utlimitnn at 99\%. These limits are independent of the new \dmnew above \val{0.1}{eV^2} (see \cref{sec:assumptions:dmsq}). The contours in \utsq vs. \umsq can be seen in \cref{fig:2p1global}.  The \umsq best fit point and limit are discussed in the next section in the analysis which focuses on that parameter.

\subsection{\FitOne analysis}\label{sec:threeponeone}

\begin{figure}
 \begin{center}
 \includegraphics[width=\fwid,clip]{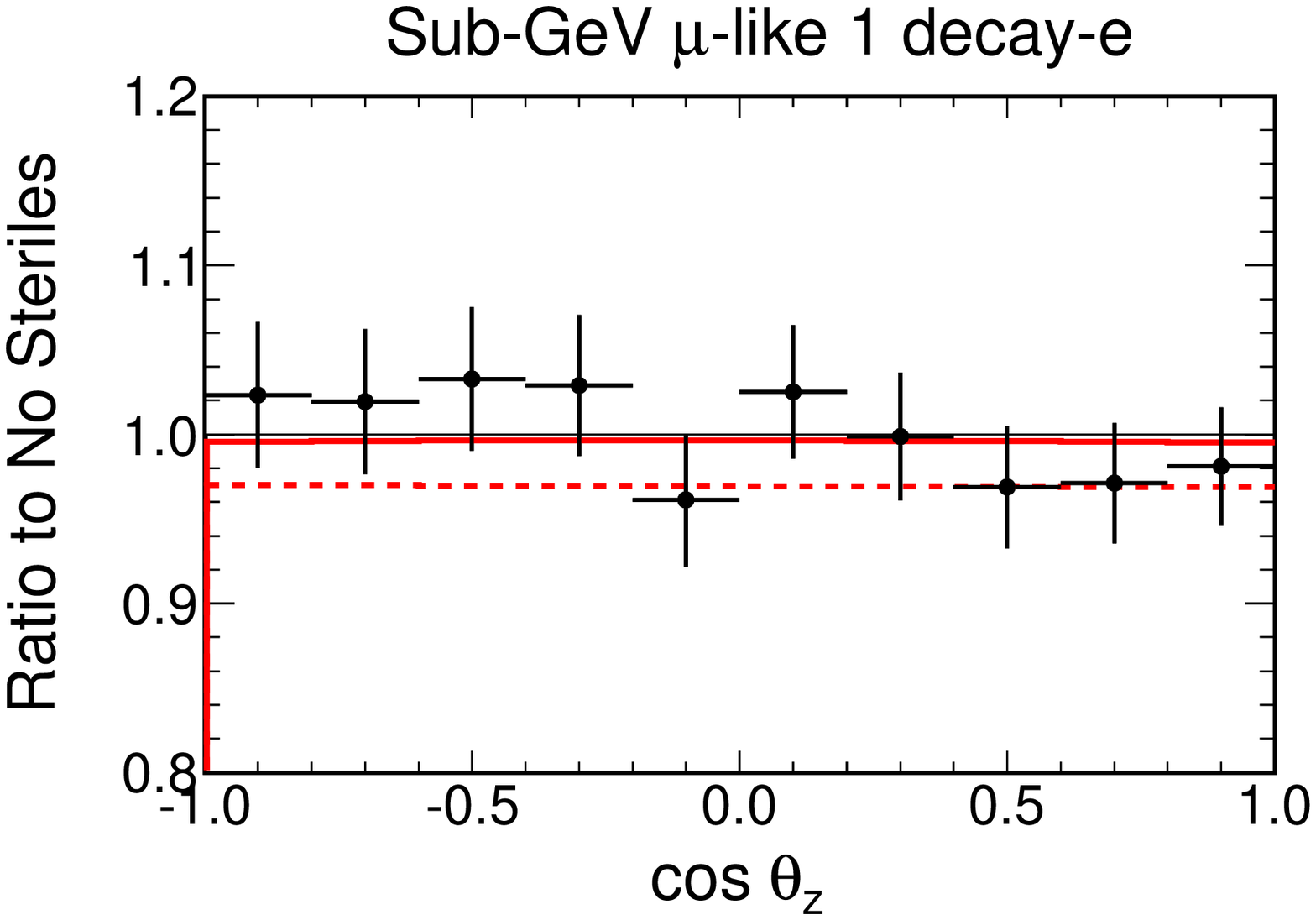}
 \includegraphics[width=\fwid,clip]{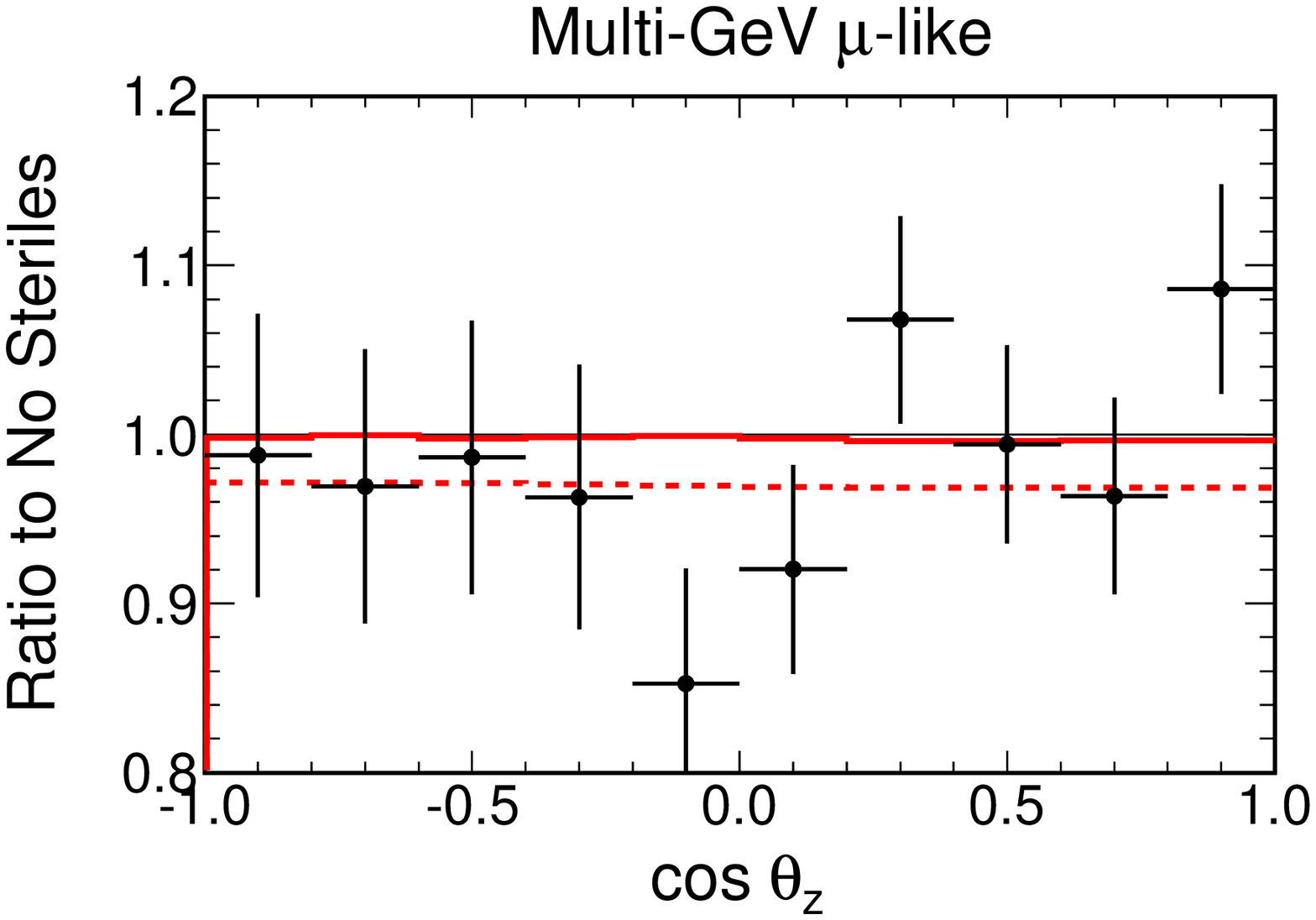}
 \includegraphics[width=\fwid,clip]{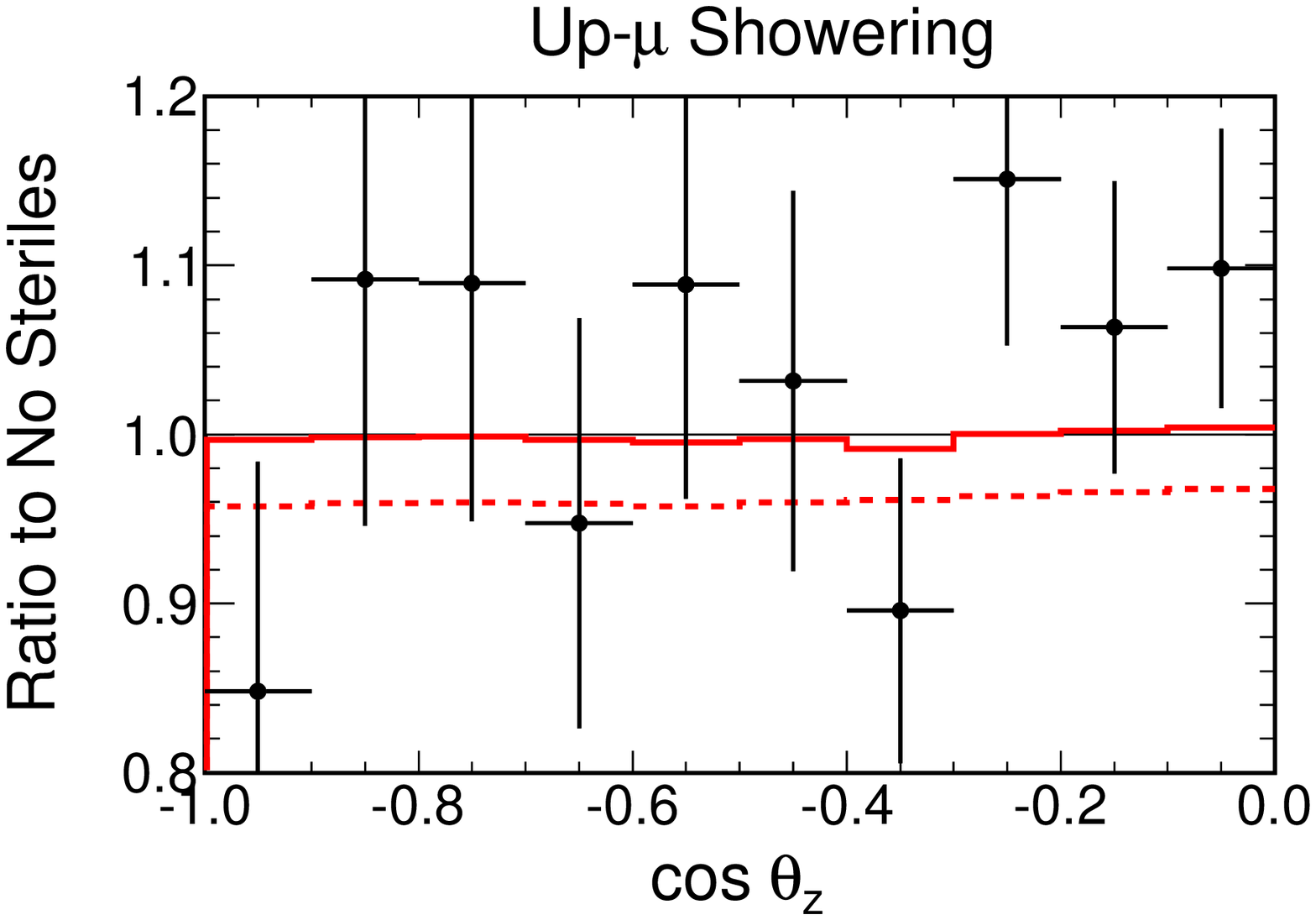}
 \caption{ (color online) Ratios to the MC prediction without sterile neutrinos, binned in zenith angle and summed across SK-I through SK-IV, for three $\mu$-like sub-samples at low (FC Sub-GeV), medium (FC Multi-GeV), and high energies (though-going \UP).  The prediction without sterile neutrinos has been fit to the data using the systematic uncertainties.  The black points represent the data with statistical error bars and the solid red line shows the MC prediction with the best fit for sterile neutrinos (\umsq = \umbest), including the best fit systematic uncertainties.  In all the samples it lines up close to unity, meaning the prediction is nearly identical to the prediction without sterile neutrinos.  The dashed red line shows the MC prediction with the same sterile component (\umsq = \umbest), but now with the same systematic uncertainty parameters as the denominator, showing the effect of just the sterile oscillations: the normalization is shifted downward by approximately 3\% in every $\mu$-like sample. 
 }
 \label{fig:zenith_3p1}
 \end{center}
\end{figure}

\begin{figure}
 \begin{center}
 \includegraphics[width=\fwid,clip]{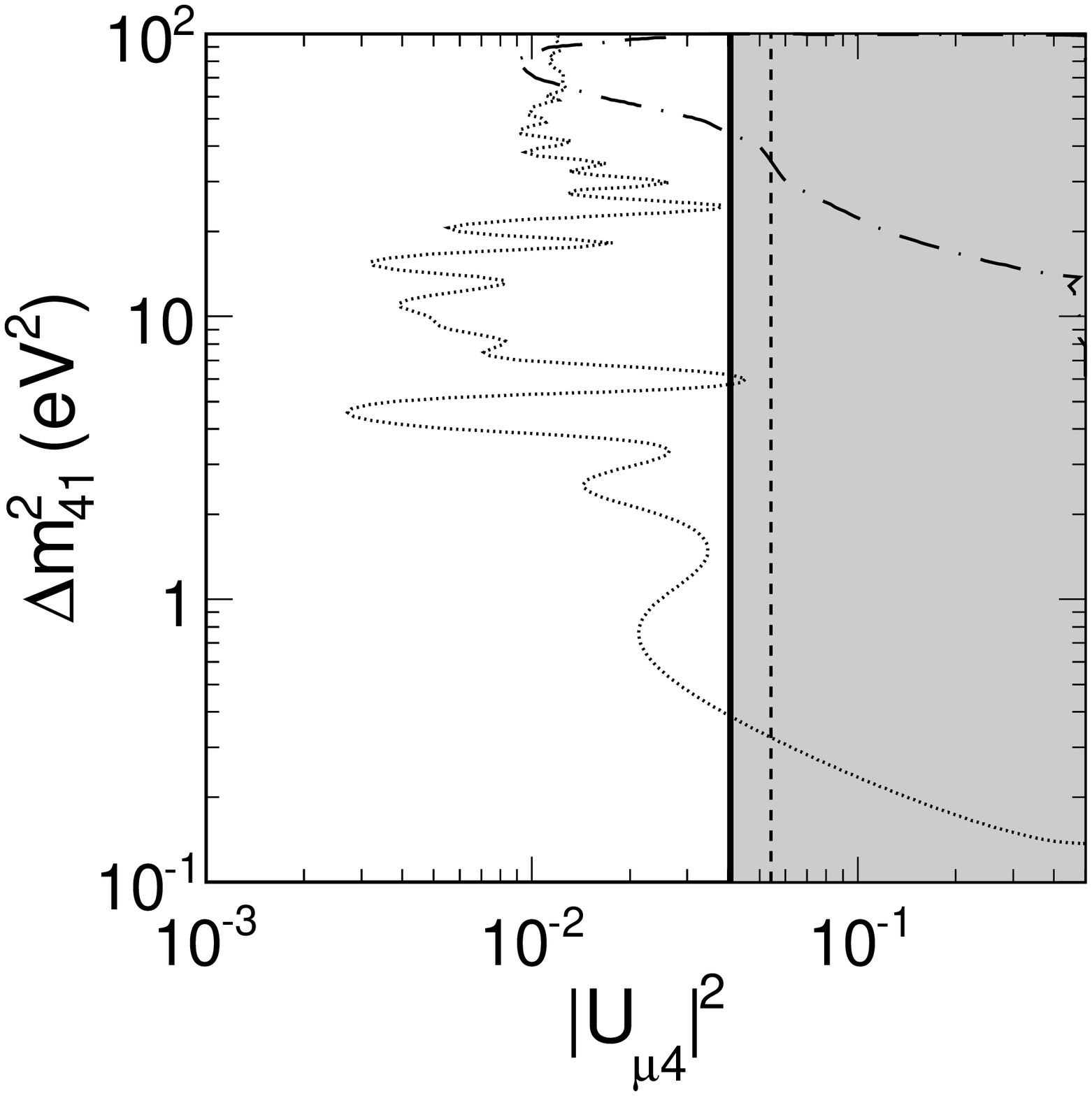}
 \caption{The 90\% and 99\% upper limits on \umsq from the \fitOne fit to \sk is shown in the solid and dashed and vertical lines, respectively.   The gray filled region is excluded at 90\%. This analysis is not sensitive to \dmnew, but the experiments who also measure \umsq are, so here the one-dimensional \sk result is shown in two dimensions.The dotted line is the 90\% limit placed by the joint analysis of MiniBooNE and SciBooNE~\cite{Cheng:2012yy} and the dot-dash line is the 90\% limit placed by the CCFR experiment~\cite{Stockdale:1984cg}. }
 \label{fig:3p1global}
 \end{center}
\end{figure}

The analysis with the \fitOne approximation fits only \umsq, the term which drives fast oscillations, creating extra disappearance at all energies and zenith angles in all $\mu$-like samples. 
The fit is performed on a one-dimensional grid of 200 points distributed logarithmically between $10^{-3}$ and $10^{-1}$. The best fit is at $\umsq = \umbest$ with $\chisq_{\rm{min}} = 532.1$ over 480 bins (goodness-of-fit $p = 0.05$).  No sterile oscillations is slightly disfavored by $\Delta \chi^2 = 1.1$.

\begin{table}
\begin{center}
  \begin{tabular}{lcc}
    \hline\hline
    Systematic Uncertainty                                               & No steriles ($\sigma$) & Best fit ($\sigma$) \\
    \hline
    $(\numu+\numubar) / (\nue+\nuebar), <\val{1}{GeV}$                   & -0.49                  & -0.13 \\
    $(\numu+\numubar) / (\nue+\nuebar), \val{1-10}{GeV}$                 & -0.50                  & -0.09 \\
    CCQE \numu / \nue                                                    &  0.36                  &  0.01 \\
    \hline\hline
  \end{tabular}
  \caption{The best fit pull values, shown for both no sterile neutrinos and the best fit point from the \fitOne analysis, of the systematics which change the most between those two points.  The values at the best sterile fit are all significantly smaller than the values assuming no sterile neutrinos, reducing the \chisq penalty term.}
  \label{tab:sysvals}
\end{center}
\end{table}

\Cref{fig:zenith_3p1} shows the best fit from this analysis in several $\mu$-like samples which closely matches the prediction without sterile neutrinos (the ratio is approximately unity across all bins).  In fact, there is no net difference in \chisq between the best fit point and the prediction without sterile neutrinos looking just at the difference between the data and the prediction in each bin.
All of the difference in \chisq at the best fit come from the reduction in the systematic penalty term from introducing a non-zero \umsq.  The dashed line in \cref{fig:zenith_3p1} shows the prediction with the best fit sterile parameter, but without separately minimizing the systematic uncertainties. It shows the effect of a non-zero \umsq in isolation: it lowers the normalization in the $\mu$-like samples by approximately 3\%.  By introducing this normalization change with the sterile oscillation parameter, several systematic error parameters can be moved closer to their nominal values, reducing the \chisq penalty term.
The reduction is concentrated in three systematic errors: the $(\numu+\numubar) / (\nue+\nuebar)$ ratio in the atmospheric flux below \val{1}{GeV} and from \val{1-10}{GeV} as well as the CCQE \numu/\nue cross-section ratio, summarized in \cref{tab:sysvals}.

All three of these systematic errors relate to the relative normalization between the $\mu$-like subsamples, which have sterile oscillations, and the $e$-like subsamples, which do not. These two flux systematics specifically affect the low-energy subsamples and the CCQE interaction mode is dominant at lower energies, so it affects the same subsamples.  
The flux uncertainty is calculated as part of the neutrino flux model, which uses direct muon flux measurements plus simulations of hadronic interactions in the atmosphere constrained by hadron production experiments~\cite{Honda:2011nf}. The uncertainty is between 2\% and 3\% in size at these energies.  The CCQE cross-section uncertainty comes from the difference between the default model in NEUT~\cite{LlewellynSmith:1971zm} and a local Fermi gas model~\cite{Nieves:2004wx} and is 1\% to 1.5\% in size.  
While the sterile oscillations create effects in basically every $\mu$-like sample, these low-energy samples are the most important in this analysis since they have the highest statistics and thus the smallest statistical uncertainties.  

We limit \umsq to less than \umlimit at 90\% and less than \umlimitnn at 99\%. These limits are independent of the new \dmnew above \val{0.1}{eV^2} (see \cref{sec:assumptions:dmsq}) and can be compared to other limits on sterile-driven \numu disappearance from short-baseline experiments in \cref{fig:3p1global}.  The limits on this parameter are dominated by the systematic uncertainties on the low-energy normalization and the sensitivity improvement with increased statistics will be relatively small unless better systematic constraints are included.  The expected sensitivity to this parameter is a limit at 0.024 at 90\%, somewhat tighter than the observed limit since it assumes a best fit with no sterile neutrino component.

\section{Extending the Analyses to Additional Sterile Neutrinos}\label{sec:extension}

The oscillation probabilities from \cref{sec:theory}, both the \fitTwo and \fitOne approximations, were developed to allow extensions to multiple sterile neutrinos.  

\subsection{Extending the \fitOne Analysis}

The most straightforward extension is with the \fitOne analysis.  Starting again with the oscillation probabilities from \crefrange{eq:pee}{eq:pmm}, we perform the substitution $\Ptab \to \Pab^{(3)}$, but leave the probabilities in terms of \dmu, recalling that $\dmu = \sum |U_{\mu i}|^2$ for $i \ge 4$:
\begin{align}
\Pee &= \Pee^{(3)} \label{eq:peen}\\ 
\Pem &= \left(1-\dmu\right) \Pem^{(3)} \label{eq:pemn}\\
\Pme &= \left(1-\dmu\right) \Pme^{(3)} \label{eq:men}\\ 
\Pmm &= \left(1-\dmu\right)^2  \Pmm^{(3)} + \sum_{i \ge 4}|U_{\mu i}|^4 \label{eq:pmmn}\\
\Pmt &= \left(1-\dmu\right)\left(1-\Pmm^{(3)}\right)\label{eq:pmtn}\\
P_{\textrm{NC}\alpha} &= P_{\alpha e} + P_{\alpha \mu} + P_{\alpha \tau} \label{eq:pncn}
\end{align}

In \cite{Maltoni:2007zf}, the authors note that these expressions are almost equivalent to \crefrange{eq:peethree}{eq:pmmthree} with $\umsq \to \dmu$, except for the constant term  $\sum |U_{\mu i}|^4$ from \cref{eq:pmmn}, which does not equal $\dmu^2$ due to potential cross terms.  Following their method, we can write the \numu survival probability as:
\begin{equation}
\Pmm = \left(1-d_\mu\right)^2  \Pmm^{(3)} + \dmu^2 (1+\xi_\mu^2)/2
\end{equation}
where $\xi_\mu$ parameterizes the second order deviation from $\dmu^2$ in the constant term introduced by additional sterile neutrinos.  Their studies show the effect of $\xi_\mu$ on the \dmu limit from atmospheric neutrinos in the context of a $5\nu$ model and show that it has no significant effect on the limit placed on \dmu~\footnote{See particularly Figs. 12 (a) and (b) of Appendix C of \cite{Maltoni:2007zf}.}.  

The independence from $\xi_\mu$ derives from how the scaling term $(1-\dmu)^2$ and constant term $\sum |U_{\mu i}|^4$ affect the oscillation probability. The primary effect we observe in atmospheric neutrinos comes from the scaling term which creates extra disappearance that is independent of baseline and energy since it scales the entire \numu survival probability.  This effect is visible almost everywhere in the atmospheric data, except where $\Pmm^{(3)} \to 0$ (see \cref{fig:lovere}).  The constant term creates an opposing, but smaller effect which reduces disappearance but it is usually overwhelmed by the scaling term.  The effect of the constant term can only be seen in the `valleys' of the oscillation probability where $\Pmm^{(3)} \to 0$.  
In atmospheric neutrinos, the bottoms of these valleys are not clearly resolved, so this effect is vanishingly small (as opposed to in long-baseline experiments which precisely measure the first oscillation minimum). Since the value of $\xi_\mu$ can be neglected when performing this fit, the oscillation probabilities in \crefrange{eq:peen}{eq:pncn} are in fact equivalent to those in \crefrange{eq:peethree}{eq:pmmthree}. 

Due to this equivalence, the limit on \umsq from the 3+1 fit shown in \cref{fig:3p1global} can be taken as the limit on \dmu in general 3+N models.

\subsection{Extending the \fitTwo Analysis}

The results from the \fitTwo analysis can also be extended, at least in an approximate way, to theories with additional sterile neutrinos.  Recall that the oscillation probabilities in \crefrange{eq:peematter}{eq:pncmatter} depend on the solutions to the two-level Hamiltonian,
\begin{equation}
\tilde{H} = H_{\rm{SM}} \pm \frac{G_FN_n}{\sqrt{2}} H_{\s}.
\end{equation}
With additional sterile neutrinos, the dependence can become quite complicated since there is a sum over multiple sterile species, $\alpha$, from \cref{eq:hamiltonian}:
\begin{equation}\label{eq:sumsterile}
    H_{\s} = 
    \sum_{\alpha = \rm{sterile}}
    \left(\begin{array}{cc}
      |\Ut_{\alpha 2}|^2 & \Ut_{\alpha2}^*\Ut_{\alpha3} \\
      \Ut_{\alpha2} \Ut_{\alpha3}^* & |\Ut_{\alpha 2}|^2
    \end{array}\right).
\end{equation}
In the most general case, $H_{\s}$ depends on $3N$ free parameters (two magnitudes and a phase difference for each sterile species $\alpha$).  However, this matrix is $2\times2$ and Hermitian, so no matter how many independent terms go into the matrix, there can only be two free parameters after diagonalization, an eigenvalue we labeled $A_s$ and a mixing angle $\theta_s$.

In \cref{sec:nonue}, we rewrote $\tilde H$ first in terms of a generic diagonalized Hermitian matrix parameterized by $A_s$ and $\theta_s$, and then calculated those parameters by explicitly diagonalizing \cref{eq:sumsterile} with only one sterile neutrino species. After that, the solutions of $\tilde H$ depend only on the two free sterile parameters in the mixing matrix, \umsq and \utsq, and thus the oscillation probabilities in \crefrange{eq:peematter}{eq:pncmatter} depend only on those parameters as well.

\begin{figure}
    \includegraphics[width=\fwid,clip]{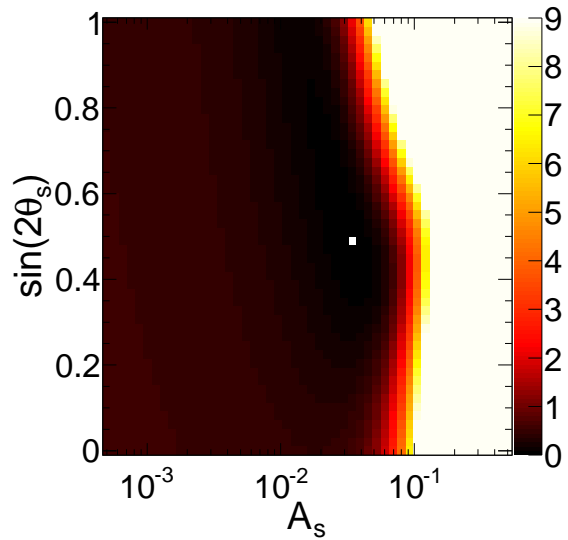}
    \caption{The $\Delta \chi^2$ from the fit to the atmospheric neutrino data in the \fitTwo approximation, plotted versus the two effective parameters, $A_s$ and $\sin(2\theta_s)$, with the third free parameter, \dmu, profiled out.}
    \label{fig:chisqgeneric}
\end{figure}

To constrain models with additional sterile neutrinos, we perform a fit using the same oscillation probabilities, but we do not solve explicitly for \crefrange{eq:as}{eq:coss}, meaning the oscillation probabilities, and hence the \chisq surface produced by the fit, depend on the two generic parameters, $A_s$ and $\theta_s$, plus \dmu on which the oscillation probabilities in \crefrange{eq:peematter}{eq:pncmatter} have an explicit dependence (we have substituted \dmu for \umsq as described in the previous section).  The values of these parameters can be calculated easily from the sterile part of the mixing matrix $U$ for any sterile neutrino theory, and they can then be used to look up the $\Delta \chi^2$ from this atmospheric fit, allowing constraints to be put on the parameters in that theory. See the supplemental material for a table containing the full three-dimensional delta log likelihood surface~\cite{Supplemental}.
As a demonstration, the $\Delta \chi^2$ surface for $A_s$ vs. $\sin(2\th{s})$ (\dmu has been profiled out) is shown in \cref{fig:chisqgeneric}.  

The $\mu \to \tau$ and hence $\mu \to s$ probabilities are approximations in this case since they neglect some potential cross-terms introduced in the sum over $\alpha$, but the fit is dominated by the \numu disappearance signal, so this approximation in the NC and $\tau$ oscillation probabilities will have little effect on the results.

\section{Conclusion}

The atmospheric neutrino data from all four periods of \superk have been fit to look for evidence of oscillations with an additional sterile neutrino. The fit was performed with two different approximations appropriate for setting limits on the two new matrix elements in the 3+1 framework to which \sk is sensitive: \umsq and \utsq. No significant evidence for fast oscillations driven by a new large \dm or of the matter effect associated with oscillations from \numu to \nus are seen. We limit the 3+1 parameters \umsq to less than \umlimit and \utsq to less than \utlimit at 90\%. While the measurement of \umsq is limited by systematic uncertainties on the neutrino flux and cross section around \val{1}{GeV}, the constraint on \utsq can potentially improve with additional atmospheric data.  Assuming only a single sterile neutrino, these new limits increase the known tension between \numu disappearance measurements and the hints seen in the \nue appearance and disappearance channels.  Since these limits are independent of the size of the new \dm, they exclude some new regions of parameter space at low mass-splittings where beam experiments are not sensitive. They can also be extended readily to 3+N models which might resolve the tensions between the three channels, and the results are provided in a format to allow tests of more general models in the supplemental materials~\cite{Supplemental}.

\section{Acknowledgments}
The authors would like to thank M. Maltoni for his help calculating and 
implementing the sterile oscillation probabilities.
The authors gratefully acknowledge the cooperation of the Kamioka 
Mining and Smelting Company. Super-K has been built and operated from 
funds provided by the Japanese Ministry of Education, Culture, Sports, 
Science and Technology, the U.S.  Department of Energy, and the 
U.S. National Science Foundation. This work was partially supported by 
the Research Foundation of Korea (BK21 and KNRC), the Korean Ministry 
of Science and Technology, the National Science Foundation of China, 
the European Union FP7 (DS laguna-lbno PN-284518 and ITN
invisibles GA-2011-289442)
the National Science and Engineering Research Council (NSERC) of Canada, 
and the Scinet and Westgrid consortia of Compute Canada.

\appendix
\crefalias{section}{asec}
\crefalias{subsection}{asec}

\section{Results in other parameterizations}\label{sec:parameterizations}

There are several mostly equivalent parameterizations that can be used for the sterile oscillation parameters in 3+1 models.  While we have chosen to present our results in terms of the magnitude of the matrix elements, we present here the limits in some other choices of parameters.

\begin{align}
\sin^2 \th{24} &=  \umsq\\
\sin^2 \th{34} &=  \utsq / (1 - \umsq) \\
\sin^2 2\th{\mu\mu} &= 4 \umsq ( 1 - \umsq ) \\
\ussq &= 1 - \umsq - \utsq
\end{align}

\begin{table}[h]
\centering
\begin{tabular}{cccc}
\hline \hline
\umsq    & $\sin^2 \th{24}$ & \th{24}     & $\sin^2 2\th{\mu\mu}$ \\ \hline
\umlimit & \umlimit         & 7.7\degrees & 0.071 \\
\hline \hline
\end{tabular}
\caption{90\% C.L.'s from the \fitOne fit.}
\label{tab:umlimits}
\end{table}

\begin{table}[h]
\centering
\begin{tabular}{cccc}
\hline \hline
\utsq    & \ussq & $\sin^2 \th{34}$ & \th{34} \\ \hline
\utlimit & 0.81  & 0.18             & 25\degrees \\
\hline \hline
\end{tabular}
\caption{90\% C.L.'s from the \fitTwo fit. The profiled value of \umsq = 0.010 for this point.}
\label{tab:utlimits}
\end{table}

\section{Assumptions in the Oscillation Model}\label{sec:assumptions}

A number of assumptions and approximations are made in order to make the 3+1 calculations easier and to allow those results to be extended to more general 3+N models.  This appendix presents the justification for the validity of three of the major assumptions.

\subsection{No sterile-electron neutrino mixing}\label{sec:assumptions:nue}

\begin{figure}[b]
    \includegraphics[width=\fwid,clip]{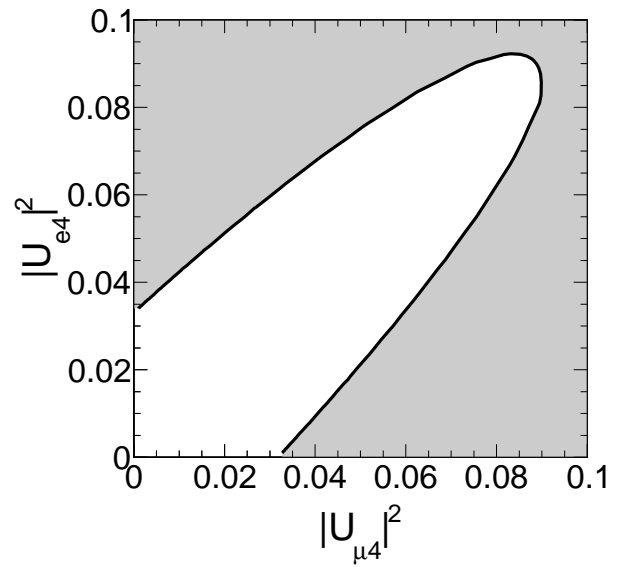}
    \caption{The 90\% sensitivity contour for the \fitOne fit with the effect \Pee from \cref{eq:peeappendix} included.  Allowing the freedom in the electron sample normalization reduces the sensitivity to \umsq as can be seen from the bowing outward on the right side of the contour.  Note that on this plot \umsq is shown in linear scale so the correlation with \uesq is clear.}
    \label{fig:dmude}
\end{figure}

Following the method in Appendix C2 of~\cite{Maltoni:2007zf}, we can approximate the primary effect of a non-zero \uesq by considering only its effect on the \nue survival probability \Pee, taken as analogous to \Pmm:
\begin{equation}\label{eq:peeappendix}
\Pee = \left(1-\uesq\right)^2 \Pee^{(3)} + |\ue|^4,
\end{equation}
where $\Pee^{(3)}$ is the standard three-flavor \nue survival probability.  When this extra free parameter is introduced, the limit on \umsq turns out to be correlated with the limit on \uesq, as shown in the sensitivity contours in \cref{fig:dmude}.  With \uesq unconstrained, the expected 90\% limit on \umsq becomes 0.067, 180\% larger than the 0.024 90\% sensitivity limit with the assumption of $\uesq = 0$.  However, once constraints from other experiments are introduced the effect is significantly reduced.  The \cite{Maltoni:2007zf} paper introduces a constraint of $\uesq < 0.012$ at the $1\sigma$ level based on a value from the Bugey~\cite{Declais:1994su} limit around $\dmnew=\val{1}{eV^2}$. Applying this constraint to this analysis leads to a 17\% change in our sensitivity.  In the low-\dmnew region, where our results are most competitive, the change is only 3\%, while at the highest \dmnew's the change can be as large as 30\%.  If instead the non-zero hints from global fits are used as constraints, the change in our limit ranges from 10\% to 40\%, with the larger effects again occurring at higher \dmnew.  A proper accounting of these constraints would require a global fit to multiple experiments introducing \dmnew and \uesq as fit parameters, which is beyond the scope of this analysis; instead we take the approach used in \cite{Maltoni:2007zf} and the atmospheric section of \cite{Kopp:2013vaa} and assume $\uesq = 0$.

\subsection{No three-flavor matter effects in the \fitTwo fit}\label{sec:assumptions:threeflavor}

The main effect of setting $\theta_{13}$ to zero in the \fitTwo fit, eliminating Multi-GeV \nue appearance, was already discussed in \cref{sec:threeponetwo}.  However, this assumption has a second effect: it eliminates the distortion in the \numu survival probability from matter effects in the Earth.  These distortion can be seen in the few-GeV region for the most upward going events ($\cz \approxeq -1$) in \cref{fig:oscillogram:no}.  

Neglecting this matter effect turns out to have little effect on the \utsq limit.  A sensitivity fit using the \fitTwo model to a MC prediction made using the full three-flavor oscillation probability which includes these distortions finds a best fit at \utsq = 0 and \umsq equal to its minimum value (it is binned in log scale and so does not go to zero).  The three flavor distortions in the \numu survival probability turn out to be relatively small (at most a few percent in the PC through-going and stopping \UP samples) and to not affect the through-going \UP samples which are distorted significantly by the sterile matter effects.

\subsection{Sterile-induced fast oscillations}\label{sec:assumptions:dmsq}

\begin{figure*}
  \subfigure[FC Sub-GeV]{
        \label{fig:msqs:subgev}
        \includegraphics[width=\fwid,clip]{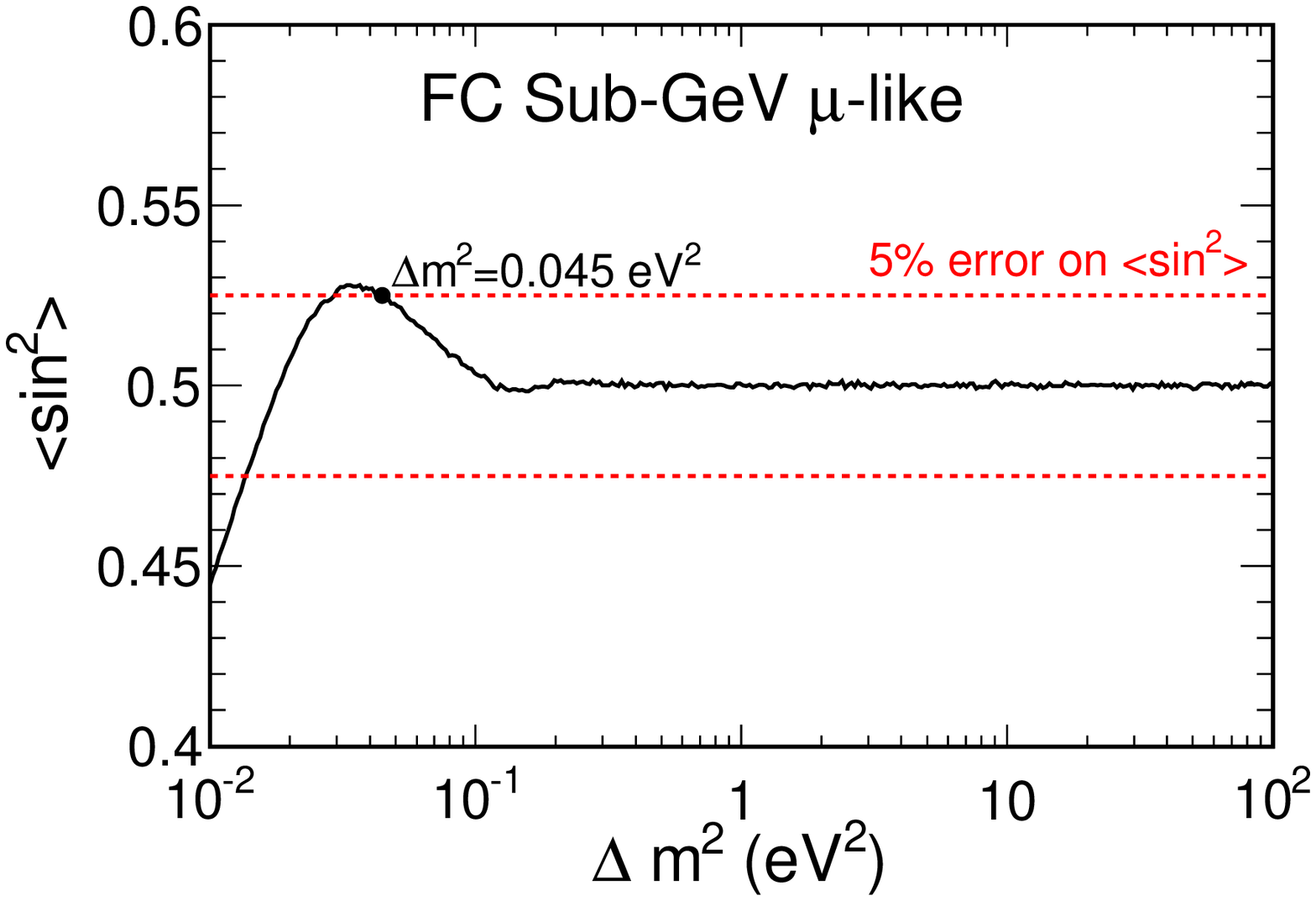}
  }
  \subfigure[FC Sub-GeV]{
        \label{fig:msqs:multigev}
        \includegraphics[width=\fwid,clip]{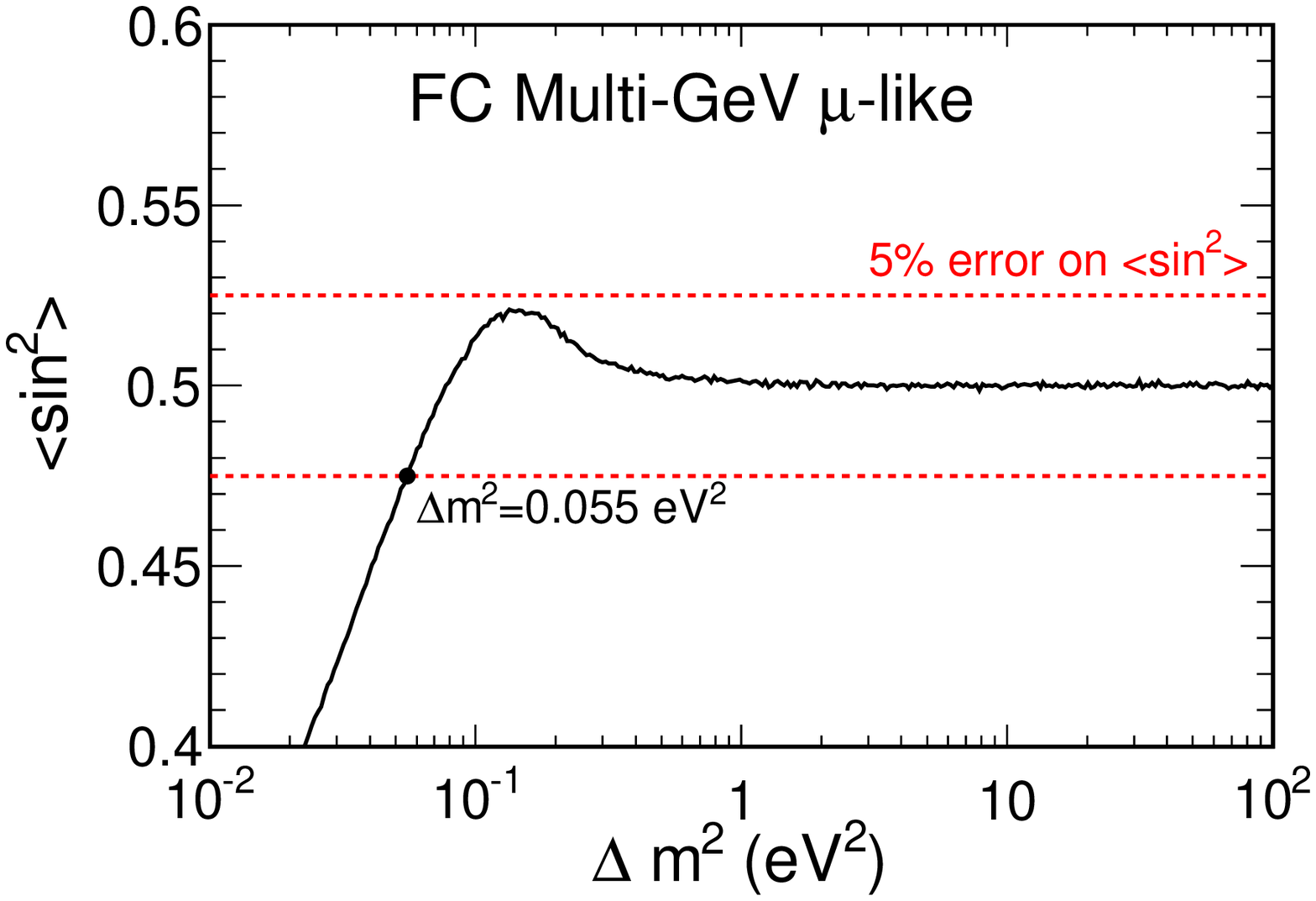}
  }
  \subfigure[PC]{
        \label{fig:msqs:pc}
        \includegraphics[width=\fwid,clip]{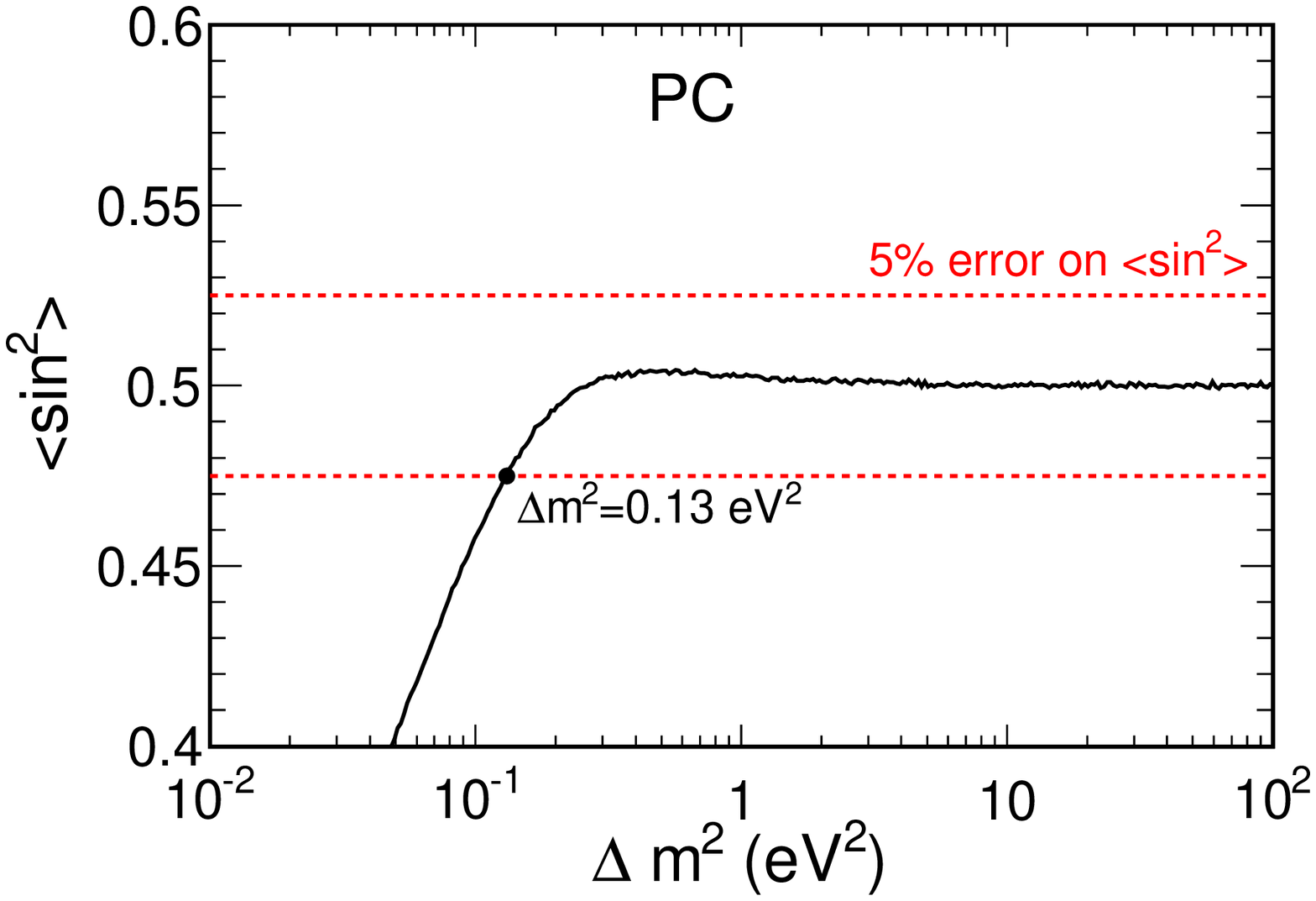}
  }
  \subfigure[\UP]{
        \label{fig:msqs:upmu}
        \includegraphics[width=\fwid,clip]{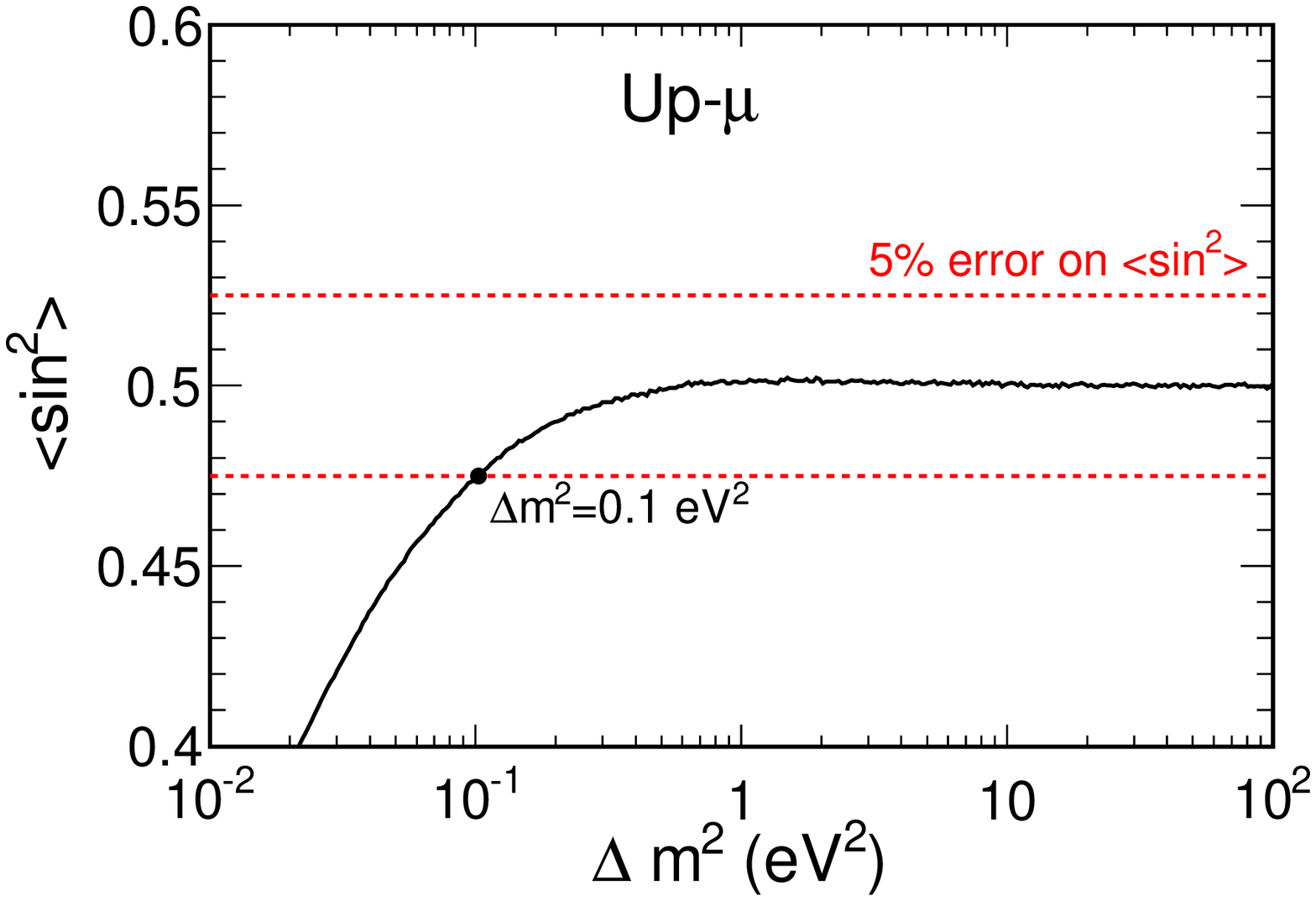}
  }
  \caption{
  The average of $\sin^2 (\dmnew L/4E)$ calculated event-by-event for a range of values of \dmnew in four SK samples: \subref{fig:msqs:subgev} FC Sub-GeV, \subref{fig:msqs:multigev} FC Multi-GeV, \subref{fig:msqs:pc} PC, and \subref{fig:msqs:upmu} \UP.  When the event-by-event average deviates significantly (here defined as 5\%) from $\Braket{\sin^2} = 0.5$, the `fast-oscillation' assumption is no longer valid.
  }
  \label{fig:msqs}
\end{figure*}

This assumption posits that the oscillations driven by \dmnew are so fast that individual oscillation periods cannot be resolved in the experiment and that functions of \dmnew can be replaced with their average values: 
\begin{align}
  \sin \left(\frac{\dmnew L}{4E}\right) &\to \Braket{\sin} =  0 \\
  \sin^2 \left(\frac{\dmnew L}{4E}\right) &\to \Braket{\sin^2} = \frac{1}{2}.
\end{align}
However, since the phase in these terms depends on $L$ and $E$ as well as \dmnew, the ranges over which they are valid could vary for the different samples used in the analysis.  For a sufficiently small \dmnew, this fast oscillation assumption will break down, and the higher the energy and shorter the path length, the larger of a value of \dmnew that is invalid.  We can estimate this lower limit by calculating the value of $\sin^2 (\dmnew L/4E)$ for many MC events in the various SK samples (FC Sub- and Multi-GeV, PC, and \UP) at a range of possible values of \dmnew.  The average is then calculated from the event-by-event values at each \dmnew and the point where the actual average deviates significantly from one half can be found. These averages vs. \dmnew for the four samples can be seen in \cref{fig:msqs}.  

Setting a threshold of 5\% error on the value of $\Braket{\sin^2}$, we find that the fast oscillation sample is valid until approximately $10^{-1}$ in all four samples. The highest limit is 0.13 in the PC sample where there are both high energies and the very short track lengths from down-going events.  

Meeting this assumption only sets the bottom of the valid \dmnew range. The upper limit on the mass for which the limits are valid is set by the requirement that the mass-splitting is sufficiently small that the neutrinos remain coherent.  A sufficiently heavy neutrino, approximately \val{1}{keV} or so, will separate from the other light neutrinos and thus not be able to participate in oscillations.

\FloatBarrier
\section{Zenith Angle and Momentum Distribution}\label{sec:zenith}

\begin{figure*}
 \begin{center}
 \includegraphics[width=\zwid,clip]{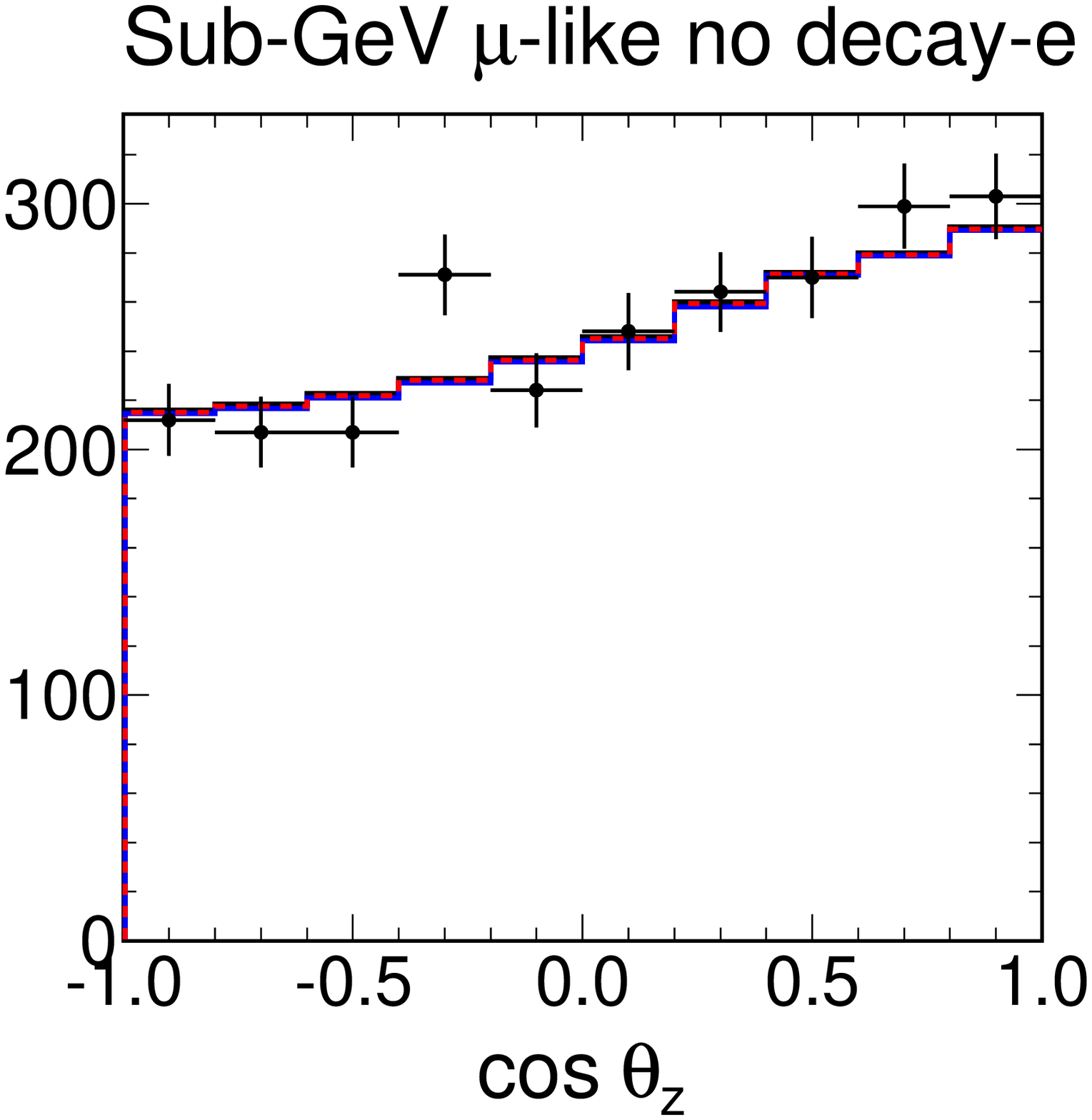}
 \includegraphics[width=\zwid,clip]{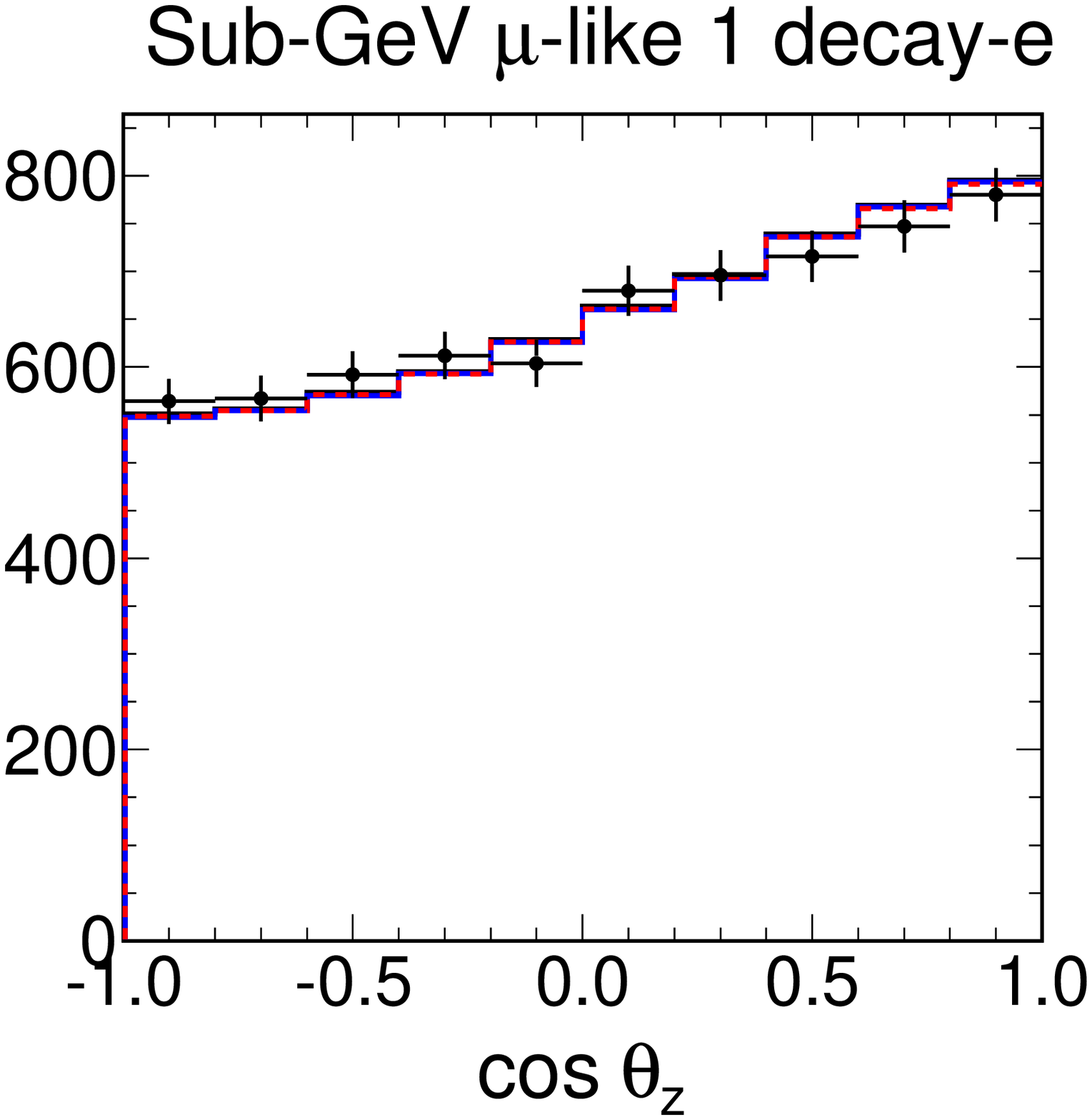}
 \includegraphics[width=\zwid,clip]{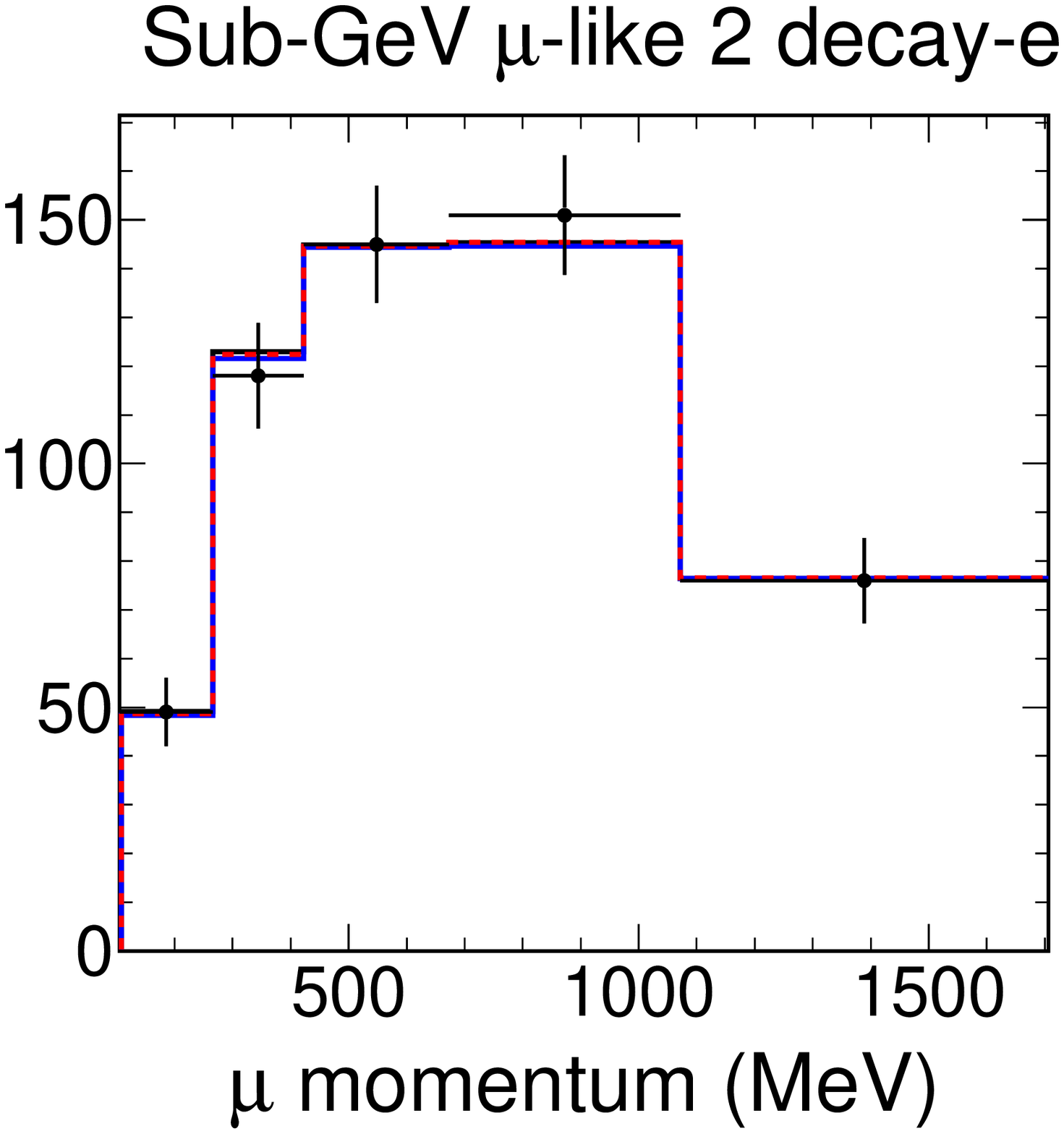} \\
 \includegraphics[width=\zwid,clip]{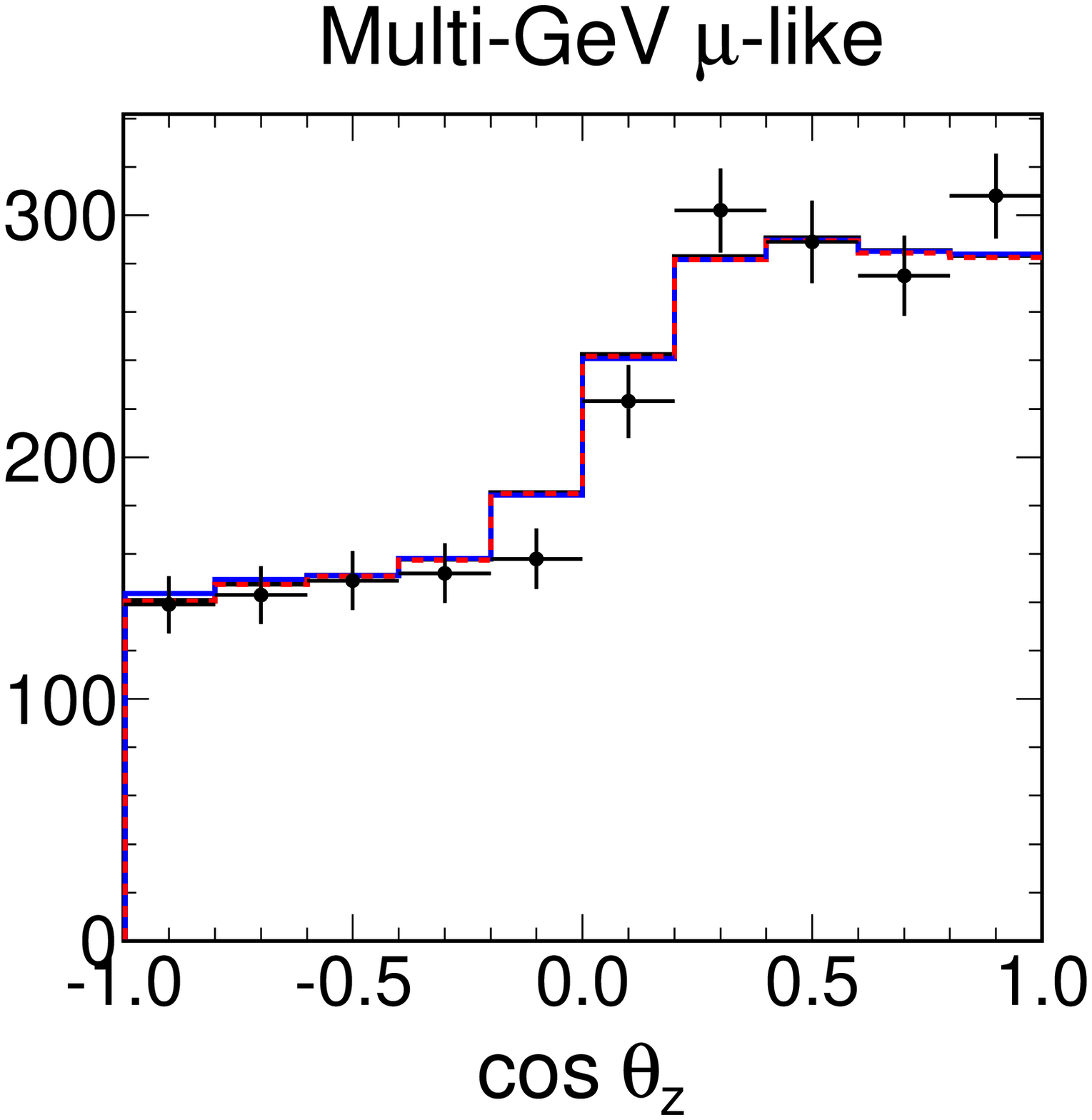}
 \includegraphics[width=\zwid,clip]{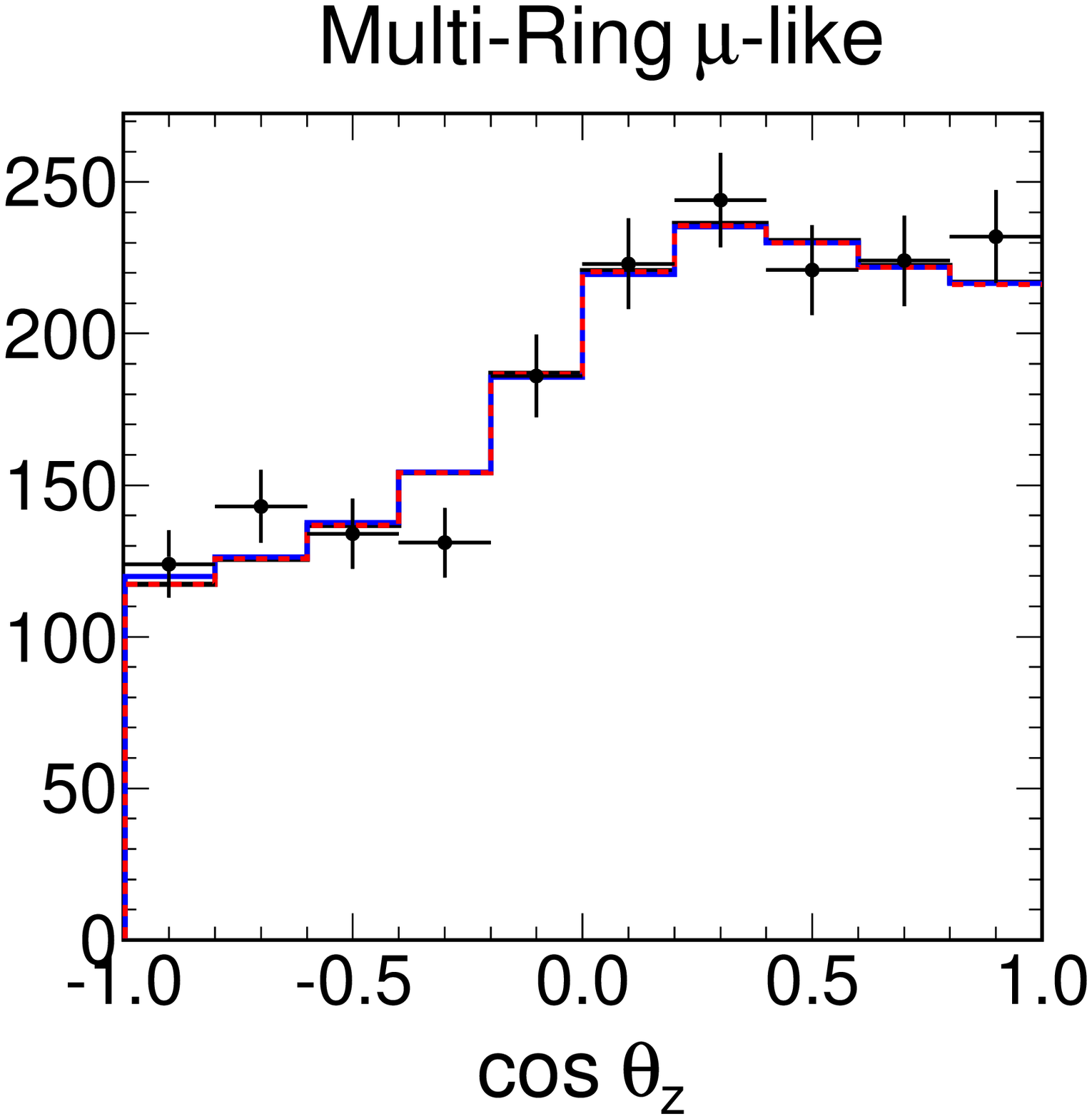}
 \includegraphics[width=\zwid,clip]{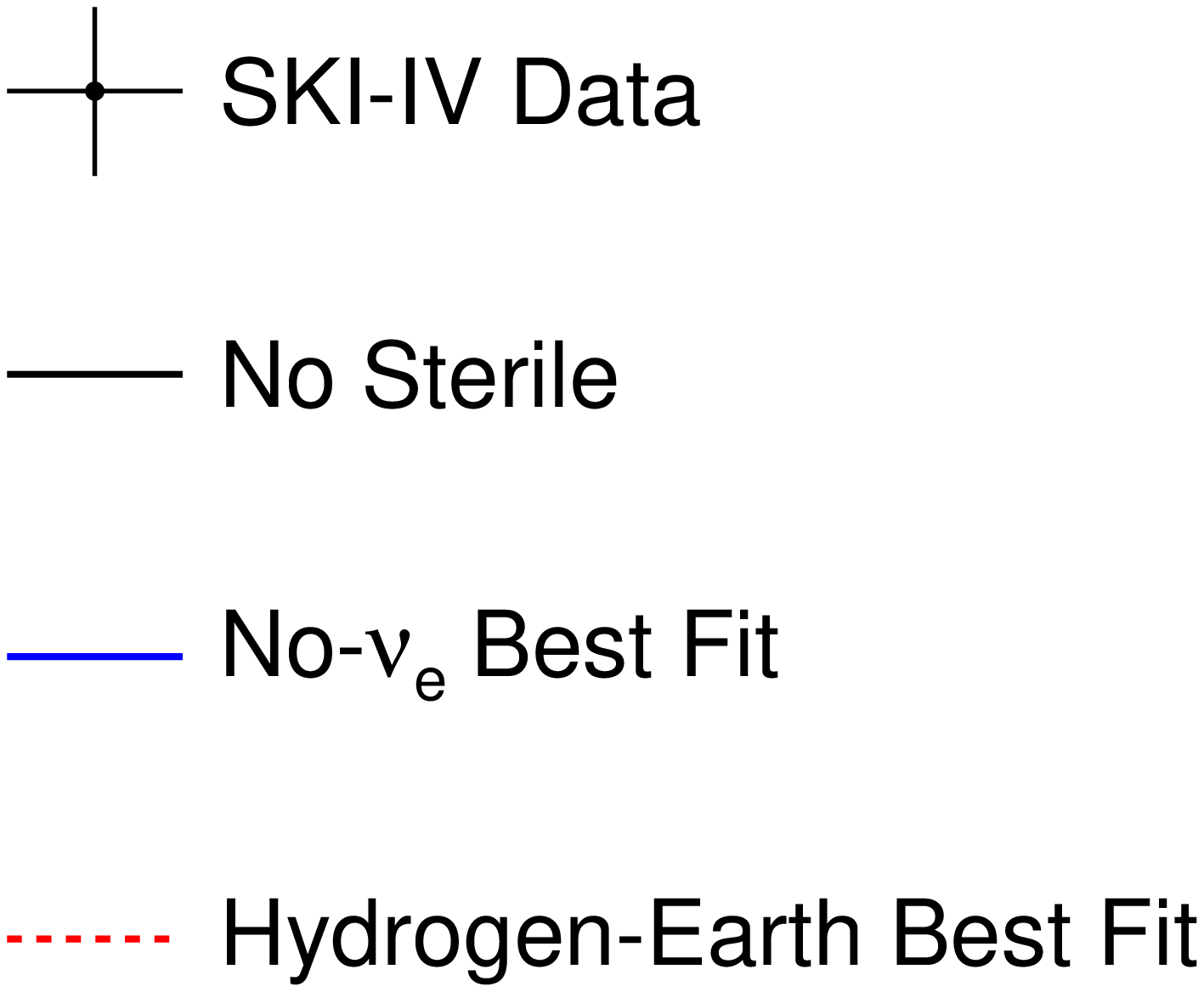} \\
 \includegraphics[width=\zwid,clip]{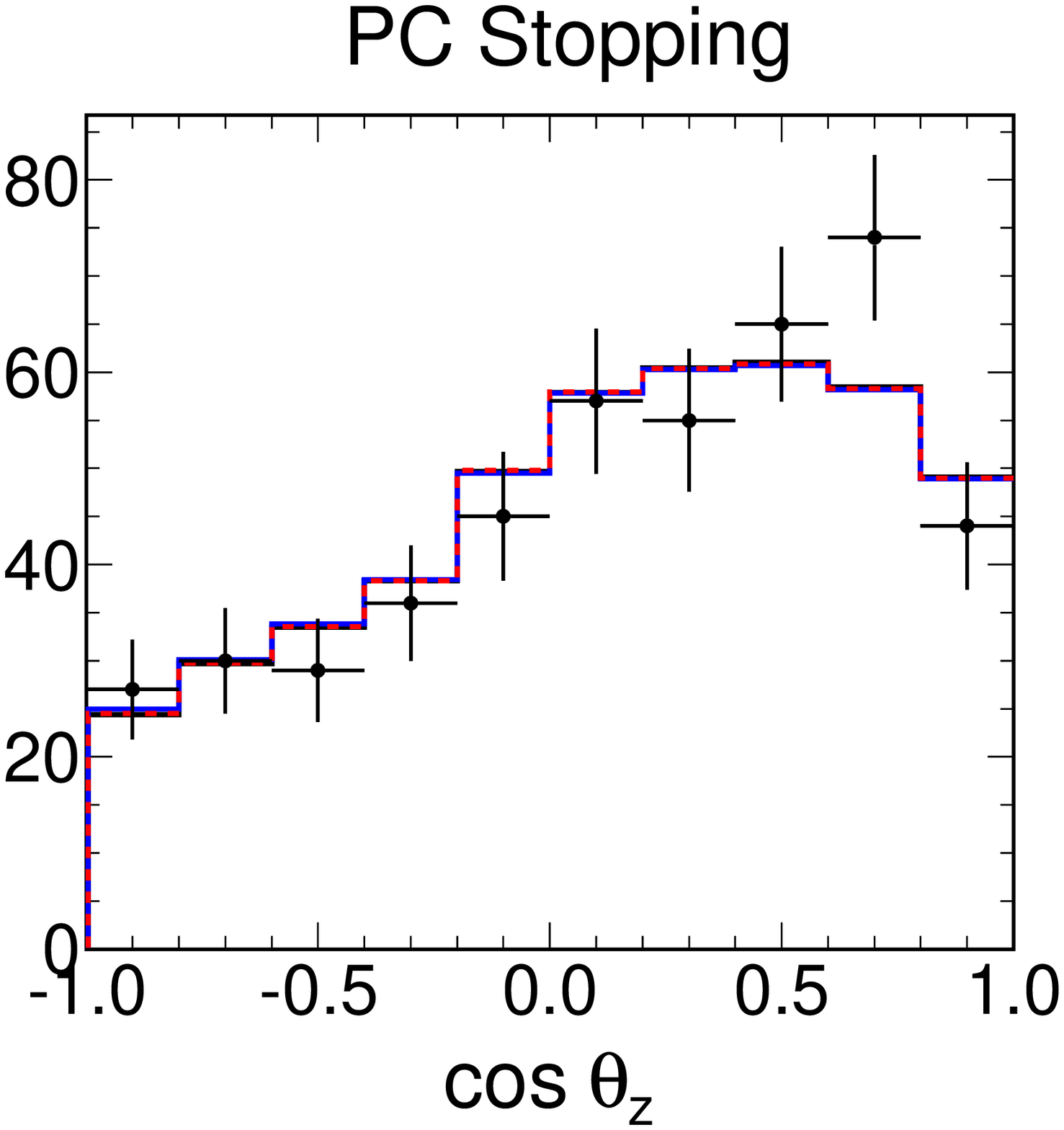}
 \includegraphics[width=\zwid,clip]{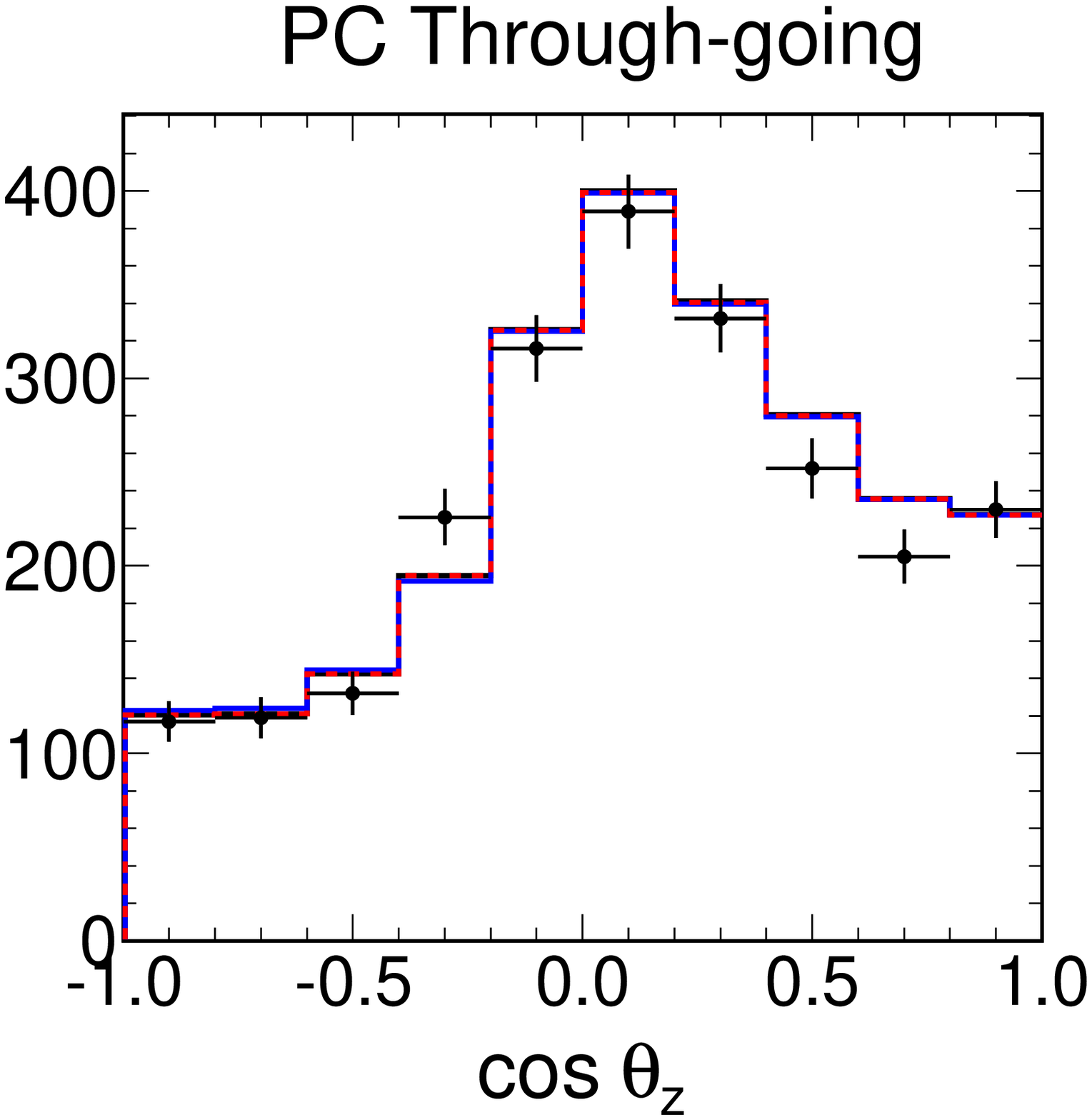}
 \includegraphics[width=\zwid,clip]{appendix_both_bestfit_blank} \\
 \includegraphics[width=\zwid,clip]{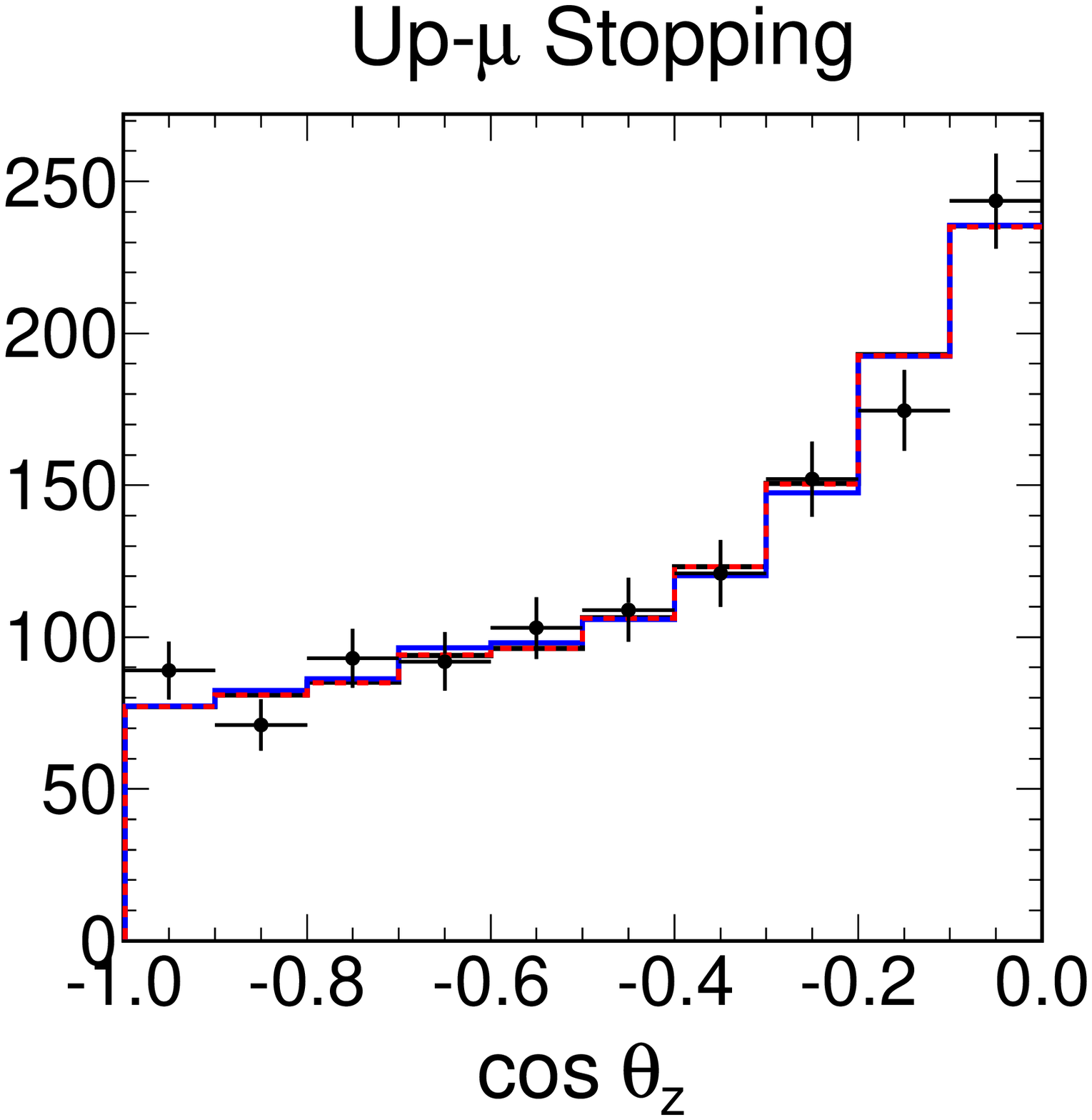}
 \includegraphics[width=\zwid,clip]{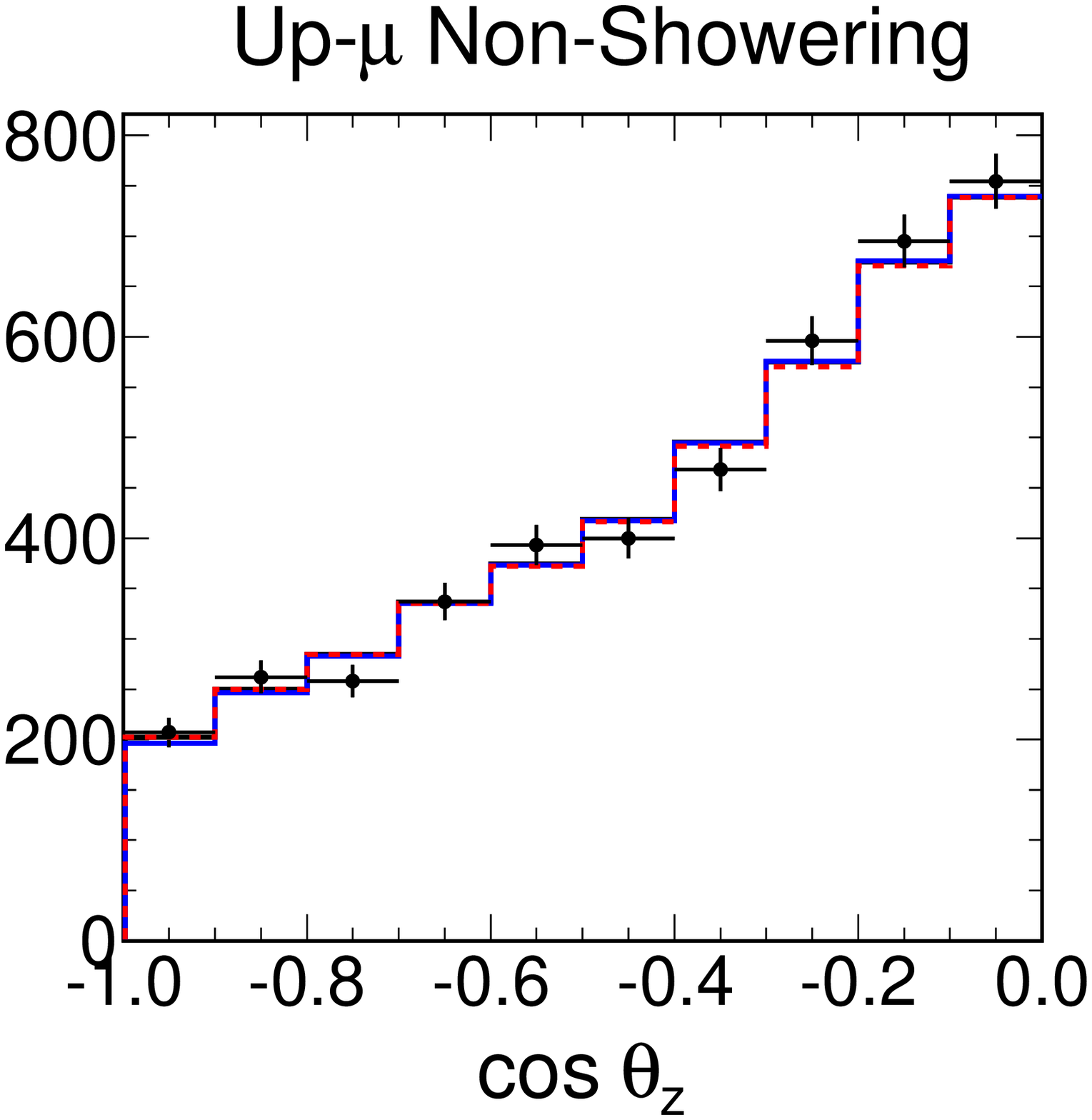}
 \includegraphics[width=\zwid,clip]{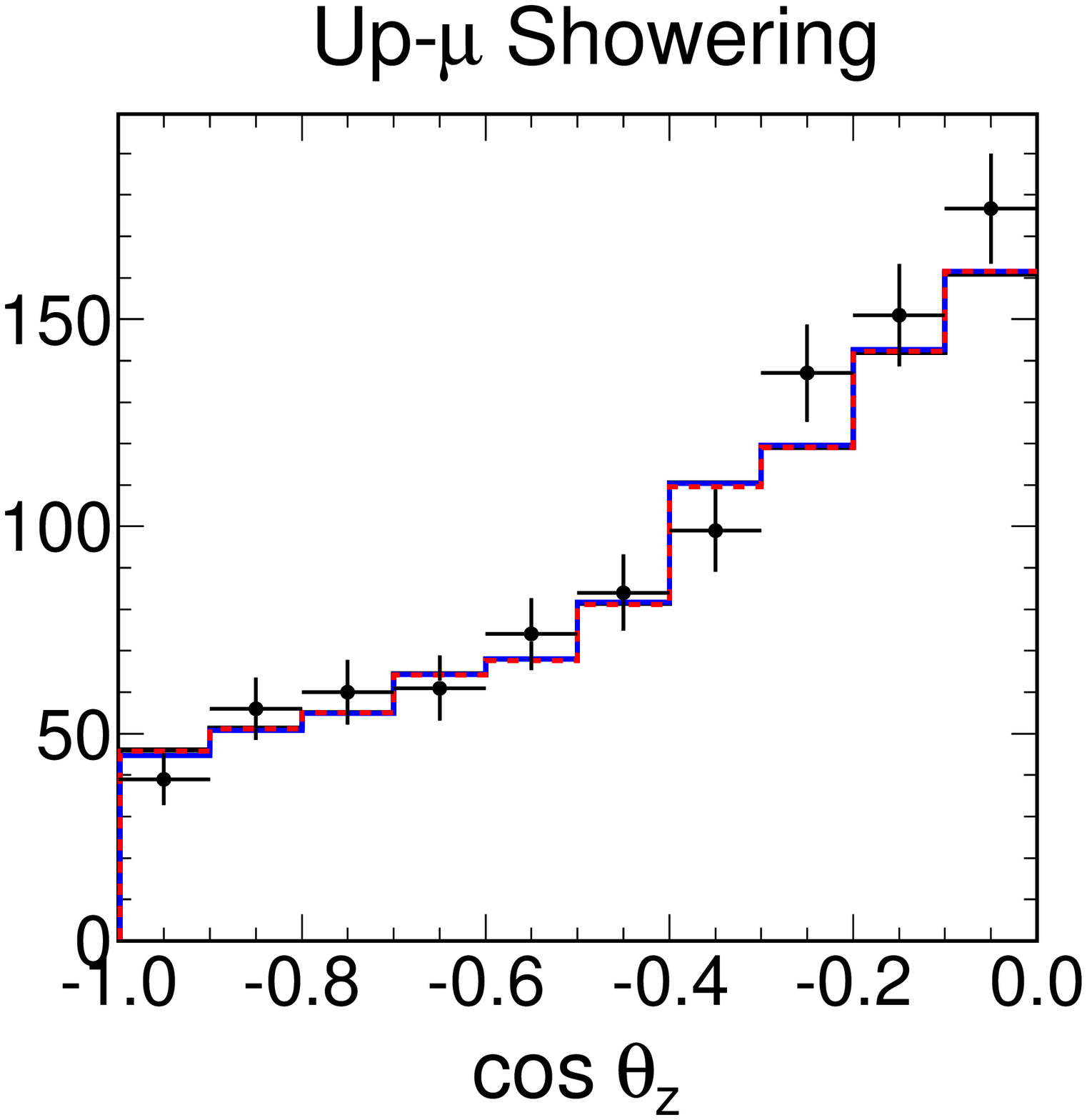}
 \caption{ (color online) Distributions of zenith angle or energy, summed across SK-I through SK-IV, for the $\mu$-like FC, PC, and \UP sub-samples. They are projected into zenith angle when binned in both angle and energy and the Sub-GeV 2 decay-e sample is binned only in momentum. The black points represent the data with statistical error bars, while the solid blue line represents the \fitTwo best fit, the dashed red line represents the \fitOne best fit, and the solid black line represents the MC prediction without sterile neutrinos.}
 \label{fig:zenith_mu}
 \end{center}
\end{figure*}

\begin{figure*}
 \begin{center}
 \includegraphics[width=\zwid,clip]{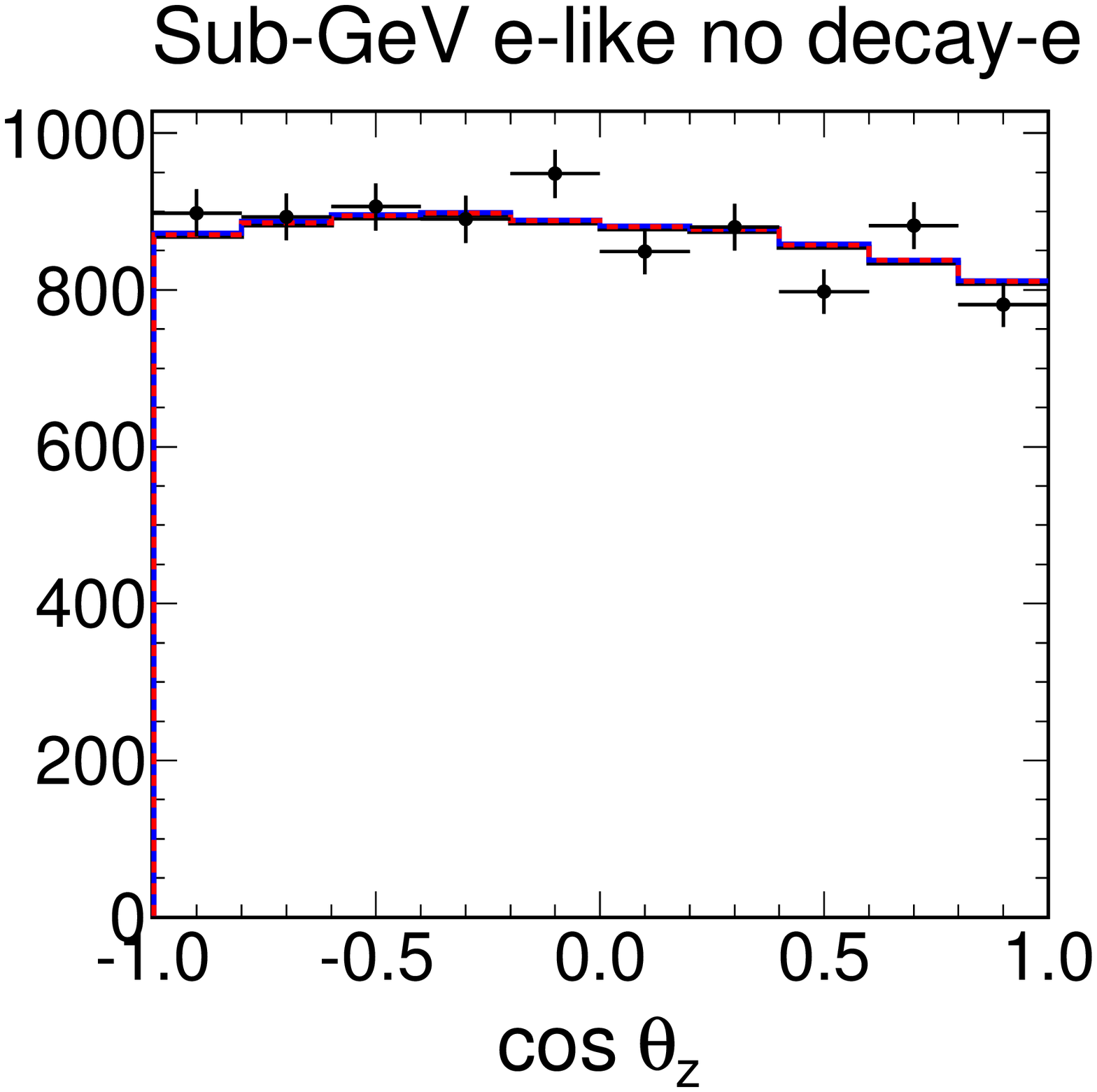}
 \includegraphics[width=\zwid,clip]{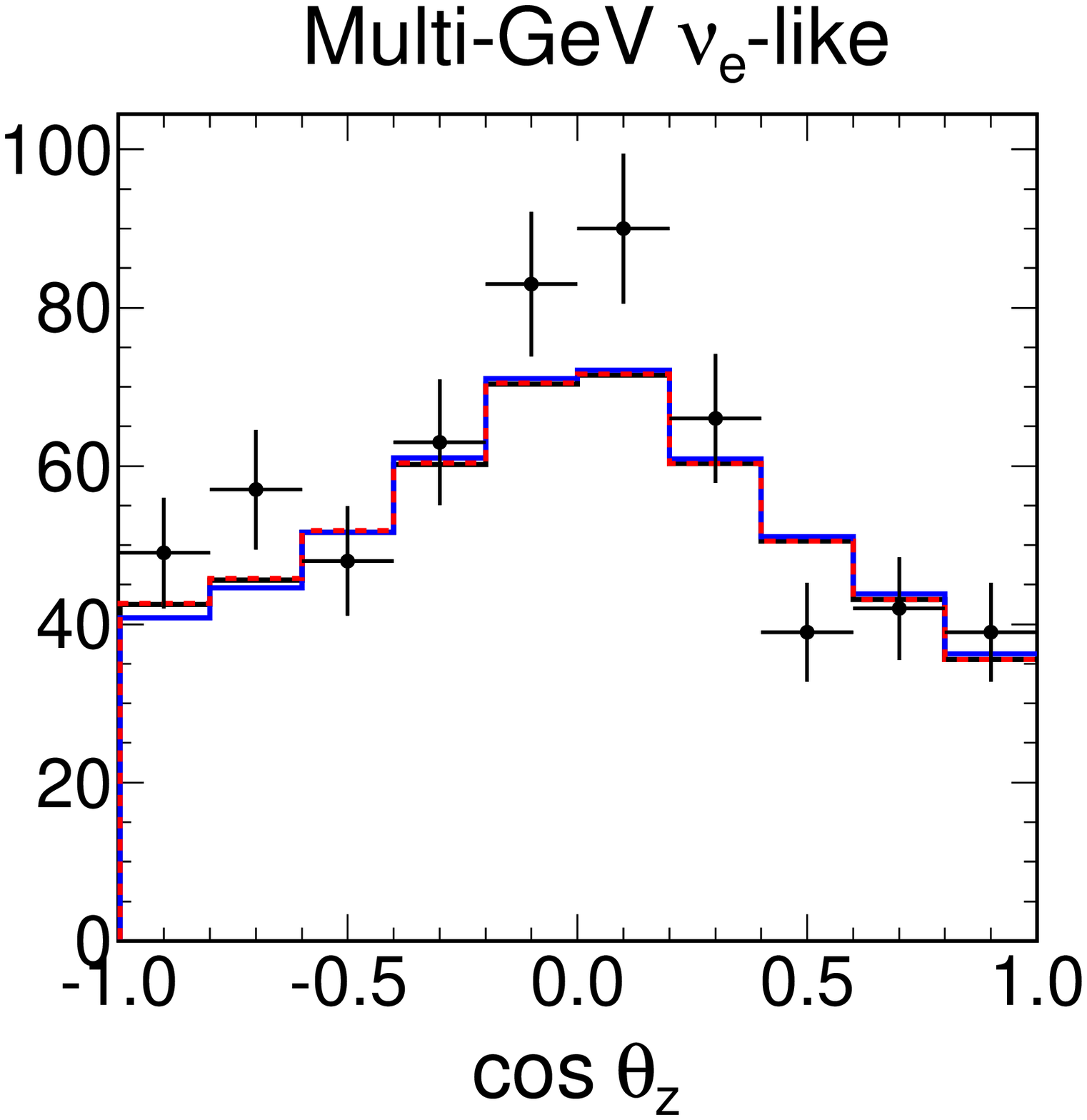}
 \includegraphics[width=\zwid,clip]{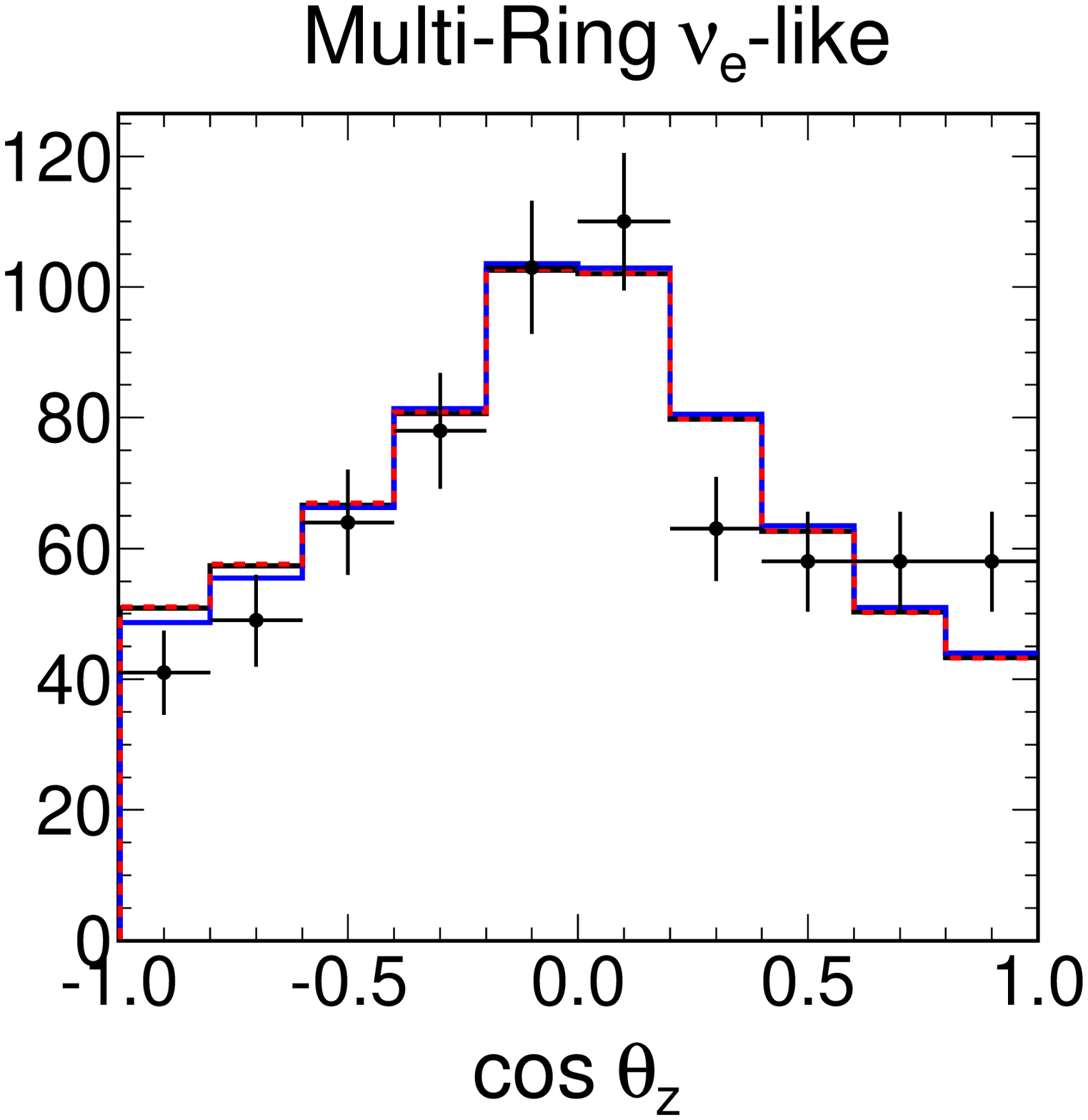} \\
 \includegraphics[width=\zwid,clip]{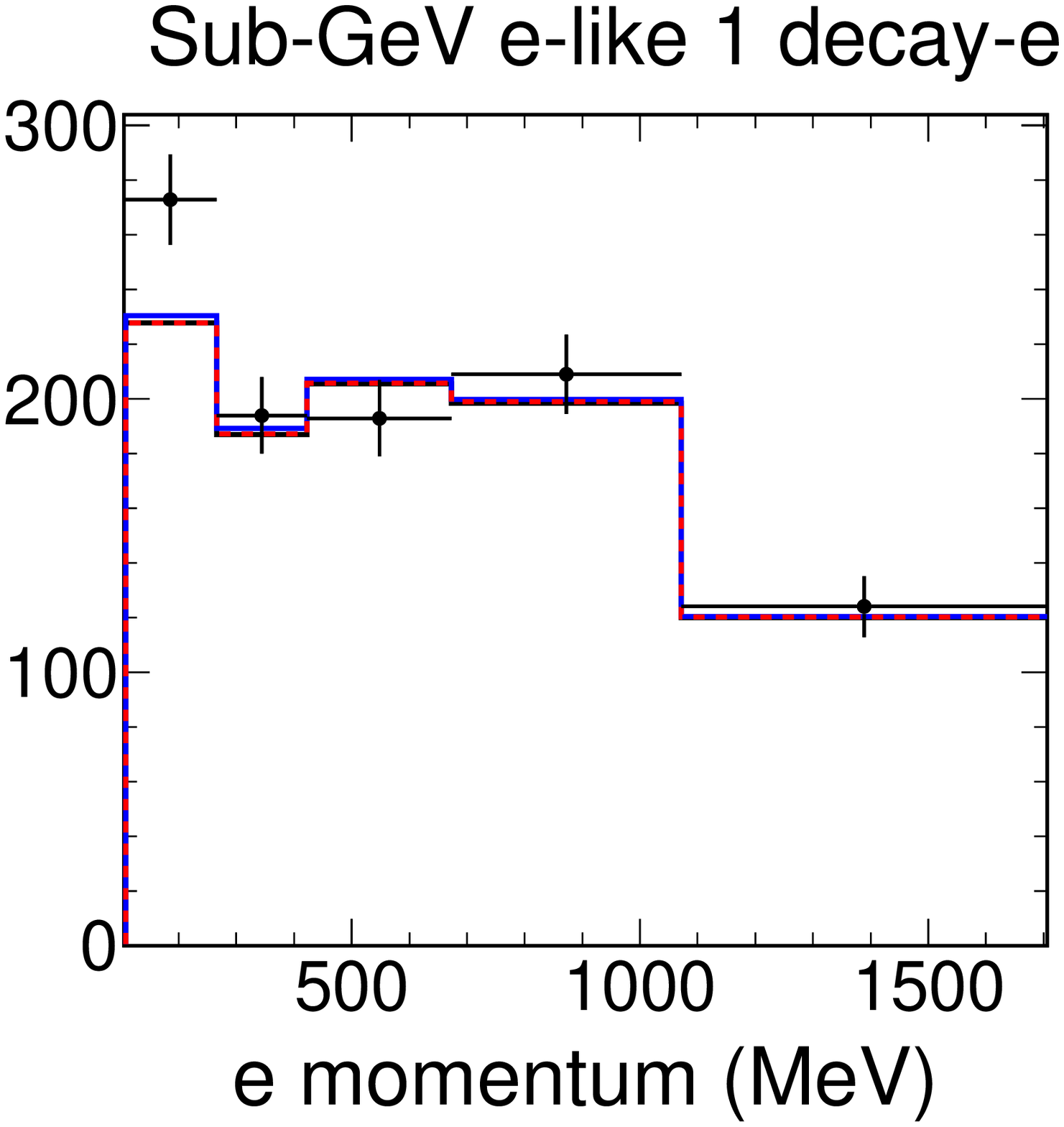} 
 \includegraphics[width=\zwid,clip]{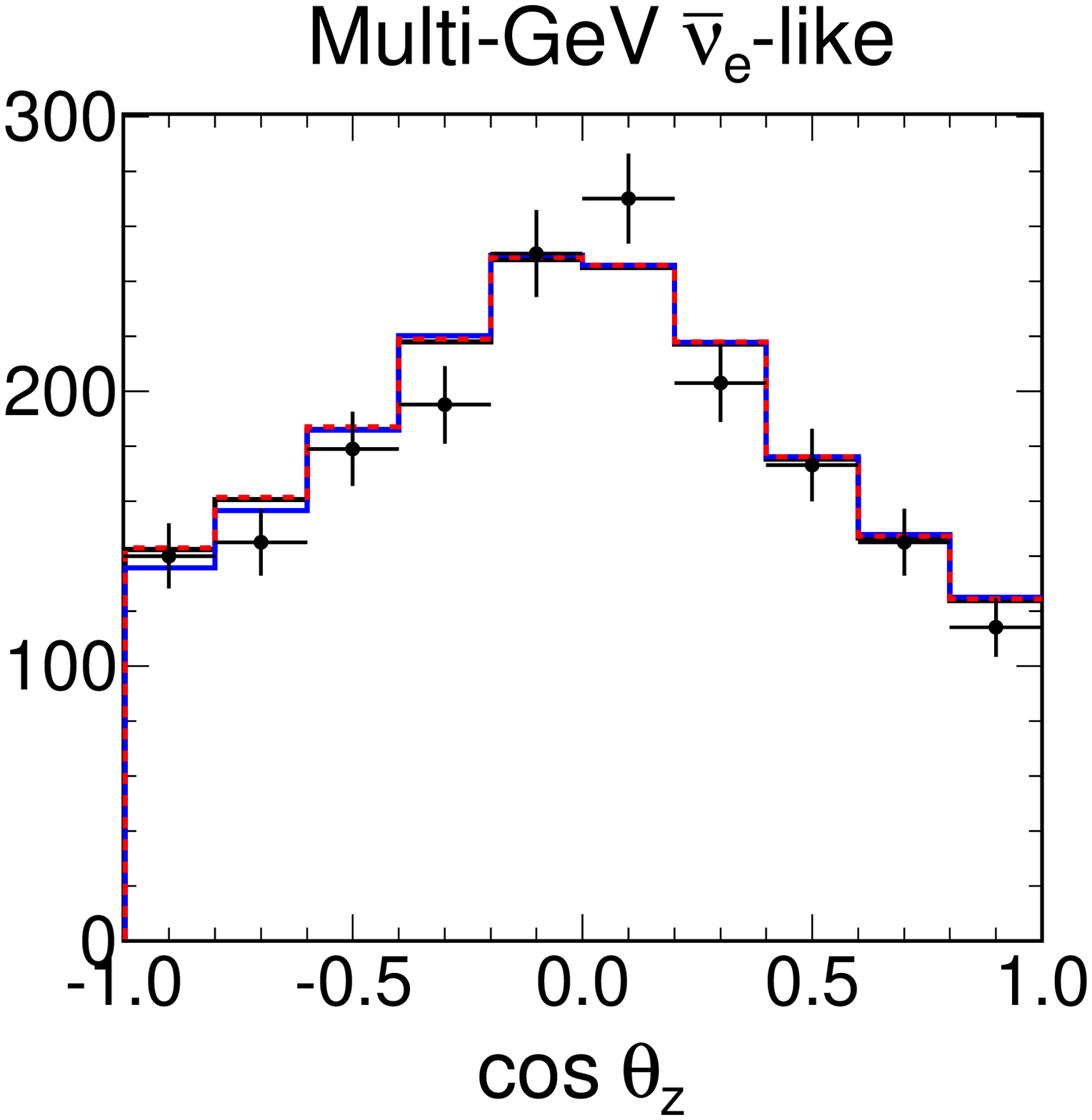}
 \includegraphics[width=\zwid,clip]{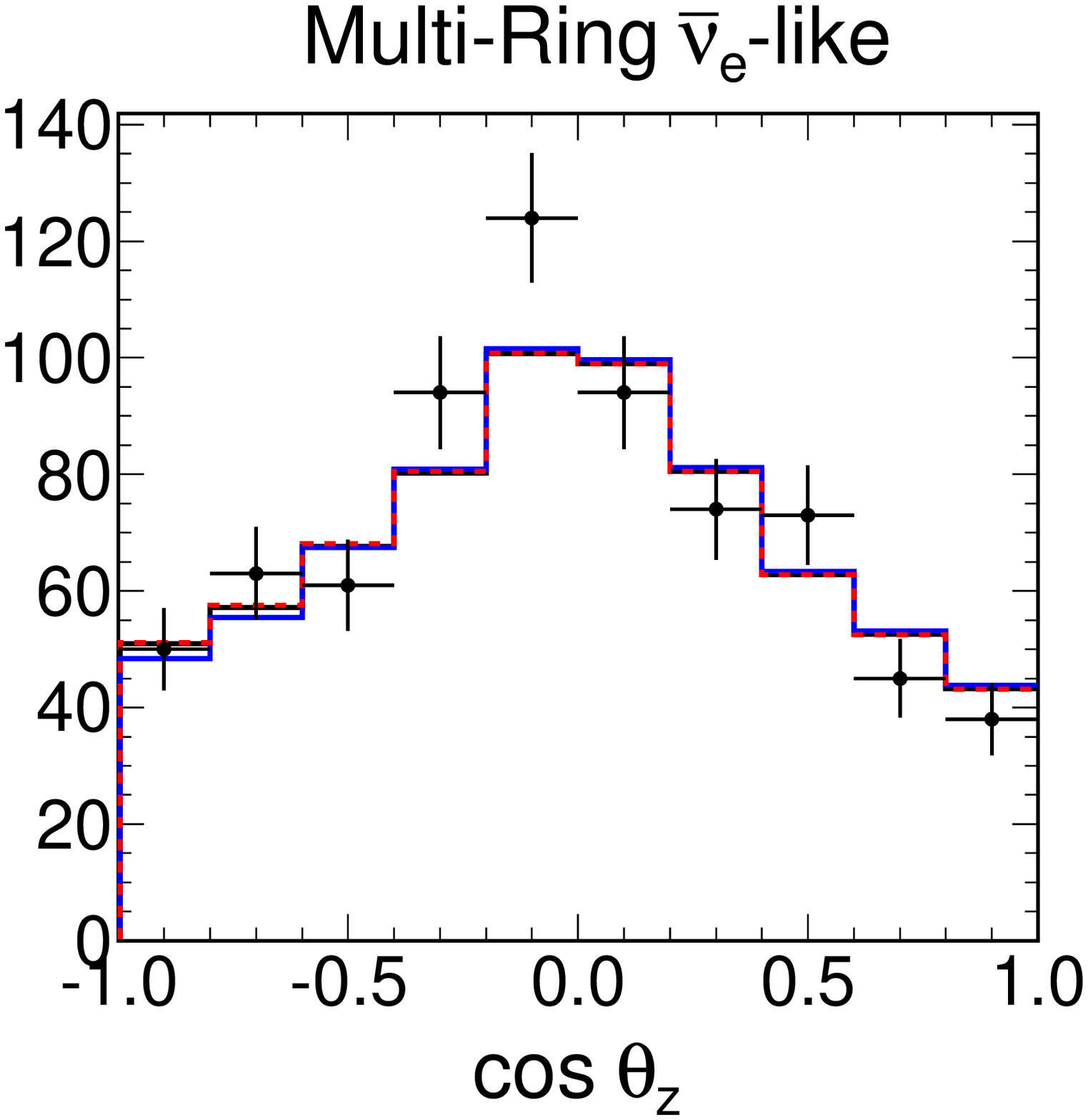}\\
 \includegraphics[width=\zwid,clip]{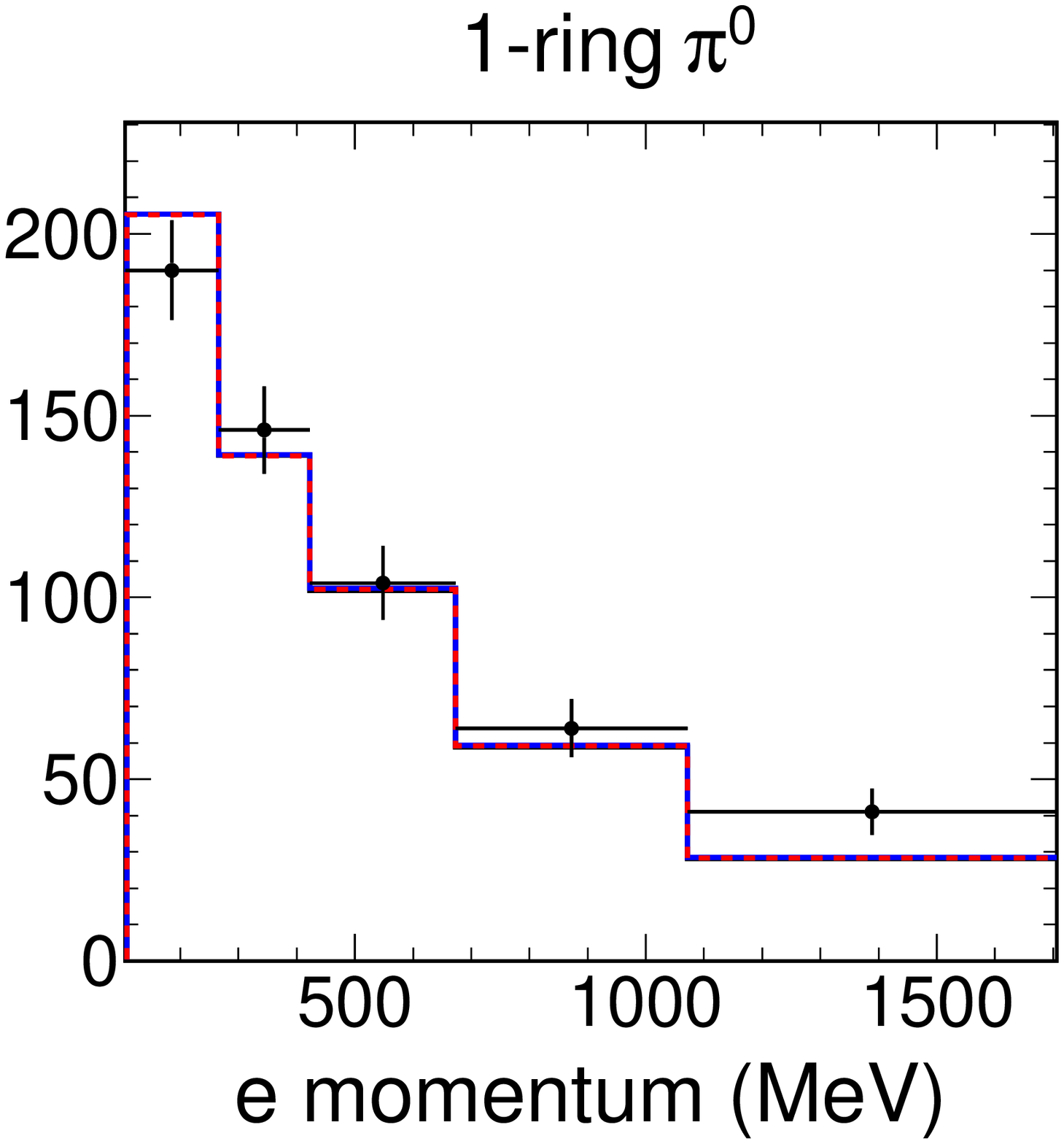}
 \includegraphics[width=\zwid,clip]{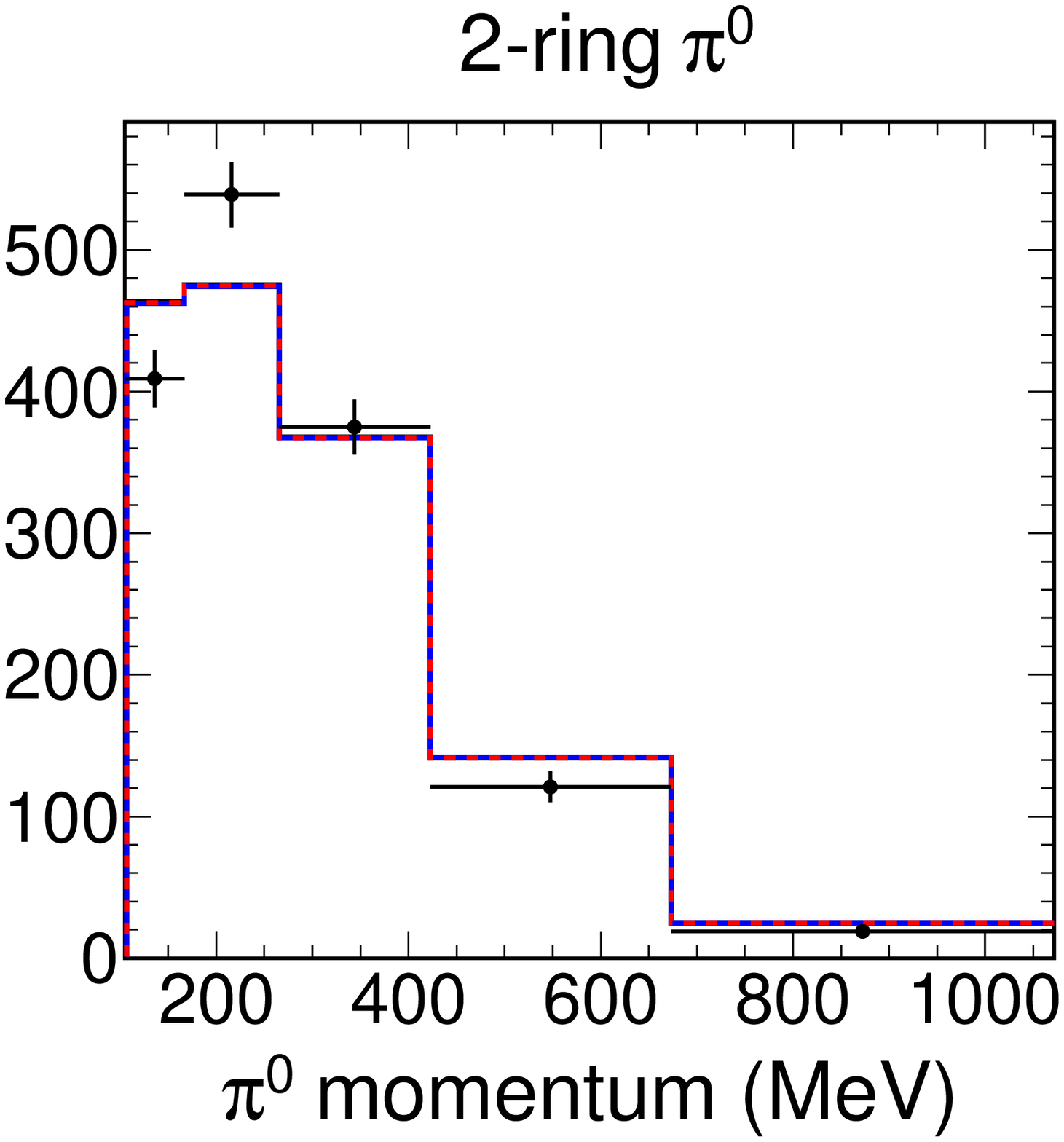}
 \includegraphics[width=\zwid,clip]{appendix_both_bestfit_legend}
 \caption{ (color online) Distributions of zenith angle or energy, summed across SK-I through SK-IV, of the $e$- and NC\pizero-like FC sub-samples. They are projected into zenith angle when binned in both.  In the fits for sterile neutrinos, these samples serve primarily to control the normalization of the atmospheric neutrino flux.  As in \cref{fig:zenith_mu}, the black points represent the data with statistical error bars, while the solid blue line represents the \fitTwo best fit, the dashed red line represents the \fitOne best fit, and the solid black line represents the MC prediction without sterile neutrinos.  The small deviation of the \fitTwo best fit in the Multi-GeV and Multi-Ring $e$-like samples is from setting \th{13} to zero.
 }
 \label{fig:zenith_e}
 \end{center}
\end{figure*}

Below are shown the zenith angle or energy distributions, summed across SK-I through SK-IV, for all the samples in the analysis.  \Cref{fig:zenith_mu} shows the $\mu$-like FC, PC, and \UP sub-samples while \cref{fig:zenith_e} shows the $e$-like and NC$\pi^0$-like samples.  For sub-samples binned in both zenith angle and energy, the projection into only zenith angle is shown.  The plots show the data represented by points with statistical error bars as well as the best fits from the two analyses (\fitTwo as solid blue and \fitOne as dashed red) as well as the MC prediction without sterile neutrinos (represented by a black line), with systematic uncertainties still fit to the data.  

The best fits generally agree quite well with the prediction without sterile neutrinos, though both fits favor a non-zero sterile oscillation component because it allows for some systematic uncertainties to fit closer to their nominal values, as discussed in \cref{sec:hydrogen}.

\FloatBarrier
\section{Systematic Uncertainties}\label{sec:systematics}

\Cref{tab:sysa,tab:sysb,tab:sysc} summarize the best fit systematic error parameters for the best fit point from the \fitOne
analysis.  The pull values for any given systematic error are expected to be normally distributed across many experiments.  For this particular data set, the pull values approximately follow a gaussian distribution, though with a width narrower than one due to the interaction between the low-energy normalization uncertainties and the sterile parameter \umsq described in \cref{sec:threeponeone}.

\renewcommand{\arraystretch}{1.0}
\newcolumntype{E}{D{.}{.}{-1}}

\begin{table*}[b!]
\centering
\begin{tabular}{lllEE}
\hline \hline
\multicolumn{3}{l}{Systematic Error} & \multicolumn{1}{c}{Fit Value (\%)} & \multicolumn{1}{c}{$\sigma$ (\%)} \\
\hline
Flux normalization                              & $E_\nu < \val{1}{GeV}$\footnote[1]{Uncertainty decreases linearly with $\log E_{\nu}$ from 25\,\%(0.1\,GeV) to 7\,\%(1\,GeV).} &                    
&     21 &    25 \\
                                                & $E_\nu > \val{1}{GeV}$\footnote[2]{Uncertainty is 7\,\% up to 10\,GeV, linearly increases with $\log E_{\nu}$ from 7\,\%(10\,GeV) to 12\,\%(100\,GeV) and then to 20\,\%(1\,TeV)} &                    
&    1.7 &    15 \\
$(\numu+\numubar)/(\nue+\nuebar)$               & \multicolumn{2}{l}{$E_\nu < \val{1}{GeV}$}            
&  -0.25 &     2 \\
                                                & $1 < E_\nu < \val{10}{GeV}$      &                    
&  -0.26 &     3 \\
                                                & $E_\nu > \val{10}{GeV}$\footnote[3] {Uncertainty linearly increases with $\log E_{\nu}$ from 5\,\%(30\,GeV) to 30\,\%(1\,TeV).} &                    
&    6.7 &     5 \\
$\nuebar/\nue$                                  & \multicolumn{2}{l}{$E_\nu < \val{1}{GeV}$}            
&    2.5 &     5 \\
                                                & $1 < E_\nu < \val{10}{GeV}$      &                    
&    2.6 &     5 \\
                                                & $E_\nu > \val{10}{GeV}$\footnote[4] {Uncertainty linearly increases with $\log E_{\nu}$ from 8\,\%(100\,GeV) to 20\,\%(1\,TeV).} &                    
&    2.6 &     8 \\
$\numubar/\numu$                                & \multicolumn{2}{l}{$E_\nu < \val{1}{GeV}$}            
&  0.021 &     2 \\
                                                & $1 < E_\nu < \val{10}{GeV}$      &                    
&    1.9 &     6 \\
                                                & $E_\nu > \val{10}{GeV}$\footnote[5] {Uncertainty linearly increases with $\log E_{\nu}$ from 6\,\%(50\,GeV) to 40\,\%(1\,TeV).} &                    
&    4.2 &    15 \\
Up/down ratio                                   & $< \val{400}{MeV}$               & $e$-like           
& -0.0037 &   0.1 \\
                                                &                                  & $\mu$-like         
& -0.011 &   0.3 \\
                                                &                                  & 0-decay $\mu$-like 
& -0.041 &   1.1 \\
                                                & $> \val{400}{MeV}$               & $e$-like           
& -0.029 &   0.8 \\
                                                &                                  & $\mu$-like         
& -0.018 &   0.5 \\
                                                &                                  & 0-decay $\mu$-like 
& -0.063 &   1.7 \\
                                                & Multi-GeV                        & $e$-like           
& -0.026 &   0.7 \\
                                                &                                  & $\mu$-like         
& -0.0074 &   0.2 \\
                                                & Multi-ring Sub-GeV               & $e$-like           
& -0.015 &   0.4 \\
                                                &                                  & $\mu$-like         
& -0.0074 &   0.2 \\
                                                & Multi-ring Multi-GeV             & $e$-like           
& -0.011 &   0.3 \\
                                                &                                  & $\mu$-like         
& -0.0074 &   0.2 \\
                                                & PC                               &                    
& -0.0074 &   0.2 \\
Horizontal/vertical ratio                       & $< \val{400}{MeV}$               & $e$-like           
&  0.011 &   0.1 \\
                                                &                                  & $\mu$-like         
&  0.011 &   0.1 \\
                                                &                                  & 0-decay $\mu$-like 
&  0.033 &   0.3 \\
                                                & $> \val{400}{MeV}$               & $e$-like           
&   0.15 &   1.4 \\
                                                &                                  & $\mu$-like         
&   0.21 &   1.9 \\
                                                &                                  & 0-decay $\mu$-like 
&   0.15 &   1.4 \\
                                                & Multi-GeV                        & $e$-like           
&   0.35 &   3.2 \\
                                                &                                  & $\mu$-like         
&   0.25 &   2.3 \\
                                                & Multi-ring Sub-GeV               & $e$-like           
&   0.15 &   1.4 \\
                                                &                                  & $\mu$-like         
&   0.14 &   1.3 \\
                                                & Multi-ring Multi-GeV             & $e$-like           
&   0.31 &   2.8 \\
                                                &                                  & $\mu$-like         
&   0.17 &   1.5 \\
                                                & PC                               &                    
&   0.19 &   1.7 \\
\multicolumn{3}{l}{K/$\pi$ ratio in flux calculation\footnote[6] {Uncertainty increases linearly from 5$\%$ to 20$\%$ between 100GeV and 1TeV.}} 
&    1.3 &    10 \\
\multicolumn{3}{l}{Neutrino path length}                                                                
&  0.094 &    10 \\
Sample-by-sample                                & \multicolumn{2}{l}{FC Multi-GeV}                      
&   -5.8 &     5 \\
                                                & PC + Stopping \UP                &                    
&   0.79 &     5 \\
\multicolumn{3}{l}{Matter effects}                                                                      
&    1.8 &   6.8 \\

\hline\hline
\end{tabular}
\caption{
Flux-related systematic errors that are common to all SK run periods. 
The flux uncertainties come from the Honda flux calculation~\cite{Honda:2011nf} and are themselves based on the external data sets used as inputs to the calculation.
The second column shows the best fit value of the systematic error parameter, $\epsilon_j$, in percent and the third column shows the estimated 1-$\sigma$ error size in percent.
\label{tab:sysa}
}
\end{table*}

\begin{table*}
\centering
\begin{tabular}{lllEE}
\hline \hline
Systematic Error & &  & \multicolumn{1}{c}{Fit Value (\%)} & \multicolumn{1}{c}{$\sigma$ (\%)} \\
\hline
$M_A$ in QE and single $\pi$                    &                                  &                    
&   -6.4 &    10 \\
\multicolumn{3}{l}{CCQE cross section\footnote[1] {Difference from the Nieves~\cite{Nieves:2004wx} model is set to 1.0}} 
&    1.8 &    10 \\
\multicolumn{3}{l}{CCQE $\bar \nu/\nu$ ratio\footnotemark[1]}                                           
&     18 &    10 \\
\multicolumn{3}{l}{CCQE $\mu/e$ ratio\footnotemark[1]}                                                  
&   0.12 &    10 \\
\multicolumn{3}{l}{Single meson production cross section}                                               
&     14 &    20 \\
\multicolumn{3}{l}{DIS cross section}                                                                   
&    2.2 &     5 \\
\multicolumn{3}{l}{DIS model comparisons\footnote[2] {Difference from CKMT~\cite{CKMT94} parametrization is set to 1.0}} 
&   -1.5 &    10 \\
\multicolumn{3}{l}{DIS $Q^2$ distribution (high W)\footnote[3] {Difference from GRV98~\cite{Gluck:1998xa} is set to 1.0}} 
&  0.003 &    10 \\
\multicolumn{3}{l}{DIS $Q^2$ distribution (low W)\footnotemark[3]}                                      
&   -3.1 &    10 \\
\multicolumn{3}{l}{Coherent $\pi$ production}                                                           
&    1.8 &   100 \\
\multicolumn{3}{l}{NC/CC}                                                                               
&    9.8 &    20 \\
\multicolumn{3}{l}{\nutau cross section}                                                                
&   -4.6 &    25 \\
\multicolumn{3}{l}{Single $\pi$ production, $\pizero/\pi^\pm$}                                          
&    -35 &    40 \\
\multicolumn{3}{l}{Single $\pi$ production, $\bar \nu_{i} /\nu_{i}$ (i=$e,\mu $)\footnote[4] {Difference from the Hernandez\cite{Hernandez07} model is set to 1.0}} 
&    -11 &    10 \\
\multicolumn{3}{l}{NC fraction from hadron simulation}                                                  
&     -3 &    10 \\
$\pi^+$ decay uncertainty Sub-GeV 1-ring        & \multicolumn{2}{l}{$e$-like 0-decay}                  
&  -0.48 &   0.6 \\
                                                & $\mu$-like 0-decay               &                    
&  -0.64 &   0.8 \\
                                                & $e$-like 1-decay                 &                    
&    3.3 &   4.1 \\
                                                & $\mu$-like 1-decay               &                    
&   0.71 &   0.9 \\
                                                & $\mu$-like 2-decay               &                    
&    4.5 &   5.7 \\
\multicolumn{3}{l}{\dmsq{32} \cite{Abe:2014ugx}}                                                        
&      2 &  3.98 \\
\multicolumn{3}{l}{\sn{23} \cite{Abe:2014ugx}}                                                          
&    2.8 &  10.9 \\
\multicolumn{3}{l}{\dmsq{21} \cite{Abe:2010hy}}                                                         
&  0.079 &  2.55 \\
\multicolumn{3}{l}{\sn{12} \cite{Abe:2010hy}}                                                           
&   0.42 &  6.89 \\
\multicolumn{3}{l}{\snt{13} \cite{PDG}}                                                                 
&  -0.55 &  10.5 \\

\hline\hline
\end{tabular}
\caption{
Neutrino interaction, particle production, and PMNS oscillation parameter systematic errors that are common to all SK run periods.  
These uncertainties come primarily from comparisons between different cross section models and external neutrino interaction measurements.  The neutrino oscillation parameter errors come from the cited measurements.
The second column shows the best fit value of the systematic error parameter, $\epsilon_j$, in percent and the third column shows the estimated 1-$\sigma$ error size in percent.}
\label{tab:sysb}
\end{table*}

\renewcommand{\tabcolsep}{0pt}

\begin{table*}
\centering
\begin{tabular}{lllEEEEEEEE}
\hline \hline
 & & 
 & \multicolumn{2}{c}{SK-I} 
 & \multicolumn{2}{c}{SK-II} 
 & \multicolumn{2}{c}{SK-III} 
 & \multicolumn{2}{c}{SK-IV} 
\\
 \multicolumn{3}{l}{Systematic Error} 
 & \multicolumn{1}{c}{Fit Value} & \multicolumn{1}{c}{$\sigma$}
 & \multicolumn{1}{c}{Fit Value} & \multicolumn{1}{c}{$\sigma$}
 & \multicolumn{1}{c}{Fit Value} & \multicolumn{1}{c}{$\sigma$}
 & \multicolumn{1}{c}{Fit Value} & \multicolumn{1}{c}{$\sigma$} 
\\
\hline
FC reduction                                    &                                  &                    
&  0.006 &   0.2 &  0.007 &   0.2 &  0.038 &   0.8 &  0.030 &   0.3 \\
\multicolumn{3}{l}{PC reduction}                                                                        
&  -0.99 &   2.4 &  -3.47 &   4.8 & -0.041 &   0.5 &  -0.24 &     1 \\
\multicolumn{3}{l}{FC/PC separation}                                                                    
& -0.027 &   0.6 &  0.081 &   0.5 &  0.003 &   0.9 & 0.0001 &  0.02 \\
\multicolumn{3}{l}{PC stopping/through-going separation (bottom)}                                       
&  -22.4 &    23 &    0.2 &    13 &   -0.2 &    12 &  -1.06 &   6.8 \\
\multicolumn{3}{l}{PC stopping/through-going separation (barrel)}                                       
&   1.88 &     7 &  -5.54 &   9.4 &   -9.0 &    29 &  -0.65 &   8.5 \\
\multicolumn{3}{l}{PC stopping/through-going separation (top)}                                          
&    8.3 &    46 &   -3.3 &    19 &   16.0 &    87 &   -3.3 &    40 \\
Non-$\nu$ background                            & \multicolumn{2}{l}{Sub-GeV $\mu$-like}                
&  0.009 &   0.1 &  0.009 &   0.1 & -0.009 &   0.1 & -0.026 &   0.1 \\
                                                & Multi-GeV $\mu$-like             &                    
&  0.036 &   0.4 &  0.009 &   0.1 & -0.009 &   0.1 & -0.026 &   0.1 \\
                                                & Sub-GeV 1-ring 0-decay $\mu$-like &                    
&  0.009 &   0.1 &  0.009 &   0.1 & -0.018 &   0.2 & -0.211 &   0.8 \\
                                                & PC                               &                    
&  0.018 &   0.2 &  0.062 &   0.7 &  -0.16 &   1.8 & -0.129 &  0.49 \\
                                                & Sub-GeV $e$-like                 &                    
&  0.016 &   0.5 &  0.003 &   0.2 & -0.003 &   0.1 & -0.000 &   0.1 \\
                                                & Multi-GeV $e$-like               &                    
&  0.003 &   0.1 &  0.002 &   0.1 & -0.013 &   0.4 & -0.000 &   0.1 \\
                                                & Multi-GeV 1-ring $e$-like        &                    
&    3.3 &    13 &  -15.0 &    38 &    5.1 &    27 &    1.1 &    18 \\
                                                & Multi-GeV Multi-ring $e$-like    &                    
&    1.1 &    12 &    2.5 &    11 &   -6.1 &    11 &    3.1 &    12 \\
\multicolumn{3}{l}{Fiducial Volume}                                                                     
&  -0.04 &     2 &   0.08 &     2 &  -0.42 &     2 &   0.40 &     2 \\
Ring separation                                 & $< \val{400}{MeV}$               & $e$-like           
&   1.07 &   2.3 &  -1.09 &   1.3 &   0.79 &   2.3 &   0.05 &   1.6 \\
                                                &                                  & $\mu$-like         
&  0.324 &   0.7 &  -1.93 &   2.3 &   1.03 &     3 &   0.09 &     3 \\
                                                & $> \val{400}{MeV}$               & $e$-like           
&  0.185 &   0.4 &  -1.43 &   1.7 &   0.44 &   1.3 &  -0.03 &     1 \\
                                                &                                  & $\mu$-like         
&  0.324 &   0.7 & -0.588 &   0.7 &  0.205 &   0.6 & -0.018 &   0.6 \\
                                                & Multi-GeV                        & $e$-like           
&   1.71 &   3.7 &  -2.18 &   2.6 &   0.44 &   1.3 &  -0.03 &     1 \\
                                                &                                  & $\mu$-like         
&   0.79 &   1.7 &  -1.43 &   1.7 &   0.34 &     1 &   0.04 &   1.2 \\
                                                & Multi-ring Sub-GeV               & $e$-like           
&  -1.62 &   3.5 &   3.19 &   3.8 &   0.44 &   1.3 &   0.06 &   1.9 \\
                                                &                                  & $\mu$-like         
&  -2.08 &   4.5 &   6.88 &   8.2 &  -0.89 &   2.6 &   0.07 &   2.3 \\
                                                & Multi-ring Multi-GeV             & $e$-like           
&  -1.44 &   3.1 &   1.59 &   1.9 &  -0.38 &   1.1 &  0.027 &   0.9 \\
                                                &                                  & $\mu$-like         
&  -1.90 &   4.1 &  0.671 &   0.8 &  -0.72 &   2.1 &  -0.07 &   2.4 \\
Particle identification (1 ring)                & Sub-GeV                          & $e$-like           
&  0.016 &  0.23 &  0.099 &  0.66 &  0.023 &  0.26 & -0.025 &  0.28 \\
                                                &                                  & $\mu$-like         
& -0.013 &  0.18 & -0.075 &   0.5 & -0.016 &  0.19 &  0.020 &  0.22 \\
                                                & Multi-GeV                        & $e$-like           
&  0.013 &  0.19 &  0.036 &  0.24 &  0.027 &  0.31 & -0.031 &  0.35 \\
                                                &                                  & $\mu$-like         
& -0.013 &  0.19 & -0.039 &  0.26 & -0.026 &   0.3 &  0.031 &  0.35 \\
Particle identification (multi-ring)            & Sub-GeV                          & $e$-like           
&  -0.31 &   3.1 &  -3.39 &     6 &   5.09 &   9.5 &   2.15 &   4.2 \\
                                                &                                  & $\mu$-like         
&  0.066 &  0.66 &   1.45 &   2.5 &  -2.79 &   5.2 &  -0.80 &   1.6 \\
                                                & Multi-GeV                        & $e$-like           
&   0.64 &   6.5 &   5.54 &   9.7 &  -2.63 &   4.9 &   1.71 &   3.3 \\
                                                &                                  & $\mu$-like         
&  -0.29 &   2.9 &  -2.24 &   3.9 &   1.43 &   2.7 &  -0.80 &   1.6 \\
\multicolumn{3}{l}{Energy calibration}                                                                  
&   0.00 &   1.1 &  -0.20 &   1.7 &   0.65 &   2.7 &  -0.36 &   2.3 \\
\multicolumn{3}{l}{Up/down asymmetry energy calibration}                                                
&  0.293 &   0.6 & -0.070 &   0.6 &   0.36 &   1.3 & -0.109 &   0.3 \\
\UP reduction                                   & \multicolumn{2}{l}{Stopping}                          
& -0.185 &   0.7 & -0.131 &   0.7 &  0.111 &   0.7 &  0.126 &   0.5 \\
                                                & Through-going                    &                    
& -0.132 &   0.5 & -0.094 &   0.5 &  0.080 &   0.5 &  0.075 &   0.3 \\
\multicolumn{3}{l}{\UP stopping/through-going separation}                                               
&  0.007 &   0.4 &  0.016 &   0.6 &  0.034 &   0.4 & -0.109 &   0.6 \\
\multicolumn{3}{l}{Energy cut for stopping \UP}                                                         
&  0.085 &   0.9 &   0.11 &   1.3 &   0.87 &     2 &   0.01 &   1.7 \\
\multicolumn{3}{l}{Path length cut for through-going \UP}                                               
&   0.86 &   1.5 &   1.50 &   2.3 &  -0.12 &   2.8 &  -1.87 &   1.5 \\
\multicolumn{3}{l}{Through-going \UP showering separation}                                              
&   3.59 &   3.4 &  -2.84 &   4.4 &   2.35 &   2.4 &  -4.88 &     3 \\
Background subtraction for \UP                  & \multicolumn{2}{l}{Stopping\footnote[1]{The uncertainties in BG subtraction for upward-going muons are only for the most horizontal bin,  $-0.1 < \cos\theta < 0$.}} 
&   10.2 &    16 &   -4.0 &    21 &   -2.2 &    20 &   -6.7 &    17 \\
                                                & Non-showering\footnotemark[1]    &                    
&   -4.0 &    18 &    0.8 &    14 &    0.6 &    24 &    1.8 &    17 \\
                                                & Showering\footnotemark[1]        &                    
&   -7.5 &    18 &  -12.9 &    14 &    2.6 &    24 &    9.6 &    24 \\
\multicolumn{3}{l}{$\nue/\nuebar$ Separation}                                                           
&  -2.67 &   7.2 &   0.08 &   7.9 &  -9.19 &   7.7 &  -4.07 &   6.8 \\
Sub-GeV 1-ring \pizero selection                & \multicolumn{2}{l}{$100 < P_e < \val{250}{MeV/c}$}    
&   3.47 &     9 &    2.9 &    10 &   2.23 &   6.3 &   1.92 &   4.6 \\
                                                & $250 < P_e < \val{400}{MeV/c}$   &                    
&   3.55 &   9.2 &    4.1 &    14 &   1.73 &   4.9 &   1.25 &     3 \\
                                                & $400 < P_e < \val{630}{MeV/c}$   &                    
&    6.1 &    16 &    3.3 &    11 &    8.4 &    24 &    5.6 &    13 \\
                                                & $630 < P_e < \val{1000}{MeV/c}$  &                    
&    5.2 &    14 &    4.8 &    16 &   2.90 &   8.2 &    7.0 &    17 \\
                                                & $1000 < P_e < \val{1330}{MeV/c}$ &                    
&    4.5 &    12 &   2.87 &   9.8 &    3.9 &    11 &    9.9 &    24 \\
\multicolumn{3}{l}{Sub-GeV 2-ring \pizero}                                                              
&   0.31 &   5.6 &  -2.42 &   4.4 &  -1.17 &   5.9 &   1.78 &   5.6 \\
\multicolumn{3}{l}{Decay-e tagging}                                                                     
&   -5.5 &    10 &   -2.7 &    10 &    1.5 &    10 &    1.1 &    10 \\
\multicolumn{3}{l}{Solar Activity}                                                                      
&    0.1 &    20 &   17.2 &    50 &    2.0 &    20 &    0.3 &    10 \\

\hline\hline
\end{tabular}
\caption{
Systematic errors that are independent in SK-I, SK-II, SK-III, and SK-IV. 
The detector uncertainties are determined using control samples like cosmic ray muons and 2-ring $\pi^0$'s, and simulation studies.
Columns labeled `fit' show the best fit value of the systematic error parameter, $\epsilon_j$, in percent and columns labeled $\sigma$ shows the estimated 1-$\sigma$ error size in percent.}
\label{tab:sysc}
\end{table*}

\clearpage

\bibliography{sterile}

\end{document}